%% file: main.tex
\documentclass[10pt,journal]{IEEEtran}

\input{setup/preamble.tex}

\input{setup/info.tex}

\begin{document}

\maketitle
\input{1_abstract.tex}
\input{2_introduction.tex}
\input{3_communication.tex}

\input{4_sensing.tex}

\input{5_computing.tex}

\input{6_conclusion.tex}

\bibliographystyle{IEEEtran}
\bibliography{reference/mybib2}

\end{document}

%% file: setup/preamble.tex
\usepackage{amssymb}
\usepackage{amsmath}
\usepackage{cite}
\usepackage{url}
\usepackage{cite,graphicx,amsmath,amssymb}
\usepackage{fancyhdr}
\usepackage{mdwmath}
\usepackage{mdwtab}
\usepackage{caption}
\usepackage{amsthm}
\usepackage{setspace}
\usepackage{hyperref}
\usepackage{algorithm}
\usepackage{algorithmic}
\usepackage{multirow}
\usepackage{makecell}
\usepackage{mathtools}
\usepackage{subcaption}
\usepackage{bm}
\usepackage{tikz}
\usepackage{xurl}
\usepackage{stfloats}
\usepackage{mathrsfs}

\usepackage{booktabs}
\usepackage{multirow}
\usepackage{siunitx}
\usepackage{balance}

\usepackage{array}
\usepackage{multirow,tabularx}

\hypersetup{colorlinks=true,
linkcolor=blue,
citecolor=blue,      
urlcolor=black,
}

\graphicspath{{./src/img/}}

\newtheorem{theorem}{Theorem}

\newtheorem{lemma}{Lemma}

\newtheorem{corollary}{Corollary}

\allowdisplaybreaks
\usepackage{caption}
\newcommand{\EmptyCircle}{%
	\raisebox{0.75mm}{\tikz[baseline]{
			\draw (0,0) circle (1.5mm);
		}%
}}

\newcommand{\CircleWithDot}{%
	\raisebox{0.75mm}{\tikz[baseline]{
			\draw (0,0) circle (1.5mm);
			\fill (0,0) circle (0.8mm);
		}%
}}

\newcommand{\FilledCircle}{%
	\raisebox{0.75mm}{\tikz[baseline]{
			\fill (0,0) circle (1.5mm);
		}%
}}

%% file: setup/info.tex
\title{Multi-Functional Programmable Metasurfaces for 6G and Beyond}

\author{
        Xu Gan,~\IEEEmembership{Member, IEEE}, Xidong Mu,~\IEEEmembership{Member, IEEE}, Yuanwei Liu,~\IEEEmembership{Fellow, IEEE}, \\ Marco Di Renzo,~\IEEEmembership{Fellow, IEEE}, Josep Miquel Jornet,~\IEEEmembership{Fellow, IEEE}, Nuria Gonz\'alez Prelcic,~\IEEEmembership{Fellow, IEEE}, \\ Arman Shojaeifard,~\IEEEmembership{Senior Member, IEEE}, and Tie Jun Cui,~\IEEEmembership{Fellow, IEEE}

\thanks{X. Gan and Y. Liu are with the Department of Electrical and Electronic Engineering, The University of Hong Kong, Hong Kong (e-mails: \{eee.ganxu, yuanwei\}@hku.hk).}
\thanks{X. Mu is with the Centre for Wireless Innovation (CWI), Queen’s University Belfast, Belfast, BT3 9DT, U.K. (e-mail: x.mu@qub.ac.uk).}
\thanks{M. Di Renzo is with Universit\'e Paris-Saclay, CNRS, CentraleSup\'elec, Laboratoire des Signaux et Syst\`{e}mes, 3 Rue Joliot-Curie, 91192 Gif-sur-Yvette, France (e-mail: marco.di-renzo@universite-paris-saclay.fr), and with King's College London, Centre for Telecommunications Research, Department of Engineering, WC2R 2LS London, United Kingdom (e-mail: marco.di\_renzo@kcl.ac.uk).}
\thanks{J. M. Jornet is with the Department of Electrical and Computer Engineering \& the Institute for the Wireless Internet of Things, Northeastern University, Boston, MA 02115, USA (e-mail: j.jornet@northeastern.edu).}
\thanks{N. G. Prelcic is with the Department of Electrical and Computer Engineering, University of California, San Diego, La Jolla, 92093 CA USA (e-mail: ngprelcic@ucsd.edu).}
\thanks{A. Shojaeifard is with the Department of Research \& Innovation, InterDigital, EC2A 3QR London, U.K. (e-mail: arman.shojaeifard@interdigital.com).}
\thanks{T. J. Cui is with the State Key Laboratory of	Millimeter Waves, Southeast University, Nanjing, 210096, China (e-mail: tjcui@seu.edu.cn).}

\vspace{-2mm}
}

%% file: 1_abstract.tex
\begin{abstract}
    The sixth-generation and beyond (B6G) networks are envisioned to support advanced applications that demand high-speed communication, high-precision sensing, and high-performance computing. To underpin this multi-functional evolution, energy- and cost-efficient programmable metasurfaces (PMs) have emerged as a promising technology for dynamically manipulating electromagnetic waves. This paper provides a comprehensive survey of representative multi-functional PM paradigms, with a specific focus on achieving \emph{full-space communication coverage}, \emph{ubiquitous sensing}, as well as \emph{intelligent signal processing and computing}. i) For simultaneously transmitting and reflecting surfaces (STARS)-enabled full-space communications, we elaborate on their operational protocols and pivotal applications in supporting efficient communications, physical layer security, unmanned aerial vehicle networks, and wireless power transfer. ii) For PM-underpinned ubiquitous sensing, we formulate the signal models for the PM-assisted architecture and systematically characterize its advantages in near-field and cooperative sensing, while transitioning to the PM-enabled transceiver architecture and demonstrating its superior performance in multi-band operations. iii) For advanced signal processing and computing, we explore the novel paradigm of stacked intelligent metasurfaces (SIMs), investigating their implementation in wave-domain analog processing and over-the-air mathematical computing. Finally, we identify key research challenges and envision future directions for multi-functional PMs towards B6G.
\end{abstract}


%% file: 2_introduction.tex
\section{Introduction} 
\IEEEPARstart{E}{merging} advanced applications are driving unprecedented requirements for multi-functional wireless networks, ranging from immersive telepresence requiring multi-gigabit data rates to cooperative unmanned aerial vehicle (UAV) swarms necessitating centimeter-level localization, and city-scale digital twins powered by real-time cloud computing~\cite{clemm2020toward,kanhere2021position,almasan2022network}. These diverse requirements have shaped the vision of sixth-generation and beyond (B6G) networks not merely as an evolution of mobile broadband, but also as a unified framework capable of delivering high-rate communication, high-precision sensing, and high-performance computing\cite{gonzalez2024integrated,letaief2021edge,d20246g}. Responding to these imperatives, the IMT-2030 framework~\cite{ITU-R-M2160-0} has established rigorous performance benchmarks, specifying peak data-rate of up to $200$ Gbit/s, positioning accuracy on the order of centimeters, and connection densities exceeding $10^6$ devices per km${}^2$. To meet these ambitious targets, conventional transceiver architectures have evolved from multiple-input multiple-output (MIMO) to massive MIMO, and now ultra-massive MIMO configurations to fully exploit spatial diversity gains~\cite{bjornson2024towards}. However, these continuous expansions of active transceiver arrays incur prohibitive hardware costs and energy consumption\cite{bjornson2015optimal}, imposing fundamental sustainability limits on the further enhancement of multi-functional system performance.

To circumvent these hardware limitations, programmable metasurfaces (PMs) have emerged as a promising solution to enhance signal propagation without a proportional increase in active radio frequency (RF) chains\cite{shojaeifard2022mimo,basar2019wireless,di2020smart}. Comprising densely integrated sub-wavelength metamaterial elements, PMs enable the precise manipulation of electromagnetic (EM) wave amplitudes and phases through dynamic electrical control. By synthesizing desired wavefront for beam steering, focusing, and interference suppression, PMs effectively render the wireless environment programmable\cite{yang2016programmable,liang2024filtering,fu2020terahertz}. Among various PM implementations, reconfigurable intelligent surfaces (RISs) have attracted significant research attention for their ability to impose adjustable phase shifts on incident signals using energy-effective passive arrays\cite{liu2021reconfigurable,pan2021reconfigurable,huang2019reconfigurable}. Recent investigations have examined the efficacy of RISs across multi-functional wireless systems\cite{9551980,10543050,9785633}. For instance, empirical studies indicate that RISs can yield power gains of $10$-$20$ dB in the office and corridor environments, with up to $35$ dB in outdoor scenarios \cite{9551980}. Furthermore, acting as anchor points and for environmental reconstruction, RISs have been exploited to simultaneously enable high-rate communications and centimeter-level positioning\cite{10543050}. In the domain of mobile-edge computing (MEC), RIS deployment has been shown to significantly enhance computation offloading capacity, delivering an $8.08$ Mb improvement for Internet-of-Things (IoT) devices~\cite{9785633}.

\subsection{Types of Advanced PMs}
\begin{figure*}[t]
	\centering
	\includegraphics[width=0.8\textwidth]{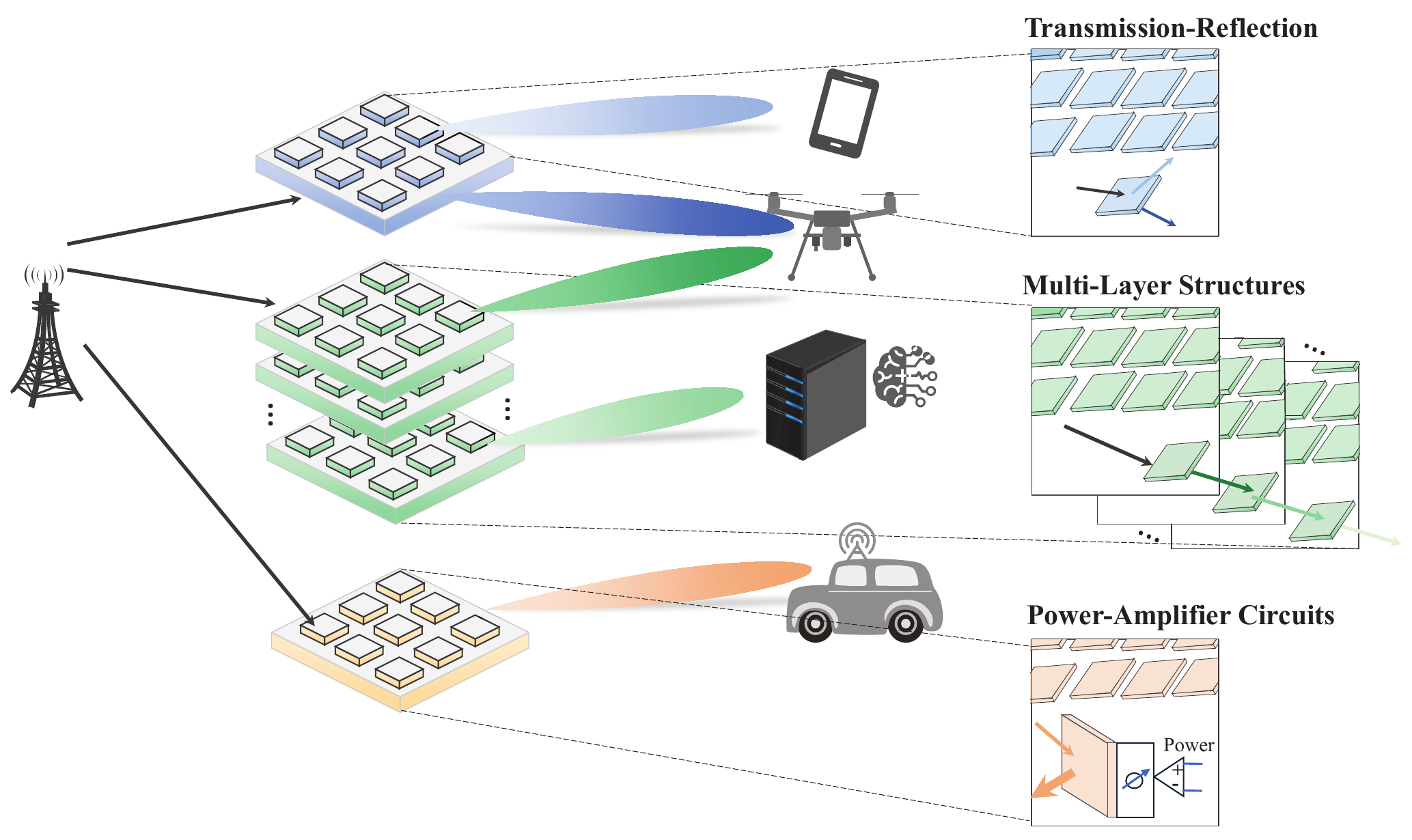}
	\caption{Three architecture of advanced PMs: simultaneous transmission and reflection, multi-layer structures, and power-amplifier circuits.}
	\label{fig:three_PM_architecture}
\end{figure*}

To further enhance signal control flexibility for multi-functional capabilities, the evolution of PMs has branched into three distinct yet complementary paradigms, as illustrated in Fig.~\ref{fig:three_PM_architecture}. These advancements include PMs with simultaneous transmission and reflection capabilities, multi-layer structures, and the integration of power amplifier components.

\subsubsection{Simultaneously Transmitting and Reflecting Surfaces (STARS)}
Conventional reflecting-only RIS implementations are fundamentally constrained by their ``half-space'' operating mechanism. STARSs overcome this limitation by enabling simultaneous signal transmission and reflection, thereby establishing a comprehensive $360^{\circ}$ smart radio environment~\cite{liu2022star,mu2021simultaneously,ahmed2023survey}. This bidirectional EM manipulation capability allows STARSs to extend coverage for users on both sides of the surface, effectively addressing the blind spots inherent in reflecting/transmitting-only RIS architectures and facilitating ubiquitous coverage for full-space multi-functional applications.

\subsubsection{Stacked Intelligent Metasurfaces (SIMs)}
Moving beyond single-layer configurations, SIMs introduce a multi-layer structure that enables deep signal manipulation through independent phase and amplitude control at each cascaded layer\cite{liu2025stacked,an2023stacked,hassan2024efficient}. This densely stacked structure significantly increases the spatial degrees of freedom (DoFs) for beamforming, enabling more precise signal steering for improved communication performance. Moreover, the multi-layer structures are capable of wave-domain analog processing. This unique characteristic facilitates advanced sensing tasks, such as direction-of-arrival (DoA) estimation\cite{an2024two}, and enables direct computation offloading to the metasurface layers, substantially reducing the latency and energy burdens on digital processors.

\subsubsection{Active RISs}
Despite their energy efficiency, conventional passive RISs performance is often bottlenecked by the severe multiplicative fading effect. Active RISs address this physical limitation by integrating amplification components into the surface elements, enabling simultaneous signal reflection and amplification\cite{zhang2022active,zhi2022active,long2021active}. The active power injection at the surface level effectively compensates for the cascaded path loss, maintaining link robustness even in harsh propagation environments. Experimental validations have substantiated the feasibility of this approach, with measured per-element reflection gains exceeding $25$ dB\cite{zhang2022active}.

\begin{table*}[t]
	\centering
	\caption{Comparison of this work with existing magazine, tutorial and survey paper.}
	\label{tab:survey}
	\resizebox{\textwidth}{!}{
		\begin{tabular}{|
				>{\arraybackslash}m{4cm}|
				>{\arraybackslash}m{4cm}|
				*{5}{ >{\centering\arraybackslash}m{0.8cm}| }  
				*{4}{ >{\centering\arraybackslash}m{0.8cm}| } >{\centering\arraybackslash}m{1.5cm}|
			}
			\hline
			\multicolumn{2}{|l|}{\textbf{Existing Works}}  & \multicolumn{5}{c|}{\textbf{Magazine Paper}} &
			\multicolumn{5}{c|}{\textbf{Tutorial and Survey}} \\
			\cline{3-12}
			\multicolumn{2}{|l|}{} & \cite{yang2023reconfigurable} & \cite{chepuri2023integrated} &
			\cite{magbool2024multi}& \cite{li2024reconfigurable} & \cite{chen2024metasurfaces} &
			\cite{liu2021reconfigurable} & \cite{li2022intelligent} & \cite{magbool2025survey} & \cite{an2025emerging}& This Work \\
			\hline
			
			\multicolumn{2}{|l|}{\textbf{Year}} &2023
			& 2023&2024 &2024 &2024 &2021 &2022 &2025
			&2025 & 2025  \\
			\hline

			\multirow{5}{4cm}{\textbf{PMs for Full-Space Communications}} &
			Operational Fundamentals & \EmptyCircle 
			& \FilledCircle  & \FilledCircle &\CircleWithDot &\EmptyCircle  & \FilledCircle   &\CircleWithDot  & \FilledCircle
			& \CircleWithDot & \FilledCircle  \\
			\cline{2-12}
			&
			Efficient Communications & \EmptyCircle 
			& \CircleWithDot  & \FilledCircle  & \CircleWithDot  & \CircleWithDot & \FilledCircle   & \FilledCircle  & \FilledCircle
			& \CircleWithDot  & \FilledCircle  \\
			\cline{2-12}
			&
			Physical Layer Security & \EmptyCircle 
			& \EmptyCircle & \CircleWithDot  & \EmptyCircle  & \CircleWithDot & \FilledCircle  & \EmptyCircle  & \FilledCircle
			& \FilledCircle   & \FilledCircle  \\
			\cline{2-12}
			&
			UAV Networks & \EmptyCircle 
			&\EmptyCircle & \EmptyCircle  &  \FilledCircle   & \EmptyCircle & \FilledCircle   &\EmptyCircle  & \FilledCircle
			& \EmptyCircle  & \FilledCircle  \\
			\cline{2-12}
			&
			Wireless Power Transfer& \EmptyCircle 
			& \EmptyCircle  & \FilledCircle & \EmptyCircle  & \CircleWithDot& \FilledCircle    & \FilledCircle & \EmptyCircle 
			& \EmptyCircle  & \FilledCircle   \\
			\hline
			
			\multirow{5}{4cm}{\textbf{PMs for Ubiquitous Sensing}} &
			Fundamental Architectures& \EmptyCircle 
			& \FilledCircle & \CircleWithDot & \FilledCircle & \FilledCircle & \EmptyCircle& \EmptyCircle  & \FilledCircle
			& \CircleWithDot & \FilledCircle    \\
			\cline{2-12}
			&
			Efficient Sensing & \EmptyCircle 
			& \FilledCircle   & \CircleWithDot &\FilledCircle &  \FilledCircle & \EmptyCircle& \EmptyCircle  &\FilledCircle
			&  \CircleWithDot& \FilledCircle   \\
			\cline{2-12}
			&
			Cooperative Sensing & \EmptyCircle 
			& \FilledCircle  &\EmptyCircle  & \CircleWithDot  & \EmptyCircle & \EmptyCircle &\EmptyCircle  &\CircleWithDot
			& \EmptyCircle & \FilledCircle   \\
			\cline{2-12}
			&
			Near-Field Sensing & \EmptyCircle 
			&\EmptyCircle & \CircleWithDot  & \EmptyCircle  &\EmptyCircle & \EmptyCircle &\EmptyCircle   &  \FilledCircle 
			& \CircleWithDot & \FilledCircle  \\
			\cline{2-12}
			&
			Multi-Band Sensing & \EmptyCircle 
			& \EmptyCircle & \EmptyCircle &\CircleWithDot  &\CircleWithDot & \EmptyCircle &\EmptyCircle  &\CircleWithDot 
			& \EmptyCircle  & \FilledCircle   \\
			\hline
			
			\multirow{3}{4cm}{\textbf{PMs for Intelligent Processing/Computing}} &
			Signal Models & \CircleWithDot 
			& \EmptyCircle &\EmptyCircle &\EmptyCircle  &\EmptyCircle &\CircleWithDot  &\CircleWithDot &\EmptyCircle 
			&\CircleWithDot & \FilledCircle   \\
			\cline{2-12}
			&
			Analog Processing & \FilledCircle 
			& \EmptyCircle  &\EmptyCircle  &\CircleWithDot &\CircleWithDot & \FilledCircle  &\FilledCircle   &\EmptyCircle 
			&  \FilledCircle & \FilledCircle   \\
			\cline{2-12}
			&
			Mathematical Computing &\FilledCircle  
			&\EmptyCircle &\EmptyCircle  &\EmptyCircle  &\EmptyCircle   &\EmptyCircle  &\FilledCircle  &\EmptyCircle 
			& \FilledCircle  & \FilledCircle  \\
			\hline
			
			\multicolumn{12}{|l|}{\EmptyCircle, \CircleWithDot \ and  \FilledCircle \ represent ``not discussed'', ``mentioned but not discussed in detail'' and ``discussed in detail'', respectively.} \\
			\hline
			
		\end{tabular}
	}
\end{table*}

\subsection{Advantages of PMs in Multi-Functional B6G}
Capitalizing on these advanced PM implementations, the resulting smart radio environment becomes capable of supporting the stringent requirements of multi-functional B6G networks. Specifically, the enhanced EM control provided by these advanced PMs translates into three core system-level benefits: extended coverage capability, robust interference mitigation, and efficient wave-domain processing. These advantages are discussed as follows.

\subsubsection{Enhancing Coverage Capability}
A fundamental advantage of PMs is their ability to establish virtual line-of-sight (LoS) links when direct transmission paths are blocked by obstacles\cite{kishk2020exploiting,singh2024performance,dardari2021nlos}, thereby significantly expanding the coverage capability for multi-functional systems. Beyond basic reflection, STARS enables simultaneous transmission and reflection, extending the controllable radio environment from the conventional $180^{\circ}$ half-space to a full $360^{\circ}$ space. This ensures consistent service for users located on both sides of the surface\cite{liu2022star}. Moreover, active RISs integrate amplification components to compensate for the severe multiplicative path loss, significantly boosting the coverage capability in scenarios with deep fading or long-distance transmission\cite{fu2024multi}.

\subsubsection{Interference Suppression in Multi-Functional Integrations}
Once ubiquitous coverage is established, the dense coexistence of multi-functional nodes inevitably leads to severe interference. PMs address this challenge by providing high DoFs for beam reconfiguration. The large aperture of PMs facilitates the generation of highly directional pencil beams\cite{ben2024design}, which establish spatially orthogonal channels. This spatial isolation effectively mitigates inter-user interference and supports efficient orthogonal multiple access (OMA) for independent multi-functional links. Furthermore, PMs can deliberately shape EM wavefronts to enhance signal separation through non-orthogonal multiple access (NOMA) with successive interference cancellation (SIC)\cite{singh2022noma,chauhan2022ris}, allowing for the effective decoupling of communication, sensing, and computing signals. For instance, recent work has demonstrated that PM-assisted frameworks can mitigate mutual cross-functional interference among these three functions through the latency-minimization optimization\cite{wu2025intelligent}.

\subsubsection{Advanced Wave-Domain Signal Processing and Computing}
Beyond manipulating signal propagation, PMs introduce a paradigm shift by performing signal processing directly in the wave domain. This capability allows PMs to perform analog beamforming in the EM field, effectively reducing the reliance on complex beamforming operations that traditionally require extensive computational resources. Furthermore, PMs can be utilized to implement analog over-the-air computing\cite{omam2025holographic,li2022intelligent,shabanpour2024multifunctional}. By mapping mathematical operations onto the EM response of the PM, computational tasks can be executed at the speed of light during signal propagation. These PM architectures significantly reduce the hardware cost and power consumption associated with high-precision analog-to-digital converters (ADCs), digital-to-analog converters (DACs), and power-intensive RF chains in conventional massive MIMO systems.

\subsection{Motivation and Contributions}
In light of the distinct PM evolution and the profound system-level benefits detailed above, PMs have indisputably established themselves as a cornerstone for underpinning the multi-functional B6G networks. Given the rapid development of PM-based technologies, a systematic review of existing PM implementations and their multi-functional applications is essential to guide future multi-functional PM research. While existing magazine papers, tutorials, and surveys have extensively covered specific aspects of PMs~\cite{yang2023reconfigurable,chepuri2023integrated,magbool2024multi,li2024reconfigurable,chen2024metasurfaces,liu2021reconfigurable,li2022intelligent,magbool2025survey,an2025emerging}, they mostly concentrate on single-function designs, lacking a holistic perspective on multi-functional PMs. To highlight the distinctiveness of this paper, Table~\ref{tab:survey} qualitatively compares our contributions with related existing works. Consequently, this paper aims to provide a comprehensive survey of multi-functional PMs, covering fundamental principles, performance analysis methodologies, and advanced system designs for wireless network functionalities. The main contributions of this work are summarized as follows:
\begin{itemize}
	\item  We elaborate on the fundamentals of STARS-enabled full-space communications, detailing their operating protocols and practical architectures. Building on this framework, we explore their application key 6G scenarios, including efficient communications, physical layer security (PLS), UAV, and simultaneous wireless information and power transfer (SWIPT).	
	
	\item We establish the fundamental architectures and basic signal models of PM-underpinned ubiquitous sensing. On this basis, we characterize the advantages of the PM-assisted architecture in near-field and cooperative sensing, while demonstrating the superiority of the PM-enabled transceiver architecture in millimeter-wave (mmWave) and Terahertz (THz) operations.

	\item We explore the emerging paradigm of SIM-enabled intelligent signal processing and computing. By deriving the signal models for multi-layer wave manipulation, we investigate the capability of SIMs to perform wave-domain analog processing and over-the-air mathematical computing, thereby offering a low-latency and energy-efficient alternative to conventional digital processors.

\end{itemize}

\begin{figure}[t]
	\centering
	\includegraphics[width=0.45\textwidth]{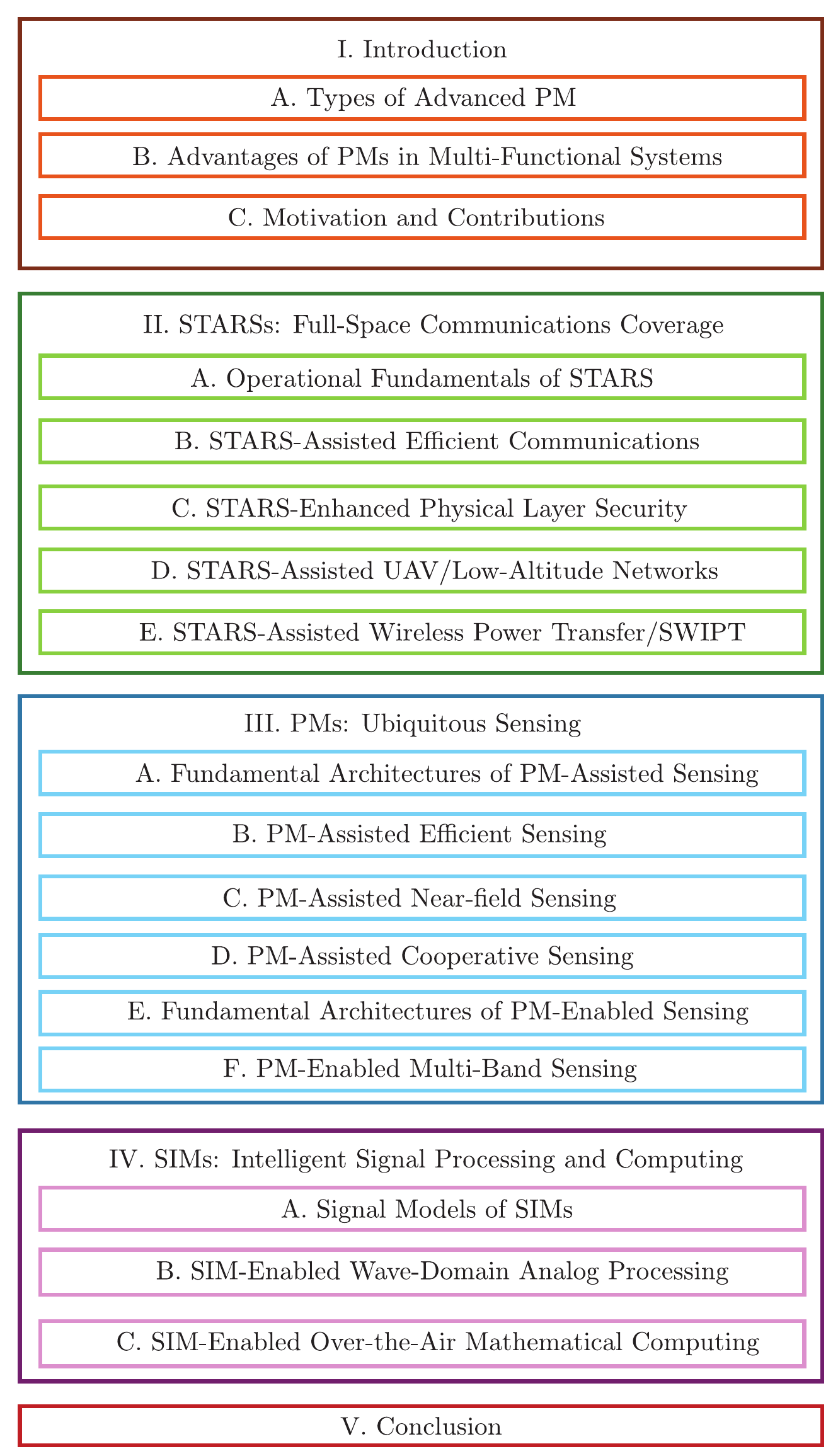}
	\caption{Organization of the survey.}
	\label{fig:organization}
\end{figure}

\begin{table}[htbp]
	\centering
	\caption{List of Acronyms}
	\label{tab:acronyms}
	\resizebox{\columnwidth}{!}{
		\begin{tabular}{|>{\arraybackslash}m{1.4cm}|>{\arraybackslash}m{7.8cm}|}
			\hline
			AF   & Ambiguity function \\ 
			BS   & Base station \\
			CRLB  & Cramér-Rao lower bound \\
			CSI  & Channel state information \\
			DNN & Deep neural network \\
			DoA & Direction-of-arrival \\
			DoF & Degree-of-freedom \\
			ES   & Energy splitting \\
			EM   & Electromagnetic \\
			FMCW & Frequency modulated continuous wave \\
			IOS & Intelligent omini-surface \\
			IoT & Internet-of-thing \\
			IRS & Intelligent reflecting surfaces \\
			ISAC & Integrated sensing and communication \\
			LoS & Line-of-sight \\
			MEC & Mobile-edge computing \\
			MIMO & Multiple-input multiple-output \\
			MmWave & Millimeter-wave \\
			ML & Machine learning \\
			MS & Mode switching \\
			NOMA & Non-orthogonal multiple access \\
			OFDM & Orthogonal frequency division multiplexing\\
			OMA & Orthogonal multiple access \\
			PLS & Physical layer security \\
			PM &  Programmable metasurface\\
			QoS & Quality-of-service \\
			RF & Radio frequency \\
			RIS & Reconfigurable intelligent surface \\
			RHS & Reconfigurable holographic surface \\
			SIC & Successive interference cancellation \\
			SIM & Stacked intelligent metasurface \\
			SINR& Signal-to-interference-plus-noise ratio\\
			STARS & Simultaneously transmitting and reflecting surface \\
			SWIPT & Simultaneous wireless information and power transfer \\
			THz & Terahertz \\
			TRIS & Transmissive RIS \\
			TS & Time switching \\
			UAV & Unmanned aerial vehicle\\
			WPT & Wireless power transfer\\
			\hline
		\end{tabular}
	}
\end{table}

\subsection{Organization}
The remainder of this paper is organized as follows. Section~\ref{sec_communication} provides a comprehensive overview of STARS, detailing its operating protocols and applications in efficient communications, PLS, UAV and low-altitude networks, and SWIPT. Section~\ref{sec_sensing} delves into PM-underpinned ubiquitous sensing, characterizing its architectural principles and performance gains across sensing efficiency, multi-node cooperation, near-field localization, and multi-band operation. Section~\ref{sec_computing} introduces the signal modeling and practical realization of SIM-based wave-domain analog processing and over-the-air computing. Finally, Section~\ref{sec_conclusion} concludes the paper. Fig.~\ref{fig:organization} summarizes the paper organization, and Table~\ref{tab:acronyms} lists the acronyms used.

%% file: 3_communication.tex
\section{STARSs: Full-Space Communication Coverage} \label{sec_communication}
With the aim of achieving full-space coverage with PMs, this section presents the fundamental principles of STARSs, including operational protocols and emerging applications in efficient communications, PLS, UAV networks, and SWIPT systems.

\begin{figure*}[htbp]
	\centering
	\begin{subfigure}[t]{0.32\textwidth}
		\centering
		\includegraphics[height=5cm]{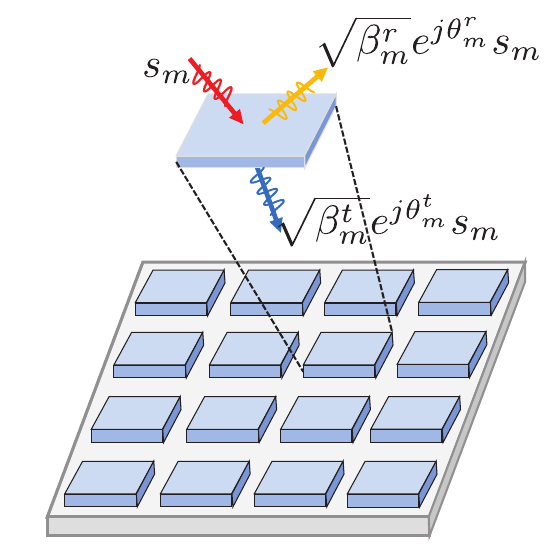}
		\caption{ES.}
		\label{fig:STARS_protocols1}
	\end{subfigure}
	\hfill
	\begin{subfigure}[t]{0.32\textwidth}
		\centering
		\includegraphics[height=5cm]{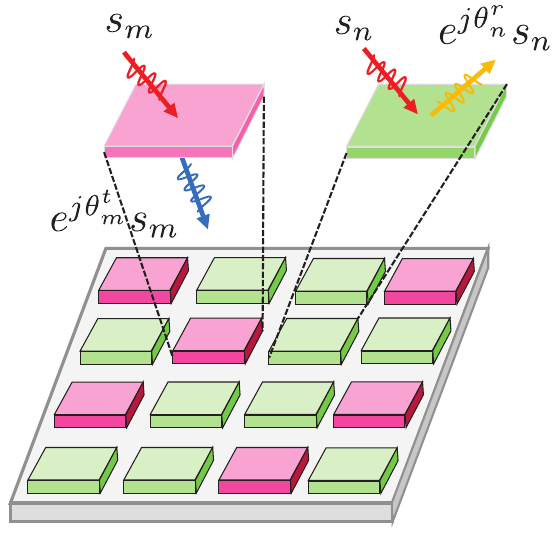}
		\caption{MS.}
		\label{fig:STARS_protocols2}
	\end{subfigure}
	\hfill
	\begin{subfigure}[t]{0.32\textwidth}
		\centering
		\includegraphics[height=5cm]{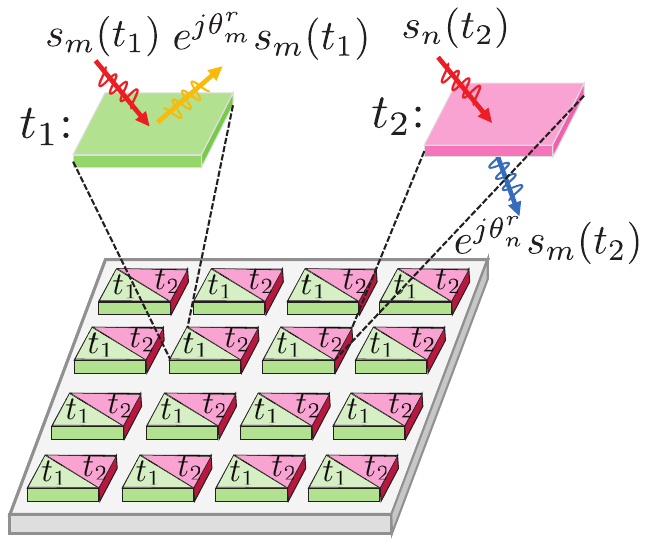}
		\caption{TS.}
		\label{fig:STARS_protocols3}
	\end{subfigure}
	\caption{Three operation protocols for employing STARS.}
	\label{fig:STARS_protocols}
\end{figure*}

\begin{table*}[htbp]
	\centering
	\caption{Summary of characteristics for three STARS operating protocols.}
	\label{tab:STARRIS_protocols}
	\resizebox{\textwidth}{!}{
		\begin{tabular}{|>{\arraybackslash}m{1.3cm}|>{\arraybackslash}m{6cm}|>{\arraybackslash}m{5cm}|>{\arraybackslash}m{8cm}|}
			\hline
			\textbf{Protocol} & \textbf{Optimization DoF} & \textbf{Advantage} & \textbf{Limitation} \\   \hline 
			ES & Amplitudes and phases of $2M$ links & Potentially highest received SINR & High-cost hardware and high-complexity algorithms\\ \hline
			MS & Phases and mode selections of $M$ links & Low-cost hardware & Inability to adjust the transmit-reflect power ratio 	\\ \hline
			TS & Phases and time allocation of $M$ links  & Low-overhead channel estimation 	& Inability to serve users on both sides simultaneously		\\ \hline 
		\end{tabular}
	}
\end{table*}

\subsection{Operational Fundamentals of STARSs}

\subsubsection{Signal Models}
The wireless signal incident on STARS elements undergo transmission, reflection, or simultaneous transmission and reflection based on their EM interaction with the metamaterial structure. As shown in Fig.~\ref{fig:STARS_protocols}. for the $m$-th element receiving incident signal $s_m$, the transmitted signal becomes $e^{j \theta_m^t} s_m$ when only transmission occurs, while the reflected signal equals $e^{j \theta_m^r} s_m$ when only reflection occurs, where $\theta_m^t$ and $\theta_m^r$ represent the phase shifts introduced during transmission and reflection, respectively. For simultaneous transmission and reflection, amplitude-based energy splitting produces transmitted and reflected signals $\sqrt{\beta_m^t} e^{j \theta_m^t} s_m$ and $\sqrt{\beta_m^r} e^{j \theta_m^r} s_m$, with energy conservation requiring $\beta_m^t + \beta_m^r = 1$\cite{mu2021simultaneously}.

\subsubsection{Operating Protocols}
Based on the element-wise signal model of STARS, three operating protocols have been developed in line with practical hardware implementations and functional requirements, namely energy splitting (ES), mode switching (MS), and time switching (TS), as shown in Fig.~\ref{fig:STARS_protocols}. The summary of characteristics for three STARS operating protocols are provided in Table~\ref{tab:STARRIS_protocols} and are detailed elaborated as follows.

$\bullet$ ES:
As illustrated in Fig.~\ref{fig:STARS_protocols1}, all elements simultaneously transmit and reflect incident signals under the ES protocol. The transmission- and reflection-coefficient matrices are given by
\begin{subequations}
	\begin{align}
		& \bm{\Theta}_t^{\text{ES}} = \text{diag}\left( \sqrt{\beta_1^t} e^{j \theta_1^t}, \sqrt{\beta_2^t} e^{j \theta_2^t}, \cdots, \sqrt{\beta_M^t} e^{j \theta_M^t} \right), \\
		& \bm{\Theta}_r^{\text{ES}} = \text{diag}\left( \sqrt{\beta_1^r} e^{j \theta_1^r}, \sqrt{\beta_2^r} e^{j \theta_2^r}, \cdots, \sqrt{\beta_M^r} e^{j \theta_M^r} \right),
	\end{align}
\end{subequations}
where $\beta_m^r,\beta_m^t \in [0,1]$, $\beta_m^r+\beta_m^t =1$, and $\theta_m^r,\theta_m^t \in [0, 2\pi)$, $\forall m$. It can be noticed that, the STARS introduces $M$ transmission paths and $M$ reflection paths. The controllable amplitude and phase shifts across these paths provide substantial DoFs for enhancing signal power and overall system performance.

$\bullet$ MS:
As depicted in Fig.~\ref{fig:STARS_protocols2}, the MS protocol partitions STARS elements into transmission-only and reflection-only groups. The corresponding transmission- and reflecting-coefficient matrices remain the same form as ES:
\begin{subequations}
	\begin{align}
		& \bm{\Theta}_t^{\text{MS}} = \text{diag}\left( \sqrt{\beta_1^t} e^{j \theta_1^t}, \sqrt{\beta_2^t} e^{j \theta_2^t}, \cdots, \sqrt{\beta_M^t} e^{j \theta_M^t} \right), \\
		& \bm{\Theta}_r^{\text{MS}} = \text{diag}\left( \sqrt{\beta_1^r} e^{j \theta_1^r}, \sqrt{\beta_2^r} e^{j \theta_2^r}, \cdots, \sqrt{\beta_M^r} e^{j \theta_M^r} \right),
	\end{align}
\end{subequations}
but with binary amplitude coefficients $\beta_m^r,\beta_m^t \in \{0,1\}$ satisfying $\beta_m^r + \beta_m^t = 1$ and phase-shift values $\theta_m^r,\theta_m^t \in [-\pi, \pi)$, $\forall m$. The MS protocol constitutes a special case of ES with discrete amplitude states. While this configuration leads to reduced DoFs, it significantly lowers hardware implementation complexity and maintains a practical performance balance.

$\bullet$ TS:
The TS protocol alternates all elements between transmission and reflection modes across different time slots as shown in Fig.~\ref{fig:STARS_protocols3}. The resultant transmission- and reflection-coefficient matrices become:
\begin{subequations}
	\begin{align}
		& \bm{\Theta}_t^{\text{TS}} = \Gamma(t) \text{diag}\left(  e^{j \theta_1^t}, e^{j \theta_2^t}, \cdots, e^{j \theta_M^t} \right), \\
		& \bm{\Theta}_r^{\text{TS}} = \left(1-\Gamma(t) \right)\text{diag}\left( e^{j \theta_1^r},  e^{j \theta_2^r}, \cdots,  e^{j \theta_M^r} \right).
	\end{align}
\end{subequations}
Then, we can use a binary variable $\Gamma(t) \in \{0,1\}$ to represent the ``on-off'' states of transmission and reflection at the $t$-th time slot. Although dynamic mode selection simplifies the beamforming design by focusing coverage on a single half-space, it introduces high switching overhead and risks service interruptions when the operational mode mismatches the transceiver locations.

\subsection{STARS-Assisted Efficient Communications}
\begin{figure}[t]
	\centering
	\includegraphics[width=0.9\linewidth]{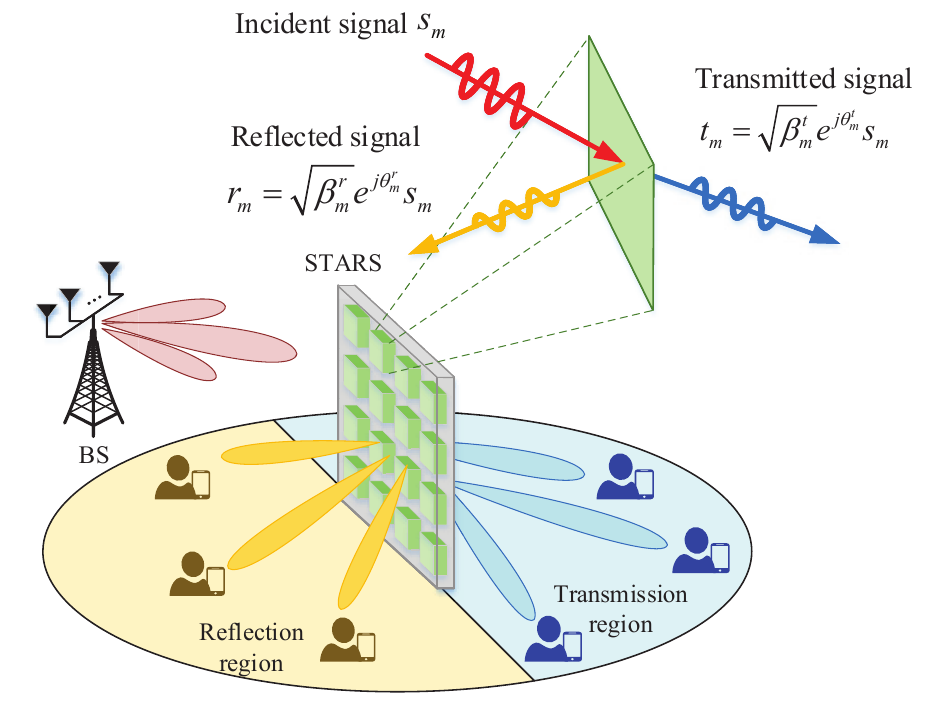}
	\caption{Illustration of STARS-assisted full-space communications.}
	\label{STARS_communication}
\end{figure}

As illustrated in Fig.~\ref{STARS_communication}, STARS enables simultaneous service provision to users in both transmission and reflection regions. This capability has motivated new analytical frameworks and beamforming designs for coverage enhancement, interference management, and spectral-energy efficiency optimization across three operating protocols, as systematically compared in Table~\ref{tab:STARS_communications}.

\begin{table*}[]
	\centering
	\caption{Summary of existing works on the STARS-assisted communications}
	\label{tab:STARS_communications}
	\resizebox{\textwidth}{!}{
		\begin{tabular}{|>{\arraybackslash}m{3.5cm}|>{\arraybackslash}m{1cm}|>{\arraybackslash}m{3cm}|>{\arraybackslash}m{2cm}|>{\arraybackslash}m{10cm}|}
			\hline
			\textbf{Performance Metrics}                               & \textbf{Ref.}                              &        \textbf{Protocols}                &  \textbf{Scenarios}                                                                                     & \textbf{Analysis/Optimization} \\ \hline
			\multirow{2}{*}{Coverage Probability}            & \cite{10685065}     & ES, TS and MS                   &  SIOC & Derive theoretical results for transmission coverage and capacity                      \\ \cline{2-5} 
			& \cite{wu2021coverage}       & ES            &  NOMA/OMA                                                                                    &  Joint optimization designs to maximize sum coverage range                  \\ \hline
			\multirow{2}{*}{Mutual Interference}         & \cite{khel2023analytical}   & ES and MS            &  THz                                                                                                   &  Derive statistical distributions of interference signals            \\ \cline{2-5} 
			& \cite{li2024star}      & ES                 &  NOMA                                                                                                    &  Joint optimization designs to enhance signal and mitigate interference                  \\ \hline
			\multirow{2}{*}{Spectral Efficiency}& \cite{papazafeiropoulos2023cooperative} & ES & mMIMO                                                                                                    & Derive closed-form downlink achievable sum spectral efficiency            \\ \cline{2-5} 
			& \cite{katwe2024spectrally}       &  ES, TS and MS      &  URLLC                                                                                                    & Joint optimization designs to maximize spectral efficiency                \\ \hline
			\multirow{2}{*}{Energy Efficiency}  & \cite{fang2022energy}  &        ES          & NOMA                                                                                                    &  Joint optimization designs to maximize energy efficiency                    \\ \cline{2-5} 
			& \cite{alishahi2025efficient}   & ES and TS         &  WPT-FL                                                                                                    & Optimize STARS modes to minimize energy consumption                    \\ \hline
	\end{tabular}	}
\end{table*}

\begin{itemize}
	\item \textbf{Coverage Probability:} The authors of \cite{10685065} developed a stochastic geometry framework for STARS-enabled simultaneous indoor-and-outdoor communications (SIOC), deriving analytical expressions for transmission coverage probability and capacity that confirmed significant coverage improvements over conventional RIS architectures. For coverage optimization, the authors of \cite{wu2021coverage} formulated sum coverage range maximization problems for both NOMA and OMA systems through joint optimization of transmission and reflection coefficients. Their proposed algorithms achieved nearly doubled coverage ranges compared to conventional RIS implementations across both multiple access schemes.
	
	\item \textbf{Mutual Interference:} To further analyze the potential of STARS for interference mitigation, the authors of \cite{khel2023analytical} derived tractable moment generating functions for interference signals using Laguerre expansions, enabling statistical characterization and subsequent analysis of capacity, outage probability, and symbol error rate. Their results revealed that inter-user interference fundamentally limits high signal-to-interference-plus-noise ratio (SINR) performance, highlighting the necessity of interference-aware STARS designs. Building on this insight, \cite{li2024star} proposed a simultaneous signal enhancement and interference mitigation framework that jointly optimized base station precoding, STARS phase shifts, and user power allocation, demonstrating superior system performance compared to signal-enhancement-only or interference-mitigation-only baseline approaches.
	
	\item \textbf{Spectral Efficiency:} In \cite{papazafeiropoulos2023cooperative}, the authors derived closed-form deterministic expressions for downlink sum spectral efficiency in cooperative RIS-STARS architectures based on large-scale statistics, accounting for practical channel estimation errors. These expressions enabled efficient gradient-based optimization of STARS configurations. As for ultra-reliable low-latency communication (URLLC) communications, the authors of \cite{katwe2024spectrally} developed a spectral-efficient design through joint robust beamforming optimization at both base station and STARS, achieving $10\%$-$15\%$ spectral efficiency improvement over conventional reflection-only RIS.
	
	\item \textbf{Energy Efficiency:} Energy efficiency optimizations have advanced through alternating optimization frameworks~\cite{fang2022energy} that jointly design base station transmit beamforming and STARS element phases for NOMA systems. Further extending this direction, the authors of \cite{alishahi2025efficient} formulated non-convex energy minimization problems for joint uplink-downlink optimization under both ES and TS protocols, enabling sustainable low-latency wireless power transfer (WPT)-federated learning (FL). 
\end{itemize}

\subsection{STARS-Enhanced PLS}
\begin{figure}[t]
	\centering
	\includegraphics[width=0.9\linewidth]{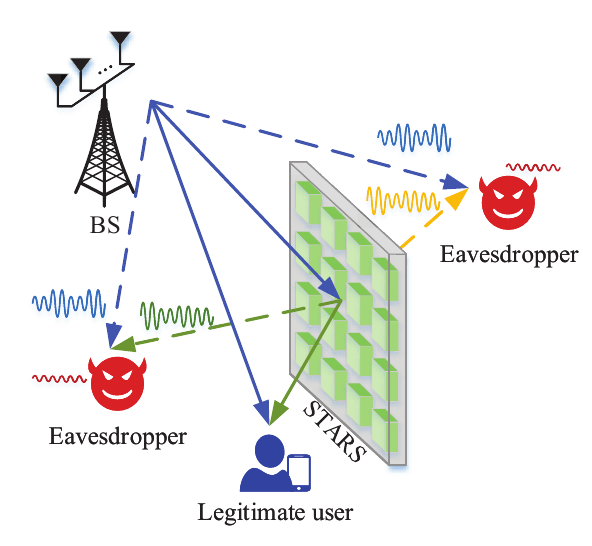}
	\caption{Illustration of STARS for PLS enhancement.}
	\label{fig:STARS_PLS}
\end{figure}

The inherent broadcast characteristics of wireless communications present fundamental challenges to transmission security and stability, as signals remain vulnerable to eavesdropping and jamming attacks.  Deploying STARS has emerged as an effective solution  that enhances channel quality for legitimate users while simultaneously degrading channel conditions for both eavesdroppers and jammers, as illustrated in Fig.~\ref{fig:STARS_PLS}. Therefore, the careful design of STARS has contributed markedly to improving PLS, which generally rely on accurate CSI of communication users and eavesdroppers. Current research on STARS-enhanced PLS can be systematically categorized according to the availability of eavesdropper CSI, encompassing scenarios with perfect CSI, imperfect CSI, statistical CSI, and complete absence of CSI, as comprehensively summarized in Table~\ref{tab:STARS_security}.

Under the assumption of perfect eavesdropper CSI availability, several studies have developed advanced security frameworks\cite{wen2024star,liang2025covert,lin2025star}. The authors of \cite{wen2024star} considered a scenario where the eavesdropper functions as a legitimate system user, enabling perfect CSI acquisition. Then, the authors proposed a novel secure communication scheme where a full-duplex legitimate receiver transmits jamming signals with STARS assistance, maximizing secrecy capacity through joint optimization of full-duplex beamforming and STARS configuration. Similarly leveraging perfect CSI, \cite{liang2025covert} derived analytical expressions for both communication outage probability and secrecy outage probability in uplink NOMA systems with semi-grant-free transmission, quantitatively evaluating the security benefits provided by STARS deployment. Further extending this direction, the authors of \cite{lin2025star} investigated STARS empowered wireless surveillance systems, designing cooperative jamming strategies with perfect information to disrupt two-way suspicious communications while leveraging full-space coverage advantages.

Recognizing the practical difficulty in obtaining accurate CSI for unauthorized nodes such as eavesdroppers and jammers, substantial research has focused on robust beamforming designs under imperfect CSI conditions\cite{zhou2023robust,pala2025robust,xiao2024star,xiao2025robust}. The authors of \cite{zhou2023robust} developed a robust beamforming framework that minimizes access point transmit power while satisfying both secrecy rate and quality of service requirements using imperfect eavesdropper CSI. Building upon this approach, \cite{pala2025robust} addressed worst-case sum secrecy rate maximization under transmit power constraints and quality of service requirements. For scenarios where only statistical CSI of eavesdropper channels is available, the authors of \cite{xiao2024star} derived closed-form expressions for secrecy rate lower bounds in joint PLS and covert communication systems. This theoretical framework was further extended in \cite{xiao2025robust} to account for eavesdropper location uncertainty across both reflection and transmission regions.

In complete absence of eavesdropper channel information, researchers have developed alternative analytical frameworks and system design methodologies\cite{xie2023physical,shang2024enhanced}. Specifically, the authors of \cite{xie2023physical} employed stochastic geometry tools to characterize secrecy outage probability for NOMA users without eavesdropper CSI. Additionally, the authors of \cite{shang2024enhanced} designed continuous artificial noise transmission from the null space of legitimate channels, enhancing security without requiring any knowledge of illegitimate channels in intelligent automated transportation (IAT) systems.

\begin{table*}[]
	\centering
	\caption{Summary of existing works on STARS-enhanced PLS across different CSI types.}
	\label{tab:STARS_security}
	\resizebox{\textwidth}{!}{
		\begin{tabular}{|>{\arraybackslash}m{2.5cm}|>{\arraybackslash}m{1cm}|>{\arraybackslash}m{3cm}|>{\arraybackslash}m{12cm}|}
			\hline
			\textbf{CSI Type}                         & \textbf{Ref.}               & \textbf{Scenario}                                        & \textbf{Analysis/Optimization}                                                                                              \\ \hline
			\multirow{4}{*}{Perfect CSI}     & \cite{wen2024star}       & Full duplex                                 & Jointly optimize FD beamforming and STARS                             \\ \cline{2-4} 
			& \cite{liang2025covert}   &  Semi-grant-free                        & Derive secrecy outage probability under user selection scheme                                                  \\ \cline{2-4} 
			& \cite{lin2025star}       & Surveillance system                   & Derive closed-form surveillance success probabilities                 \\ \cline{2-4} 
			& \cite{dong2024star}      & MIMO                                         & Joint optimization designs  to maximize the secrecy rate            \\ \hline
			\multirow{3}{*}{Imperfect CSI}   & \cite{zhou2023robust}    & PLS                                             & Robust beamforming designs for anti-jamming and anti-eavesdropping \\ \cline{2-4} 
			& \cite{pala2025robust}    & DRL                                          & Robust beamforming designs to maximize the worst-case sum secrecy rate                                             \\ \cline{2-4} 
			& \cite{ye2025physical}    &  IoT                                         & Joint optimization designs to maximize sum rate constrained by Eve' leakage rate  \\ \hline
			\multirow{2}{*}{Statistical CSI} & \cite{xiao2024star}      & MmWave                                       & Derive closed-form expressions for wardens' minimum detection error probability                                  \\ \cline{2-4} 
			& \cite{xiao2025robust}    & PLS                                             & Derive average security rate under the uncertainty of eavesdropper's location        \\ \hline
			\multirow{2}{*}{No CSI}          & \cite{xie2023physical}   & NOMA                                         & Derive the secrecy outage probability and average secrecy capacity                \\ \cline{2-4} 
			& \cite{shang2024enhanced} & IAT & Propose the security index modulation to improve security performance                             \\ \hline
	\end{tabular}}
\end{table*}

\subsection{STARS-Assisted UAV/Low-Altitude Networks}

\begin{figure*}[htbp]
	\centering
	\begin{subfigure}[t]{0.31\textwidth}
		\centering
		\includegraphics[height=4.1cm]{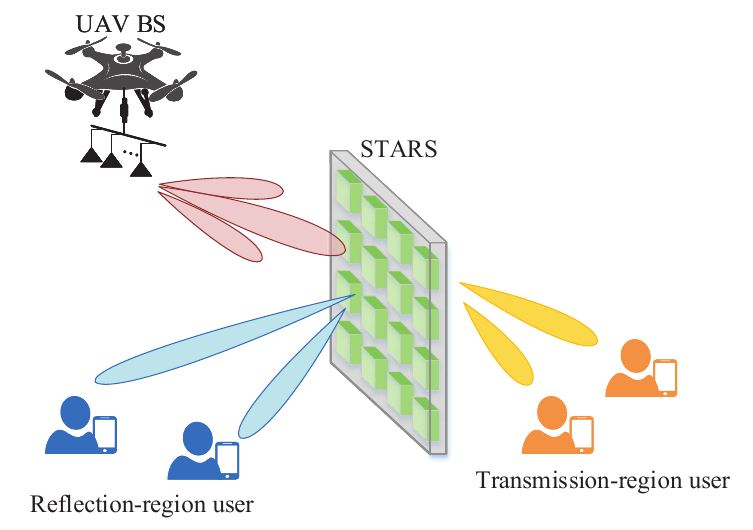}
		\caption{STARS-assisted UAV-BS communications.}
		\label{fig:STARS_UAV1}
	\end{subfigure}
	\hfill
	\begin{subfigure}[t]{0.34\textwidth}
		\centering
		\includegraphics[height=4.1cm]{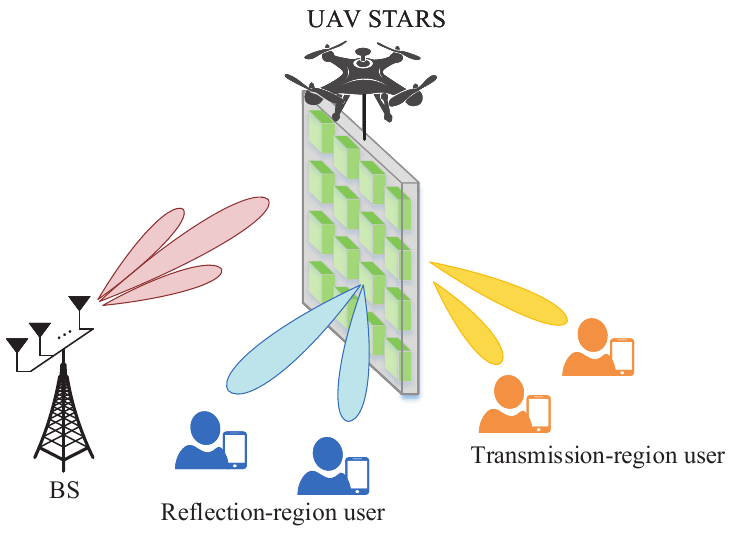}
		\caption{UAV-mounted STARS-assisted communications.}
		\label{fig:STARS_UAV2}
	\end{subfigure}
	\hfill
	\begin{subfigure}[t]{0.31\textwidth}
		\centering
		\includegraphics[height=4.1cm]{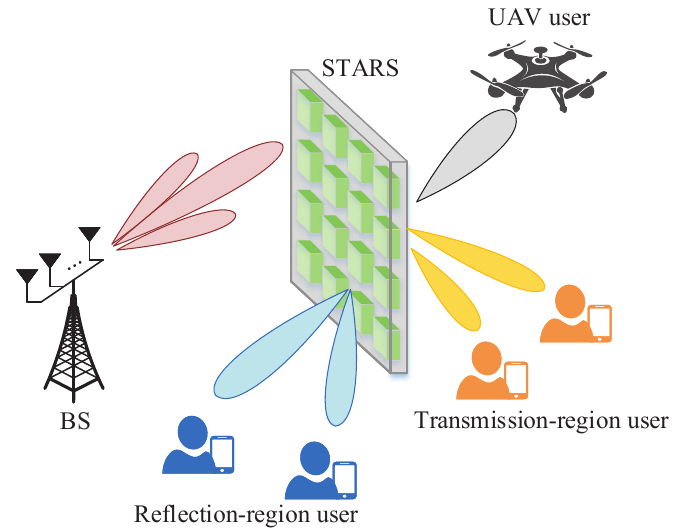}
		\caption{STARS-assisted UAV-user communications.}
		\label{fig:STARS_UAV3}
	\end{subfigure}
	\caption{Illustration of three STARS applications in UAV communications.}
	\label{fig:STARS_UAV}
\end{figure*}

The expanding low-altitude economy demands advanced wireless infrastructure capable of supporting UAVs in highly dynamic aerial environments\cite{song2025trustworthy}. Realizing intelligent UAV services represents a promising direction for future wireless systems, where cost-effective STARS offer significant advantages over conventional active communication components. As depicted in Fig.~\ref{fig:STARS_UAV}, UAV applications within STARS assisted networks can be categorized into three primary architectural roles according to their functional positioning: UAV BS, UAV-mounted STARS, and UAV user, as systematically summarized in Table~\ref{tab:STARS_UAV}.

The flexible deployment capabilities of UAVs enable their utilization as mobile BSs equipped with active transmit antennas, establishing temporary communication hotspots in conjunction with STARS when fixed infrastructure is unavailable. Initial investigations in \cite{zhang2022joint} and \cite{zhao2022simultaneously} formulated sum-rate maximization problems through joint optimization of UAV trajectories, active beamforming parameters, and STARS passive beamforming configurations, demonstrating substantial performance improvements through coordinated design. This framework was extended to multiple STARS deployments in~\cite{lei2023noma}, where a UAV BS served multiple users via NOMA technology, with users typically connecting to their nearest STARS. The challenging throughput maximization problem was effectively addressed using a Lagrange reward constrained proximal policy optimization algorithm. Multi-UAV coordination has further emerged as a critical evolution for low-altitude networks, with \cite{deng2025energy} deploying multiple UAV BSs to serve user equipment in obstructed urban environments assisted by multiple STARS, proposing a novel framework to maximize both energy efficiency and total throughput. More complex multi-functional networks were investigated in \cite{li2024energy}, where multiple UAV BSs and STARS collaboratively provided integrated communication and wireless charging services through a distributed scheduling algorithm that jointly optimized UAV trajectories, association variables, charging time allocation, and STARS coefficient matrices to maximize system energy efficiency.

Mounting STARS on UAV platforms significantly extends communication coverage areas for existing BSs. The authors of \cite{zhao2025aerial} deployed an active STARS on a UAV to enhance connectivity between BSs and IoT devices while supporting multiple devices through NOMA, maximizing system sum rate through joint optimization of active STARS beamforming, UAV trajectory, and power allocation. This approach was adapted for security applications in \cite{guo2023secure}, where a UAV-mounted STARS improved secrecy energy efficiency by protecting transmissions against eavesdroppers. The operational flexibility of UAV-mounted STARS further enabled integration with satellite networks, as demonstrated in \cite{lukito2024integrated} through coordinated optimization of UAV flight paths, STARS phase shifts, and device power allocation, achieving enhanced energy efficiency for low-Earth orbit (LEO) satellites with $67\%$ higher sum rates compared to conventional approaches. For disaster response scenarios with unavailable terrestrial infrastructure,the authors of \cite{khan2025efficient} established a long-distance relay system where observation UAVs collected user information, forwarded through UAV-mounted STARS to emergency BSs, significantly improving video streaming utility through optimized UAV positioning and resource allocation. In contrast, research on UAVs functioning as communication users served by both BSs and STARS remains limited, with UAVs more frequently investigated as sensing targets as demonstrated in~\cite{zhou2024near} and \cite{cai2025energy} , where joint BS and STARS optimization schemes simultaneously enhanced communication services while improving sensing performance for moving UAV targets.

\begin{table*}[t]
	\centering
	\caption{Summary of existing works on STARS-assisted UAV communications.}
	\label{tab:STARS_UAV}
	\resizebox{\textwidth}{!}{
		\begin{tabular}{|>{\arraybackslash}m{2.5cm}|>{\arraybackslash}m{1cm}|>{\arraybackslash}m{4.5cm}|>{\arraybackslash}m{11cm}|}
			\hline
			\textbf{UAV Type}                          & \textbf{Ref.}                   & \textbf{Scenario}                                                                                                 & \textbf{Techniques}                                                                                                                                                                              \\ \hline
			\multirow{4}{2cm}{UAV BS}          
			& \cite{zhao2022simultaneously} & One UAV BS & Joint optimization designs to maximize system sum rate                                                                      \\ \cline{2-4} 
			& \cite{lei2023noma}            & One UAV BS              & Propose the optimization algorithm to maximize system throughput                                                                                                                                                       \\ \cline{2-4} 
			& \cite{deng2025energy}         & Multiple UAV BSs                                & Centralized training to maximize energy efficiency and throughput                                                                                       \\ \cline{2-4} 
			& \cite{li2024energy}           & Multiple UAV BSs                  & Propose a distributed scheduling algorithm to maximize energy efficiency                                                                                                         \\ \hline
			\multirow{4}{2.5cm}{UAV-mounted STARS}  & \cite{zhao2025aerial}         & One UAV-mounted STARS                                                                    & Joint optimization designs to maximize system sum rate                                                                  \\ \cline{2-4} 
			& \cite{guo2023secure}          & One UAV-mounted STARS                   & Joint optimization designs to maximize secrecy energy efficiency                                                                     \\ \cline{2-4} 
			& \cite{lukito2024integrated}   &  One UAV-mounted STARS                                     & Joint optimization designs to maximize energy efficiency                                                     \\ \cline{2-4} 
			& \cite{khan2025efficient}      & One UAV-mounted STARS                                    & Joint optimization designs to augment the average video streaming utility                         \\ \hline
			\multirow{2}{2cm}{UAV User}                                          & \cite{zhou2024near}           & One UAV target                                         & Joint optimization designs to minimize CRLBs of sensing the UAV      \\ \cline{2-4} 
			& \cite{cai2025energy}          & Multiple UAV targets                                    & Joint optimization designs to minimize average power consumption
			\\ \hline
	\end{tabular}}
\end{table*}

\subsection{STARS-Assisted WPT/SWIPT}
\begin{figure}[t]
	\centering
	\includegraphics[width=0.8\linewidth]{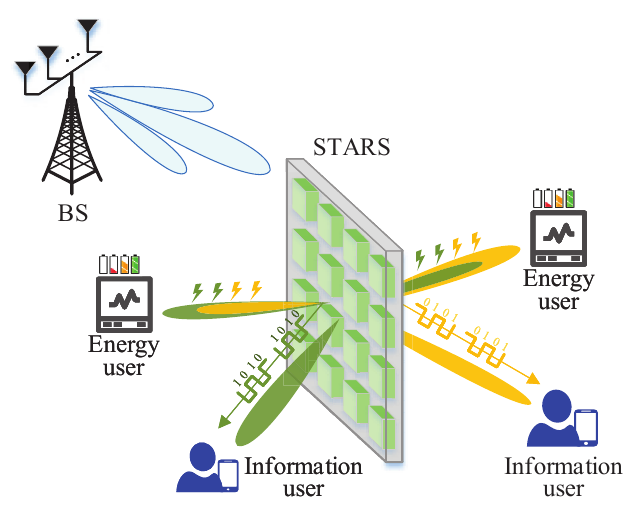}
	\caption{Illustration of STARS for SWIPT.}
	\label{fig:STARS_SWIPT}
\end{figure}

The integration of SWIPT with STARS is anticipated to deliver dual benefits of enhanced communication rates and improved power harvesting efficiency, as illustrated in Fig.~\ref{fig:STARS_SWIPT} and summarized in Table~\ref{tab:STARS_SWIPT}. This synergistic integration leverages the fundamental characteristic that both functionalities utilize EM waves for information and power transfer, while STARS provides unprecedented control over EM wave propagation. To provide a comprehensive overview, existing literature on STARS-assisted SWIPT frameworks can be broadly categorized into passive and active architectures for information receiver (IR) and energy receiver (ER).

\begin{table*}[]
	\centering
	\caption{Summary of existing works on STARS-assisted wireless power transfer.}
	\label{tab:STARS_SWIPT}
	\resizebox{\textwidth}{!}{
		\begin{tabular}{|>{\arraybackslash}m{3cm}|>{\arraybackslash}m{1cm}|>{\arraybackslash}m{6.5cm}|>{\arraybackslash}m{11cm}|}
			\hline
			\textbf{STARS Type} &
			\textbf{Ref.} &
			\textbf{Scenarios} &
			\textbf{Analysis/Optimization} \\ \hline
			\multirow{6}{3cm}{Passive STARS} &
			\cite{xie2024star} &
			Reflection IRs, transmission ERs &
			Derive power and information outage probabilities \\ \cline{2-4} 
			&
			\cite{yaswanth2024toward} &
			Reflection-transmission IRs, reflection ERs &
			Joint optimization to minimize transmit power \\ \cline{2-4} 
			&
			\cite{zhu2023robust} &
			Reflection and transmition IRs and ERs  &
			Resource allcation to maximize the minimum data rate and harvested power \\ \cline{2-4} 
			&
			\cite{luo2024robust} &
			Reflection IR, transmission ER &
			Joint optimization of BS transmit beamforming and artificial noise \\ \cline{2-4} 
			&
			\cite{chen2025star} &
			Far-field IRs, near-field ERs &
			Joint optimization to maximize the weighted harvested power \\ \cline{2-4} 
			&
			\cite{xie2025simultaneous} &
			Indoor IRs, outdoor ERs &
			Design fly-hover-broadcast and path discretization protocols \\ \hline
			\multirow{4}{3cm}{Active STARS} &
			\cite{zhu2025star} &
			Reflection and transmition IRs and ERs &
			Joint optimization to minimize power consumption \\ \cline{2-4} 
			&
			\cite{yang2024joint} &
			Reflection and transmission IRs and ERs &
			Achieve an optimal trade-off between data rate and energy efficiency \\ \cline{2-4} 
			&
			\cite{gao2024power} &
			Transmition IRs, reflection ERs &
			Maximize spectral efficiency and reduce overall system energy consumption \\ \cline{2-4} 
			&
			\cite{faramarzi2025energy} &
			Reflection and transmition IRs and ERs &
			Joint optimization to maximize energy efficiency \\ \hline
	\end{tabular}}
\end{table*}

Exploiting the passive STARs, the authors of \cite{xie2024star} investigated a system where reflection and transmission users harvested and stored energy in buffers as supplementary power sources, with partial utilization for uplink transmissions. Within this STARS assisted SWIPT framework, the authors derived closed-form expressions for power outage probability, information outage probability, sum throughput, and joint outage probability over Nakagami-m fading channels. Their theoretical analysis and simulation results demonstrated that the proposed system achieved superior performance compared to conventional RIS schemes. Building on this foundation, the authors of \cite{yaswanth2024toward} minimized total power consumption through joint beamforming design at the BS and STARS while satisfying minimum rate requirements and energy harvesting thresholds. Recognizing that optimization performance depends on CSI quality, \cite{zhu2023robust} and \cite{luo2024robust} developed robust design methods accounting for CSI errors. Specifically, the authors of \cite{luo2024robust} examined two CSI error models, considering both bounded errors and statistical errors, while studying robust SWIPT performance through joint optimization of transmit beamforming, artificial noise, and STARS amplitude and phase configurations. Meanwhile, the authors of \cite{zhu2023robust} simultaneously maximized the minimum data rate and minimum harvested power to investigate fundamental rate-energy trade-offs, revealing that channel uncertainty affected information users more significantly than energy users. Further advancing this research direction, the authors of \cite{chen2025star} explored a STARS assisted hybrid far-field and near-field SWIPT framework where energy users resided in the near-field region while information users operated in the far-field region. This study maximized the weighted harvested power for near-field energy users subject to quality constraints for far-field communication users, demonstrating that near-field beamforming concentrated energy to substantially improve harvesting performance. Additionally, when STARS was deployed on building surfaces, autonomous aerial vehicles could radiate energy-carrying information signals to multiple outdoor energy receivers and indoor information receivers without entering restricted indoor airspace\cite{xie2025simultaneous}.

To address the multiplicative fading effect inherent in passive elements, active STARS implementations have been proposed to provide signal amplification with independent phase control, thereby significantly enhancing SWIPT performance~\cite{zhu2025star,yang2024joint,gao2024power,faramarzi2025energy}. The authors of \cite{zhu2025star} conducted a comprehensive comparison between active and passive STARS configurations, formulating a power consumption minimization problem through joint beamforming design at the access point and STARS while satisfying QoS requirements. Their findings revealed that active STARS achieved superior performance with smaller aperture sizes, though this advantage gradually diminished with increasing aperture dimensions. Further advancing this research direction, the authors of \cite{yang2024joint} investigated sum achievable rate maximization in active STARS assisted SWIPT systems through joint optimization of base station transmit beamforming and the STARS coefficient and amplification matrices, demonstrating a $16.5\%$ performance improvement over conventional active RIS designs and a $114\%$ enhancement compared to passive STARS implementations. Subsequently, the authors of \cite{gao2024power} expanded the optimization objectives to simultaneously maximize spectrum efficiency while reducing overall system energy consumption, while \cite{faramarzi2025energy} developed an energy efficiency maximization framework that specifically addressed substantial energy demands in SWIPT systems through optimized power output reduction at the BS, confirming that active STARS configurations could achieve lower energy consumption while maintaining high energy efficiency.

%% file: 4_sensing.tex
\section{PM-Underpinned Ubiquitous Sensing}\label{sec_sensing}
After discussing the fundamentals and emerging applications of STARS for achieving full-space coverage, this section explores PM-underpinned ubiquitous sensing. We first introduces the PM-assisted sensing architecture, where PMs modulate the wireless channel to facilitate environmental sensing. Within this architecture, we systematically characterize the performance gains in sensing efficiency, as well as the enhancements in near-field localization and multi-node cooperation. Furthermore, we advance to the PM-enabled sensing architecture, which employs the PM directly as a transmitter or receiver. Under this transceiver-based framework, we analyze the superior performance of PMs in multi-band operation, specifically mmWave and THz spectrums.

\subsection{Fundamental Architectures of PM-Assisted Sensing}

\begin{figure}[htbp]
	\centering
	\begin{subfigure}[b]{0.24\textwidth}
		\centering
		\includegraphics[height=4cm]{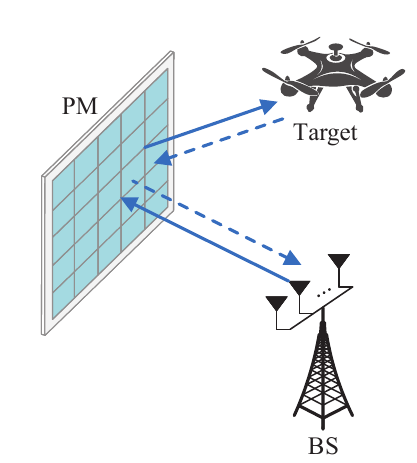}
		\caption{PM-assisted monostatic sensing.}
	\end{subfigure}
	\hfill
	\begin{subfigure}[b]{0.24\textwidth}
		\centering
		\includegraphics[height=3.2cm]{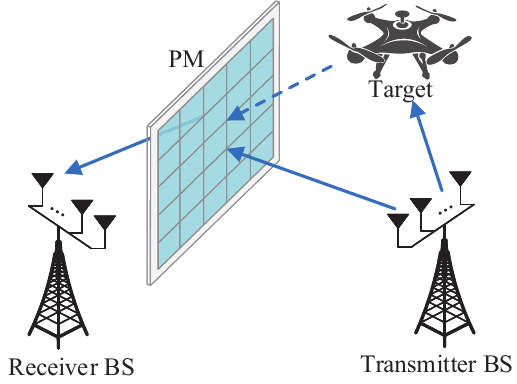}
		\caption{PM-assisted bistatic sensing.}
	\end{subfigure}
	
	\vspace{0.5cm}
	\begin{subfigure}[b]{0.24\textwidth}
		\centering
		\includegraphics[height=3.8cm]{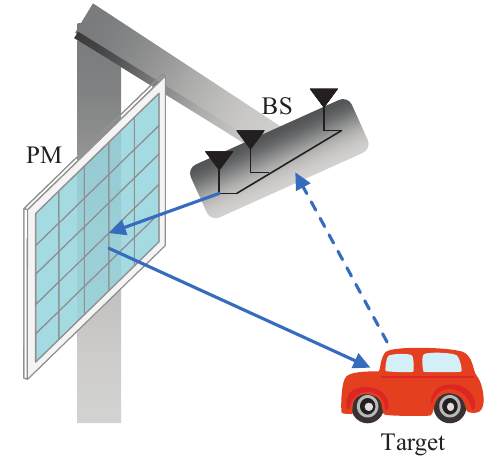}
		\caption{Illuminated-PM-assisted sensing.}
	\end{subfigure}
	\hfill
	\begin{subfigure}[b]{0.24\textwidth}
		\centering
		\includegraphics[height=4.2cm]{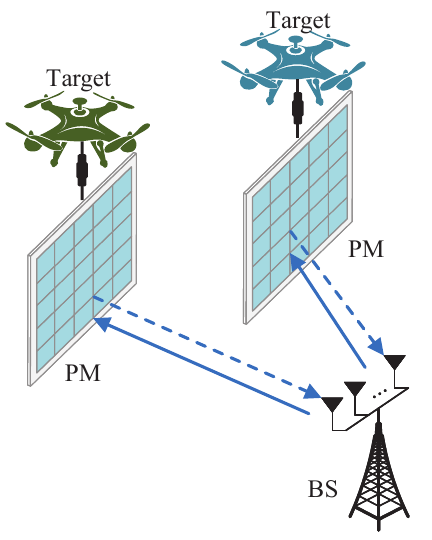}
		\caption{Target-mounted PM sensing.}
	\end{subfigure}	
	\caption{Four architectures of PM-assisted sensing.}
	\label{fig:PM_assisted_sensing_architectures}
\end{figure}

As illustrated in Fig.~\ref{fig:PM_assisted_sensing_architectures}, we consider fundamental architectures of PM-assisted sensing, where we consider PM-assisted monostatic and bistatic sensing, illuminated-PM-assisted sensing, and target-mounted PM sensing. The essence of different architectures lies in the variation of sensing channel modeling. In particular, we will derive the expressions for the channel from transmitter to the $l$-th target and subsequently to the PM, where both transmitter and PM comprise uniform planar arrays in three-dimensional Cartesian coordinates. 

Let $\mathbf{p}_T = (0,0,0)$, $\mathbf{p}_l = (x_l^T,y_l^T,z_l^T)$ and $\mathbf{p}_P = (x^P,y^P,z^P)$ denote position coordinates of the transmitter, $l$-th target, and PM, respectively. The channels from transmitter to the $l$-th target and from the $l$-th target to the PM are modeled as $\mathbf{g}_l \in \mathbb{C}^{M \times 1}$ and $\mathbf{t}_l^T \in \mathbb{C}^{N \times 1}$, where $M$ and $N$ represent the number of transmitter antennas and PM elements, respectively. The effective angles of departure from the transmitter are represented as $u_l^x$, $u_l^y$, and $u_l^z$ along the $x$-, $y$- and $z$-axes, respectively, calculated as
\begin{subequations}
	\begin{align}
		& u_l^x =  \cos\phi_l^T \cos \theta_l^T, \\
		&u_l^y =\cos \phi_l^T \sin \theta_l^T, \\
		& u_l^z = \sin \phi_l^T,
	\end{align}
\end{subequations}
where $\phi_l^T$ and $\theta_l^T$ represent elevation and azimuth angles to the $l$-th target. The channel from transmitter to the $l$-th target is given by
\begin{equation}\label{s4_gl}
	\mathbf{g}_l = \sqrt{C_0 (D_l^T)^{-\alpha}} e^{-j 2\pi \frac{D_l^T}{\lambda}} \mathbf{a}_T(u_l^x,u_l^y,u_l^z),
\end{equation}
where $C_0$ is the path-loss coefficient at 1 meter reference distance, $\alpha$ is the path-loss exponent, $\lambda$ denotes the wavelength, $D_l^T$ is the distance between transmitter and the $l$-th target, and $D_l^T/\lambda$ represents propagation time. The $m$-th element of the array steering vector is expressed as $\exp\!\left(\! -j \frac{2\pi}{\lambda}  (u^x \Delta x_m^T \!+\! u^y \Delta y_m^T\! +\! u^z \Delta z_m^T) \right)$, where $ \Delta x_m^T$, $ \Delta y_m^T$ and $ \Delta z_m^T$ are relative distances from the $m$-th transmit antenna to the transmitter reference point $\mathbf{p}_T$ along respective coordinate axes. Similarly, the channel from the $l$-th target to the PM is characterized as
\begin{equation}\label{s4_tl}
	\mathbf{t}_l = \sqrt{C_0 (D_l^P)^{-\alpha}} e^{-j 2\pi \frac{D_l^P}{\lambda}} \mathbf{b}_P(v_l^x,v_l^y,v_l^z),
\end{equation} 
where $D_l^P$ is the distance between the $l$-th target and the PM. Parameters $v_l^x$, $v_l^y$ and $v_l^z$ denote effective angles of arrival at the PM, given by
\begin{subequations}
	\begin{align}
		& v_l^x = \frac{2\pi}{\lambda} \Delta d_P \cos\phi_l^P \cos \theta_l^P, \\
		& v_l^y = \frac{2\pi}{\lambda} \Delta d_P \cos \phi_l^P \sin \theta_l^P, \\
		& v_l^z = \frac{2\pi}{\lambda} \Delta d_P \sin \phi_l^P,
	\end{align}
\end{subequations}
where $\Delta d_P$ is PM element spacing, and $\phi_l^P$ and $\theta_l^P$ are elevation and azimuth angles from the $l$-th target. The $n$-th element of the PM array steering vector is expressed as $\exp\left( -j \frac{2\pi}{\lambda} (v^x \Delta x_n^P \!+\!u^y \Delta y_n^P\!+\! u^z \Delta z_n^P) \right)$, where $\Delta x_n^P$, $\Delta y_n^P$ and $\Delta z_n^P$ are relative distances from the $n$-th PM element to the PM reference point $\mathbf{p}_P$ along respective coordinate axes.  Based on the derived channels in Eqs.~\eqref{s4_gl} and \eqref{s4_tl}, the composite channel from transmitter to PM via $L$ targets is represented as
\begin{equation}
	\begin{aligned}
		\mathbf{H} =& C_0 \sum\limits_{l=1}^L \beta_l (D_l^T D_l^P)^{-\frac{\alpha}{2}} e^{-j 2\pi (D_l^T+D_l^P)} \\ & \times \mathbf{a}_T(u_l^x,u_l^y,u_l^z) \mathbf{b}_P^T(v_l^x,v_l^y,v_l^z)
	\end{aligned}
\end{equation}
where $\beta_l$ is the radar cross section of the $l$-th target. Then, the mathematical relationships between effective angles of departure and target position coordinates are derived as $u_l^x = \frac{x_l^T}{D_l^T}$, $u_l^y = \frac{y_l^T}{D_l^T}$ and $u_l^z = \frac{z_l^T}{D_l^T}$. In a similar way, relationships between effective angles of arrival and target position coordinates are given by $v_l^x = \frac{x^P - x_l^T}{D_l^P}$, $v_l^y = \frac{y^P - y_l^T}{D_l^P}$ and $z_l^x = \frac{z^P - z_l^T}{D_l^P}$.

Hence, the channel expression can consequently be reformulated as
\begin{equation}\label{s4_H}
	\begin{aligned}
		&\mathbf{H} = C_0 \sum\limits_{l=1}^L \beta_l (D_l^T D_l^P)^{-\frac{\alpha}{2}} e^{-j 2\pi (D_l^T+D_l^P)} \\ & \times \mathbf{a}_T\!\left(\!\frac{x_l^T}{D_l^T},\frac{y_l^T}{D_l^T},\frac{z_l^T}{D_l^T}\! \right)\! \mathbf{b}_P^T\!\left(\!\frac{x^P - x_l^T}{D_l^P},\frac{y^P - y_l^T}{D_l^P},\frac{z^P - z_l^T}{D_l^P}\!\right).
	\end{aligned}
\end{equation}

Leveraging the channel models derived from the PM-assisted sensing architecture, the ability of PMs to control and manipulate the wireless channel environment can be leveraged to significantly enhance key performance metrics for efficient sensing. Moreover, this analytical framework provides a rigorous basis for exploring the fundamental performance limits in advanced scenarios, specifically facilitating the investigation of high-resolution near-field localization and distributed multi-node cooperative sensing.

\subsection{PM-Assisted Efficient Sensing}
Based on the signal models and channel formulations established above, this section comprehensively examines the performance limits and efficiency of PM-assisted sensing systems. The analysis encompasses key performance indicators including ambiguity function, detection probability, estimation accuracy, and other relevant metrics. These indicators are analyzed across a range of operational scenarios, with a detailed summary provided in Table~\ref{tab:sensing_performance}.

\begin{table*}[]
	\centering
	\caption{Summary of performance metrics for PM-assisted sensing systems}
	\label{tab:sensing_performance}
	\resizebox{\textwidth}{!}{
		\begin{tabular}{|>{\arraybackslash}m{4cm}|>{\arraybackslash}m{1cm}|>{\arraybackslash}m{5cm}|>{\arraybackslash}m{12cm}|}
			\hline
			\textbf{Performance Metric} &
			\textbf{Ref.} &
			\textbf{Scenario} &
			\textbf{Characteristics/Techniques} \\ \hline
			\multirow{2}{4cm}{Detection probability} &
			\cite{buzzi2021radar} &
			SISO bistatic sensing &
			Derive detection probability for RIS-assisted closely- and widely-spaced radar \\ \cline{2-4} 
			&
			\cite{buzzi2022foundations} &
			MIMO monostatic/bistatic sensing &
			Derive detection probability for RIS-assisted radar with/without LoS links \\ \hline
			\multirow{3}{4cm}{Estimation accuracy} &
			\cite{chen2023doa} &
			MIMO bistatic sensing &
			Design RIS to minimize the derived CRLB of DoA estimation\\ \cline{2-4} 
			&
			\cite{liu2025ris} &
			SIMO bistatic sensing &
			Design RIS to minimize the derived BCRLB constrained by communication \\ \cline{2-4} 
			&
			\cite{wang2023target} &
			MIMO bistatic sensing &
			Joint estimation for the location and orientation of target-mounted IRS \\ \hline
			\multirow{2}{4cm}{Sensing resolution} &
			\cite{chen2025integrated} &
			SISO monostatic sensing &
			Employ illuminated-RIS-assisted high-resolution scene depth estimation \\ \cline{2-4} 
			&
			\cite{wang2024ris} &
			MIMO monostatic sensing &
			Propose a novel RIS space-time beamforming to improve the sensing resolution \\ \hline
			Coverage probability &
			\cite{gan2025modeling} &
			MIMO monostatic sensing &
			Examine the beneficial impact of RISs on the coverage rate \\ \hline
			Mutual information &
			\cite{li2024joint} &
			MIMO monostatic sensing &
			Joint optimization to maximize sensing mutual information\\ \hline
		\end{tabular}
	}
\end{table*}

\begin{itemize}
	\item \textbf{Ambiguity Function (AF):} AF quantifies the correlation between a transmitted waveform and its time/frequency-shifted or delay/Doppler-shifted versions, providing a compact characterization of sensing signal response in temporal and frequency domains. In particular, the authors of \cite{jin2023ris} employed the AF to quantify sensing accuracy while implementing RIS beamforming in sensing-assisted communication schemes. Their approach ensured eavesdroppers received signals with deliberate amplitude and phase distortion, demonstrating improved bit error rate performance and enhanced PLS. Additionally, the authors of \cite{zou2025target} modeled received signals through AF representations with matched filters incorporating biased range-Doppler cells. Based on this formulation, they developed both non-coherent and coherent detectors, deriving corresponding detection and false alarm probabilities as functions of the AF, followed by RIS phase shift optimization according to the Neyman-Pearson criterion. 
	
	\item \textbf{Detection Probability:} Detection probability represents the likelihood of correct target identification under specified false alarm constraints, given prevailing noise and interference conditions. In \cite{buzzi2021radar}, the authors derived closed-form expressions for detection probability in systems incorporating closely-spaced and widely-spaced single-input single-output radar and RIS configurations under different beampattern configurations. Their analysis demonstrated that RIS elements significantly enhance detection probability in radar applications, with the widely-spaced scenario effectively realizing a low-cost bistatic radar architecture where the RIS provides two-fold diversity gain through different observation angles. Building upon this foundation, the authors of \cite{buzzi2022foundations} extended the SISO framework to MIMO radar configurations, investigating detection probability for both monostatic and bistatic RIS-assisted radar systems with and without direct LoS paths to potential targets. They subsequently formulated RIS phase shift optimization problems aimed at maximizing detection probability for specified locations under fixed false alarm probability constraints. The theoretical results clearly illustrated performance benefits from RIS integration and elucidated key parameter inter-dependencies. Further extending this methodology, the authors of \cite{zhang2024target} derived detection probabilities for both RIS-enabled and reconfigurable holographic surface (RHS)-enabled radar systems, exploring how these technologies can be optimally deployed in different operational scenarios for target detection applications.

	\item \textbf{Estimation Accuracy:} Estimation accuracy quantifies bias, variance, and root mean square error in sensing parameters including delay, Doppler shift, direction-of-arrival/angle-of-arrival, and position coordinates. The Cramér-Rao lower bound (CRLB) characterizes fundamental sensing limits by encapsulating information content in received signals and array geometry configurations. The authors of \cite{chen2023doa} derived the CRLB for DoA estimation in RIS-assisted systems to assess ultimate estimation accuracy, designing RIS phase shifts through manifold optimization methods to minimize the CRLB in their proposed scheme. To address practical implementation challenges, \cite{liu2025ris} adopted Bayesian CRLB formulations to evaluate and optimize sensing performance when estimated parameters are unknown but can be characterized through prior distributions. Furthermore, the authors of \cite{wang2023target} solved two optimization problems of estimation the location and orientation of target-mounted intelligent reflecting surface (IRS), which were verified to achieve the estimation accuracy close to theoretical bound.

	\item \textbf{Sensing Resolution:} The authors of \cite{chen2025integrated} adopted illuminated-RIS architecture for environment sensing. In particular, the authors proposed an interaction codebook which enabled desirable sensing grids of RIS's beams for high-resolution depth map construction. The authors of \cite{wang2024ris} established that sensing channel rank fundamentally determines achievable resolution, demonstrating that time-domain RIS adjustments can enhance resolution capabilities. They developed a novel RIS space-time beamforming scheme that obtained time-domain beamforming gains to effectively improve sensing resolution.

	\item \textbf{Other Performance Metrics:} Beyond conventional metrics, comprehensive sensing performance evaluation requires consideration of additional indicators including coverage probability, resolution, and mutual information. The authors of \cite{gan2025modeling} derived joint communication and sensing coverage probabilities in distributed RIS-assisted millimeter-wave ISAC networks, accounting for blockage effects through stochastic geometry analysis that examined how RIS density impacts coverage rates for both functionalities. Mutual information provides another valuable metric, representing uncertainty reduction about target response matrices given transmitted and received echo signals. Leveraging this interpretation, \cite{li2024joint} formulated a joint base station and RIS beamforming optimization problem that maximized sensing mutual information subject to QoS constraints for communication users.
\end{itemize}

\subsection{PM-Assisted Near-field Sensing}
\begin{figure}[t]
	\centering
	\includegraphics[width=\linewidth]{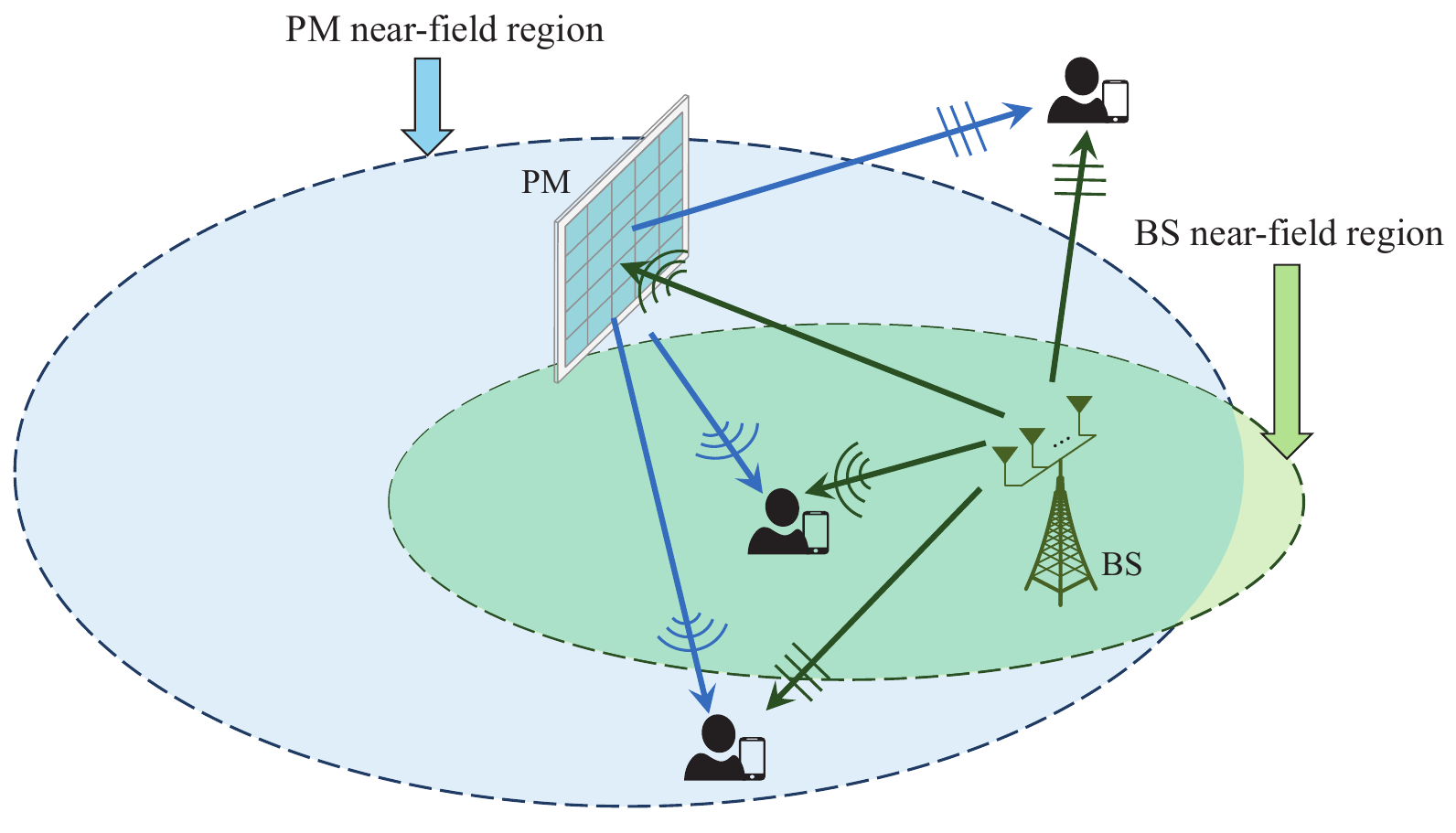}
	\caption{Illustration of near-field regions of the BS and PM.}
	\label{fig:sensing_near_filed}
\end{figure}

The conventional demarcation between far-field and near-field propagation regions is defined by the Fresnel distance, beyond which EM waves can be accurately approximated as planar waves, and within which the spherical wave model must be employed to avoid significant modeling errors\cite{wang2025near,bjornson2020power}. This critical boundary distance exhibits direct proportionality to the antenna aperture dimensions. Consequently, in systems employing large-aperture PMs, targets are more likely to reside within the near-field region, as depicted in Fig.~\ref{fig:sensing_near_filed}. The near-field channel inherently encapsulates coupled range and angle information, enabling highly precise sensing capabilities. The following sections provide a comprehensive review of PM-assisted sensing methodologies, examining implementation architectures and performance characteristics across near-field and hybrid-field operational environments, with key comparisons summarized in Table~\ref{tab:sensing_near_field}.

\begin{table*}[]
	\centering
	\caption{Summary of existing works on PM-assisted near-field and hybrid-field sensing.}
	\label{tab:sensing_near_field}
	\resizebox{\textwidth}{!}{
		\begin{tabular}{|>{\arraybackslash}m{2cm}|>{\arraybackslash}m{1cm}|>{\arraybackslash}m{5cm}|>{\arraybackslash}m{12cm}|}
			\hline
			\textbf{Field Type}                  & \textbf{Ref.}                            & \textbf{Scenario}                                      &    \textbf{Charcteristic/Techinique}                     \\ \hline
			
			\multirow{4}{*}{Near-field} & \cite{dardari2021nlos}           & SISO bistatic sensing          & Propose localization algorithm to localize user under NLoS conditions                                                                       \\ \cline{2-4} 
			& \cite{zhou2024near}        & MIMO monostatic sensing                              & Joint optimization designs to minimize the CRLB of target's distance and AoA                                           \\ \cline{2-4} 
			& \cite{liu2025reconfigurable}       & SIMO monostatic sensing          & Design a beam training scheme to estimate target 3D position with low complexity                                             \\ \cline{2-4} 
			& \cite{yuan2024near}             & SIMO monostatic sensing          &Design KF-based channel tracking to sense time-varying channels                                                                                    \\ \hline
			\multirow{4}{*}{Hybrid-field}  & \cite{zhu2025hybrid}                 & MIMO monostatic sensing        & Propose a low-complexity method to localize multi-user in near-field or far-field                \\ \cline{2-4} 
			& \cite{li2023sensing}                  & MIMO monostatic sensing       & Localizing multi-user accounting for hybrid-field beam squint effects  \\ \cline{2-4} 
			& \cite{wu2025hybrid}                & MIMO monostatic sensing              & Present a hybrid channel model by dynamically adapting weighting factors    \\ \cline{2-4} 
			& \cite{yoo2025ris}                     & MIMO monostatic sensing         & Joint optimization to enhance ISAC performance for hybrid-field industrial IoTs     \\ \hline
		\end{tabular}
	}
\end{table*}

\begin{itemize}
	\item \textbf{Near-field Sensing:} Near-field sensing architectures leverage the substantial physical dimensions of PMs to enable spherical wave modeling and create novel localization opportunities. The authors of \cite{dardari2021nlos} proposed two localization algorithms that exploited large-aperture RISs to capitalize on near-field channel characteristics, demonstrating successful user equipment localization with single-antenna BSs under NLoS conditions. Remarkably, their algorithms maintained high localization accuracy despite increasing RIS obstruction levels. In a complementary approach, \cite{liu2025reconfigurable} implemented large-scale RISs for near-field localization through beam training schemes incorporating approximated near-field channel models, significantly reducing training overhead. Their methodology utilized training results to estimate covariance matrices containing array spatial correlation information, enabling decoupling of three-dimensional target position parameters. Further advancing this domain, the authors of \cite{zhou2024near} investigated large-scale STARS assisted near-field ISAC systems, jointly designing communication beamformers, sensing signal covariance matrices, and surface parameters to minimize CRLB for joint distance and angle-of-arrival estimation. To solve computational complexity challenges inherent in spherical wavefront models, the authors of \cite{yuan2024near} developed a Kalman filter (KF)-based framework implementing large-scale RIS assisted channel tracking that leveraged temporal correlation in time-varying channels to predict subsequent CSI, effectively managing near-field channel estimation complexity.

	\item \textbf{Hybrid/Mixed-field Sensing:} Practical multi-RIS assisted wireless systems present significant sensing challenges due to coexisting near-field and far-field users requiring different channel models. Addressing this hybrid-field localization problem, the authors of \cite{zhu2025hybrid} developed a two-phase methodology that initially estimated relative user positions to individual RIS elements before fusing results for final localization, progressively narrowing candidate location sets to reduce computational complexity. Beyond fundamental wavefront differences, wide-bandwidth operations introduce distinct beam squint effects that further complicate channel modeling. The authors of \cite{li2023sensing} confronted these challenges through a joint channel and location sensing scheme employing frequency selective polar-domain redundant dictionaries, where RISs functioned as anchors constraining user positions along hyperbolic coordinates to improve channel and position recovery fidelity. In healthcare monitoring and positioning applications for the Internet of Everything, the authors of \cite{wu2025hybrid} introduced a hybrid channel model incorporating dynamically adjusted weighting factors that balanced near-field and far-field route gains, enabling practical trade-offs between positioning precision and monitoring coverage while achieving $45\%$ positioning accuracy improvement and $42\%$ monitoring reliability enhancement compared to conventional methods. Within ISAC frameworks where sensing echoes and communication waveforms arrive concurrently across field regions, \cite{yoo2025ris} implemented NOMA with cooperative bandwidth splitting strategies, formulating joint optimization of RIS phase shifts, bandwidth allocation, and receive beamforming to maximize sensing accuracy measured through CRLB minimization while satisfying industrial IoT data rate requirements and resource constraints.
\end{itemize}

\subsection{PM-Assisted Cooperative Sensing}
\begin{figure*}[htbp]
	\centering
	\begin{subfigure}[t]{0.32\textwidth}
		\centering
		\includegraphics[height=4cm]{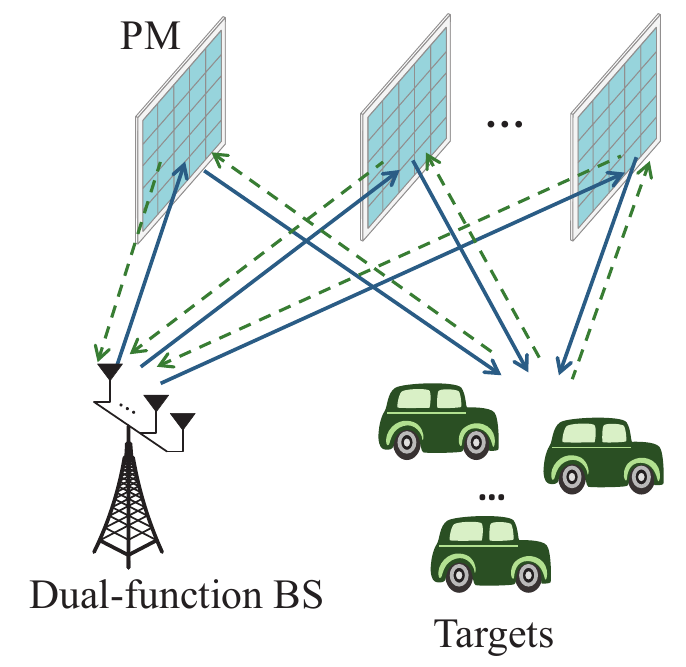}
		\caption{PM-assisted multi-BS sensing.}
		\label{fig:sensing_cooperative1}
	\end{subfigure}
	\hfill
	\begin{subfigure}[t]{0.32\textwidth}
		\centering
		\includegraphics[height=3.8cm]{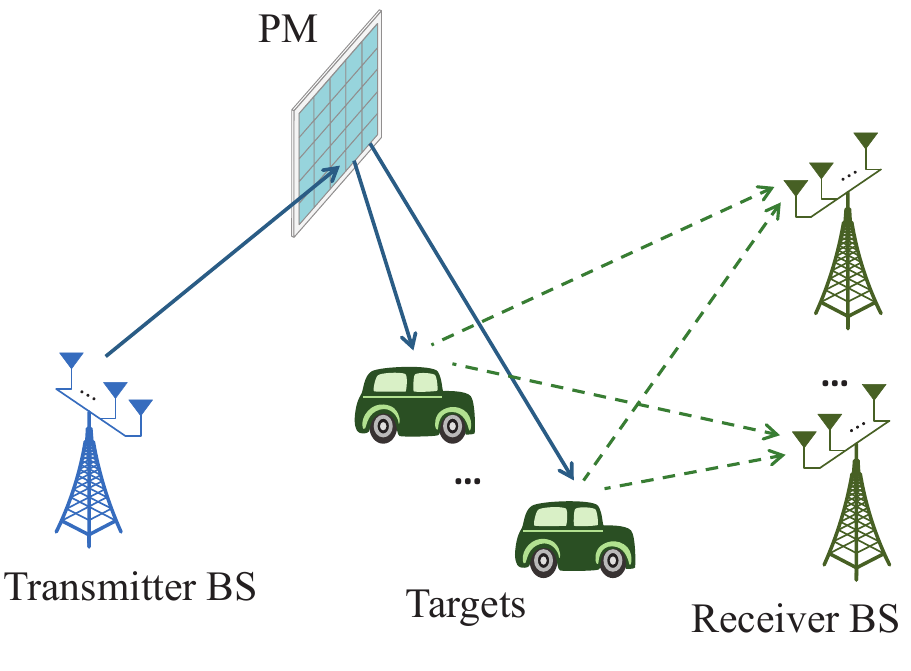}
		\caption{Multi-PM-assisted BS sensing.}
		\label{fig:sensing_cooperative2}
	\end{subfigure}
	\hfill
	\begin{subfigure}[t]{0.32\textwidth}
		\centering
		\includegraphics[height=3.5cm]{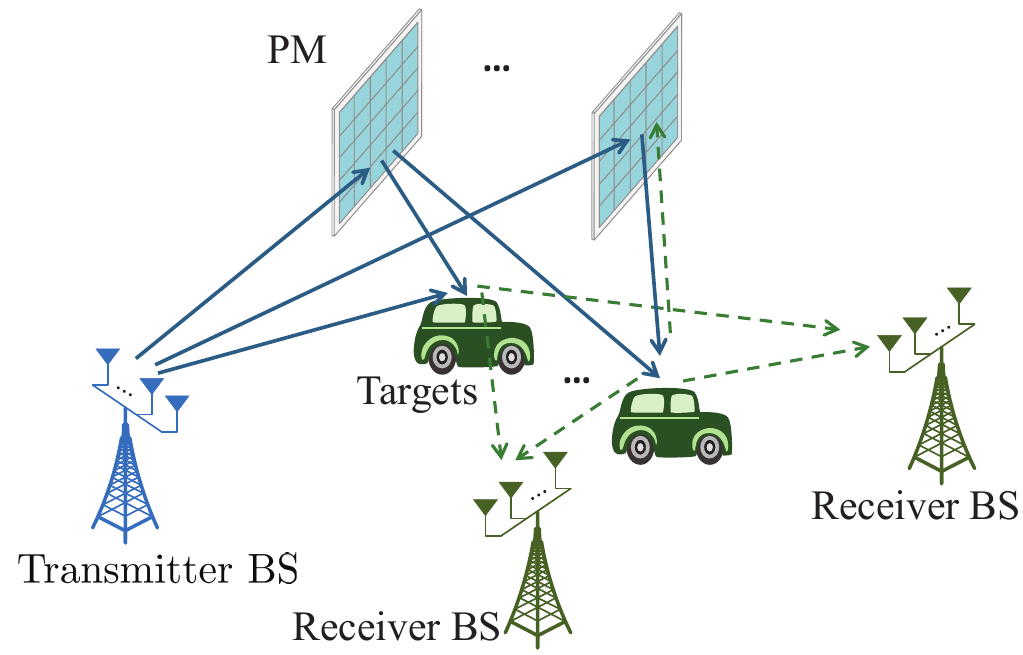}
		\caption{Multi-PM-assisted multi-BS sensing.}
		\label{fig:sensing_cooperative3}
	\end{subfigure}
	\caption{Three typical multi-BS, multi-PM cooperative sensing scenarios.}
	\label{fig:sensing_cooperative}
\end{figure*}

The performance of wireless sensing can be fundamentally enhanced by incorporating spatial diversity through multiple sensing nodes. The deployment of BSs and/or multiple PMs enables the reception of target echoes from varied spatial sensing, effectively creating multi-view observations as in Fig.~\ref{fig:sensing_cooperative}. Owing to these significant advantages, various cooperative sensing architectures have attracted considerable research attention, with detailed comparisons provided in Table~\ref{tab:sensing_cooperative}.

\begin{table*}[]
	\centering
	\caption{Summary of existing works on multi-BS, multi-PM cooperative sensing systems.}
	\label{tab:sensing_cooperative}
	\resizebox{\textwidth}{!}{
		\begin{tabular}{|>{\arraybackslash}m{3cm}|>{\arraybackslash}m{1cm}|>{\arraybackslash}m{4.5cm}|>{\arraybackslash}m{12cm}|}
			\hline
			\textbf{Cooperation Type} &
			\textbf{Ref.} &
			\textbf{Scenario} &
			\textbf{Characteristic/Technique} \\ \hline
			\multirow{4}{3cm}{PM-Assisted Multi-BS Sensing} &
			\cite{kong2024signal} &
			One target-mounted IRS &
			Joint optimization to improve sensing performance of the legal receiver \\ \cline{2-4} 
			&
			\cite{zhang2024intelligent} &
			Multiple passive targets  &
			Jointly optimization to guarantee $360^{\circ}$ ISAC coverage  \\ \cline{2-4} 
			&
			\cite{yang2024ris} &
			One passive target &
			Joint optimization designs to minimize the transmit power \\ \cline{2-4} 
			&
			\cite{salem2024integrated} &
			One passive target &
			Maximize weighted sum of the communication and radar SINRs \\ \hline
			\multirow{4}{3cm}{Multi-PM-Assisted BS Sensing} &
			\cite{sankar2023beamforming} &
			Multiple passive targets &
			Deploy and design RISs to maximize the worst-case target illumination power \\ \cline{2-4} 
			&
			\cite{liaquat2025improving} &
			One passive target &
			Analyze the power received in dual RIS configurations for monostatic sensing \\ \cline{2-4} 
			&
			\cite{li2025multi2} &
			One passive target &
			Energy-efficient optimization to minimize the multi-RIS size-to-coverage sum ratio \\ \cline{2-4} 
			&
			\cite{saigre2023self} &
			Multiple passive targets &
			Convergence of localization, sensing and communications for self-adaptive RISs\\
			\hline
			\multirow{4}{3cm}{Multi-PM-Assisted Multi-BS Sensing} &
			\cite{wang2022location} &
			One active target &
			Derive the fundamental limits of localization accuracy of two CISs\\ \cline{2-4} 
			&
			\cite{pang2023cellular} &
			One passive target &
			Propose a three-stage sensing protocol to estimate the location of the target  \\ \cline{2-4} 
			&
			\cite{keykhosravi2021semi} &
			Multiple target-mounted RISs &
			Develop a low-complexity estimator of users' 3D position with sub-meter accuracy \\ \cline{2-4} 
			&
			\cite{li2025multi} &
			One passive target &
			Joint optimization to maximize the minimal sensing performance \\
			\hline
		\end{tabular}
	}
\end{table*}
\begin{itemize}
	\item \textbf{PM-Assisted Multi-BS Sensing:} In bistatic ISAC networks compromised by adversarial BSs, the authors of~\cite{kong2024signal} investigated the benefits of IRS optimization when mounted directly on target surfaces. Their work focused on sensing suppression in scenarios where adversarial BSs attempted to illicitly capture target-reflected information, demonstrating significant signal enhancement and suppression capabilities through target-mounted IRS in environments containing both legitimate and adversarial BSs. Beyond security considerations, the authors of \cite{zhang2024intelligent} addressed the requirement for $360^{\circ}$ ISAC coverage using an intelligent omni-surface (IOS) to enhance sensing signals for two receive BSs deployed in respective reflective and transmissive areas. Their approach maximized the minimum SINR for multi-target sensing while ensuring multi-user, multi-stream communications through joint optimization of BS and IOS parameters. Furthermore, multi-cell infrastructure offers substantial potential for overcoming limitations of single ISAC BS configurations. the authors of~\cite{yang2024ris} leveraged RISs to assist joint data transmission and cooperative target sensing across multiple BSs. Extending this concept to cell-free architectures, the authors of~\cite{salem2024integrated} formulated a weighted sum maximization problem for communication and radar SINRs in RIS-assisted full-duplex cell-free MIMO systems. Their solution jointly optimized communication and radar receive beamformers alongside RIS reflection coefficients, significantly enhancing integrated cooperative sensing and communication performance.
	
	\item \textbf{Multi-PM-Assisted BS Sensing:} The authors of \cite{sankar2023beamforming} implemented two dedicated RISs that separately served communication users and sensing targets, maximizing worst-case target illumination power while guaranteeing desired communication SINR requirements. Their work confirmed dual-RIS configurations significantly enhance ISAC performance, particularly when users and targets reside outside the transmitter's direct LoS. However, architectures employing limited RISs may fail in scenarios where reliable links cannot be established. To addressing this limitation, the authors of \cite{liaquat2025improving} derived comprehensive analytical expressions for received power, SINR, and path loss in monostatic radar systems assisted by multiple RISs. Their analysis revealed substantial NLoS target detection improvements, quantifying SINR enhancements up to $14.42$ dB. Advancing ISAC functionalities for air-ground integrated networks, the authors of \cite{li2025multi2} utilized ray-tracing simulations calibrated to real-world mmWave environments. They formulated and solved an energy-efficiency-driven optimization problem minimizing the multi-RIS size-to-coverage ratio while considering joint beamforming strategies at BSs and RISs. Furthermore, the authors of \cite{saigre2023self} investigated multi-RIS assistance in rich-scattering environments, establishing that comprehensive context-awareness through integrated localization and sensing constitutes a prerequisite for RIS-empowered communications. Their prototypical case study illustrated essential operational procedures for self-adaptive RIS deployment under rich scattering conditions, demonstrating effective convergence of localization, sensing, and communication functionalities.
	
	\item \textbf{Multi-PM-Assisted Multi-BS Sensing:} The integration of multiple PMs within multi-BS sensing frameworks represents a significant advancement in achieving superior spatial diversity and cooperative gains. Specifically, the authors of \cite{wang2022location} established a general signal model for continuous intelligent surface assisted localization and communications across near-field and far-field regimes, deriving theoretical localization limits for two continuous intelligent surfaces (CISs) and demonstrating that phase-optimized surfaces substantially improve accuracy. Building upon cooperative operation principles, the authors of \cite{pang2023cellular} deployed one IRS beside each BS and introduced a three-stage sensing protocol that controlled the activation states of individual surfaces to extract essential angle information, subsequently estimating target positions through least-squares methods. To solve synchronization and data association challenges in systems with one transmitter and multiple asynchronous receivers, the authors of \cite{keykhosravi2021semi} equipped each user with a RIS and utilized time-of-arrival measurements to develop a low-complexity estimator achieving the CRLB, alongside a novel phase profile design that eliminated inter-path interference. Furthermore, the authors of \cite{li2025multi} managed collaboration and interference among multiple ISAC BSs through strategically deployed STARSs, jointly optimizing active and passive beamforming at both BSs and STARSs to maximize minimal sensing performance while satisfying communication requirements, demonstrating effective interference suppression and cooperative gains in multi-base station, multi-PM systems.
\end{itemize}

\subsection{Fundamental Architectures of PM-Enabled Sensing}

\begin{figure}[t]
	\centering
	\begin{subfigure}[b]{0.4\textwidth}
		\centering
		\includegraphics[height=3.2cm]{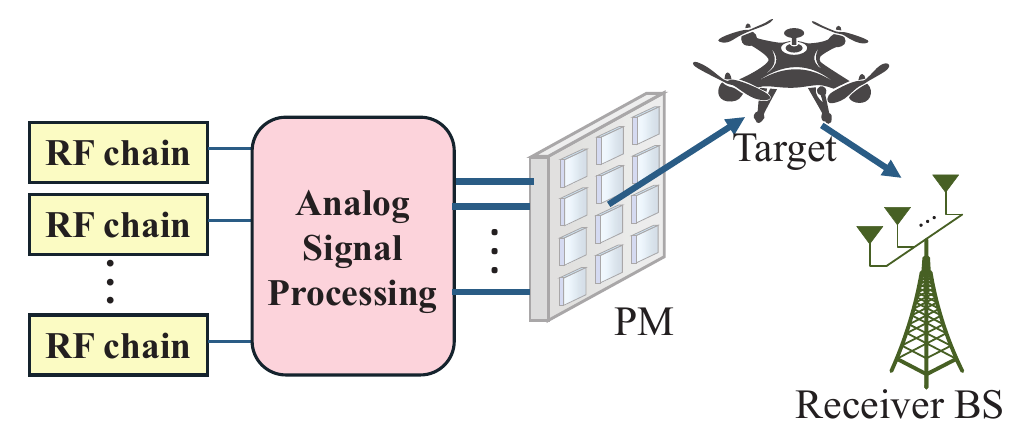}
		\caption{Radiation-PM transmitter.}
	\end{subfigure}
	\hfill
	\begin{subfigure}[b]{0.4\textwidth}
		\centering
		\includegraphics[height=4cm]{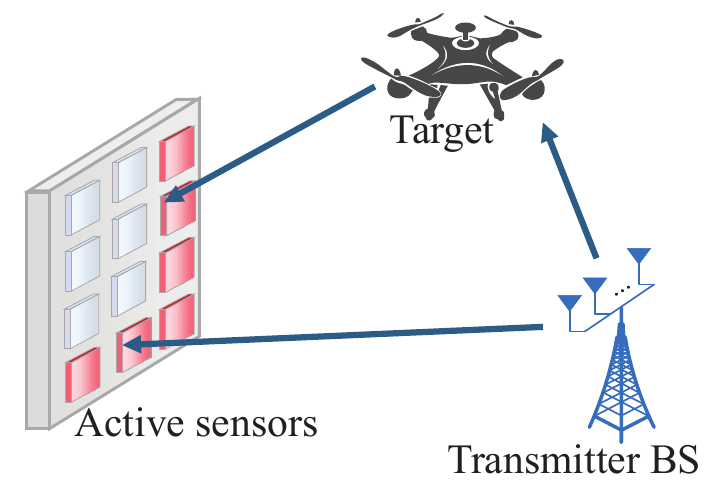}
		\caption{Active-PM receiver.}
	\end{subfigure}

	\caption{Two architectures of PM-enabled sensing.}
	\label{fig:PM_enabled_sensing_architectures}
\end{figure}

Distinct from the assisted framework where the PM functions primarily for channel reconfiguration, the PM-enabled architecture redefines the surface itself to serve as an intelligent transceiver. As illustrated in Fig.~\ref{fig:PM_enabled_sensing_architectures}, we consider two fundamental architectures of PM-enabled sensing, where radiation-PM transmitter and active-PM receiver are highlighted. Specifically, the radiation-PM transmitter leverages a feed mechanism to excite the PM, dynamically synthesizing directional transmit waveforms for target illumination. Besides, the active-PM receiver functions as an integrated aperture that directly captures the electromagnetic echoes into the sensing circuitry. In particular, we consider three fundamental signal modelings for transmitter applications: (Fig.~\ref{fig:sensing_array1}) phased-array radar, (Fig.~\ref{fig:sensing_array2}) MIMO-array radar, and (Fig.~\ref{fig:sensing_array3}) hybrid MIMO-phased array radar. Let $\bm{\phi}_t = [\phi_1(t), \phi_2(t),\cdots,\phi_R(t)]^T$ denote the transmit waveform vector, where $R$ represents the number of RF chains. This vector satisfies the orthogonality condition $\int_{T} \bm{\phi}(t) \bm{\phi})_t^H = \mathbf{I}_M$, where $T$ is the width of radar pulse. The feed signals for architectures (\ref{fig:sensing_array1})-(\ref{fig:sensing_array3}) at time instant $t$ are formulated as follows, with $P_T$ representing transmit power and $M$ denoting the number of feed elements:

\begin{figure*}[htbp]
	\centering
	\begin{subfigure}[t]{0.32\textwidth}
		\centering
		\includegraphics[height=3.2cm]{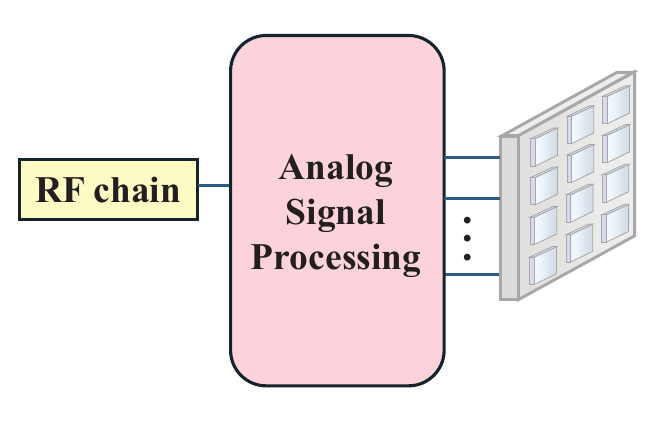}
		\caption{Phase-array radar.}
		\label{fig:sensing_array1}
	\end{subfigure}
	\hfill
	\begin{subfigure}[t]{0.32\textwidth}
		\centering
		\includegraphics[height=3.2cm]{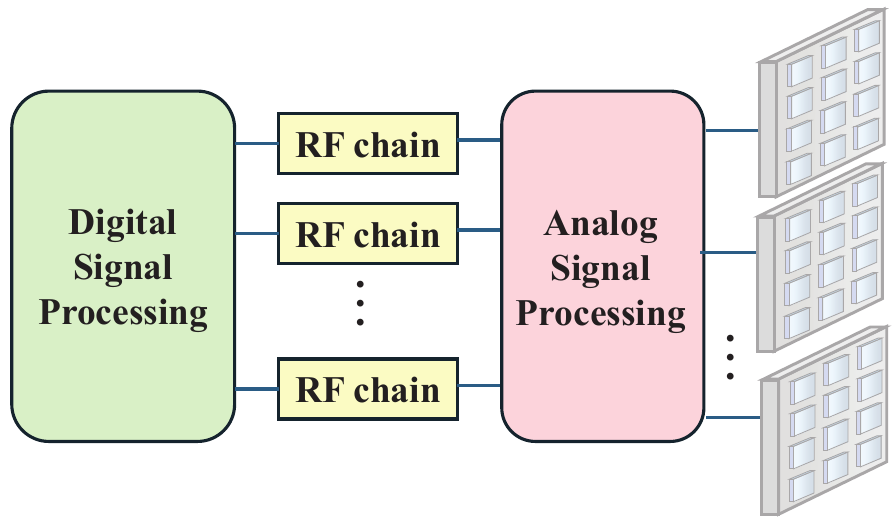}
		\caption{MIMO-array radar.}
		\label{fig:sensing_array2}
	\end{subfigure}
	\hfill
	\begin{subfigure}[t]{0.32\textwidth}
		\centering
		\includegraphics[height=3.2cm]{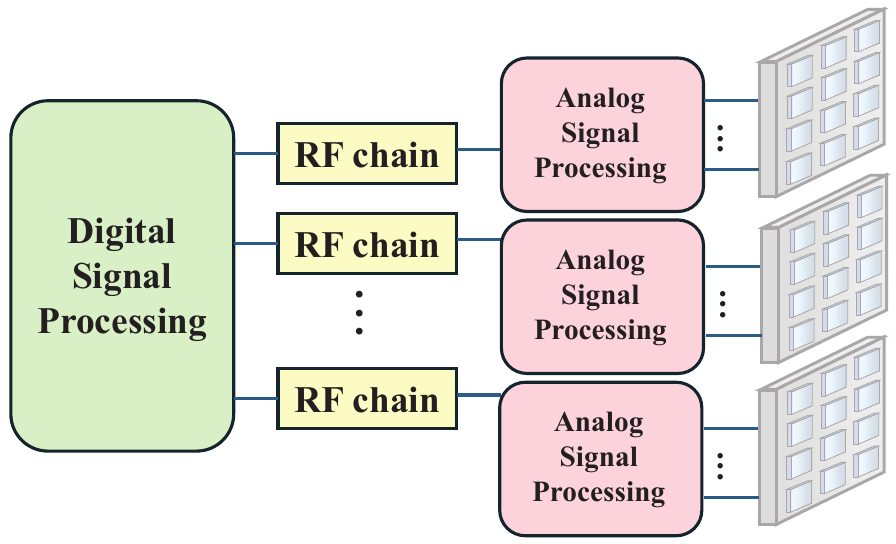}
		\caption{Hybrid MIMO-phase radar.}
		\label{fig:sensing_array3}
	\end{subfigure}
	\caption{Three sensing architectures of PM-enabled transmitter.}
	\label{fig:sensing_array}
\end{figure*}

\begin{itemize}
	\item Phase-array Radar: $\mathbf{s}^{\text{(a)}}(t) = \sqrt{\frac{P_T}{M}} \phi_1(t) \mathbf{f}^{\text{(a)}}$, where $R=1$ and $\mathbf{f}^{\text{(a)}} \in \mathbb{C}^{M \times 1}$ represents the unit-norm analog beamforming vector.
	
	\item MIMO-array Radar: $\mathbf{s}^{\text{(b)}}(t) = \sqrt{\frac{P_T}{M}}\sum\limits_{r=1}^R \phi_r(t) \mathbf{f}^{\text{(b)}}_r$, where $\mathbf{f}^{\text{(b)}}_r \in \mathbb{C}^{M \times 1}$ denotes the unit-norm analog beamforming vector corresponding to the $r$-th waveform in full-connected architectures.
	
	\item Hybrid MIMO-phase Radar: $\mathbf{s}_r^{\text{(c)}}(t) = \sqrt{\frac{P_T }{M_r}} \phi_r(t) \mathbf{f}^{\text{(c)}}_r$, where $\mathbf{f}^{\text{(c)}}_r \in \mathbb{C}^{M_r \times 1}$ represents the unit-norm analog beamforming vector for the $r$-th waveform, with $\sum_{r=1}^R M_r = M$. The composite signal vector is expressed as $\mathbf{s}^{\text{(c)}}(t) = [(\mathbf{s}_1^{\text{(c)}})^T(t),(\mathbf{s}_2^{\text{(c)}})^T(t),\dots,(\mathbf{s}_R^{\text{(c)}})^T(t)]^T$.	
\end{itemize}

\begin{figure}[t]
	\centering
	\includegraphics[width=0.9\linewidth]{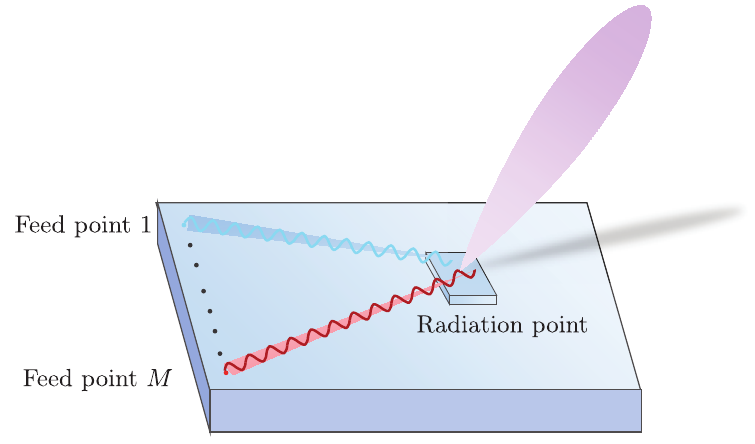}
	\caption{Illustration of the geometric relationship between the radiation point and $M$ feed points.}
	\label{fig:sensing_feed_point}
\end{figure}

To mathematically characterize the signal formulations for the three PM transmitters (\ref{fig:sensing_array1})-(\ref{fig:sensing_array3}) based on $\mathbf{s}^{\text{(a)}}(t)$-$\mathbf{s}_r^{\text{(c)}}(t)$, we examine the relationship between PM transmitted signals and their feed points and radiation elements as in Fig.~\ref{fig:sensing_feed_point}. More particularly, considering propagation from feed signals $\mathbf{s}(t) \in \mathbb{C}^{M \times 1}$ at feed point positions $\mathbf{p}^f = [\mathbf{p}_1^f,\mathbf{p}_2^f,\cdots,\mathbf{p}_M^f]$ to the $n$-th PM element, the transmitted signal $\mathbf{t}$ is computed as follows.
\begin{itemize}
	\item Phase-array Radar: \\ $[\mathbf{t}^{\text{(a)}}]_n = \exp\left( -j \mathbf{k}_s \|\mathbf{p}_1^f - \mathbf{p}_n^r\|  \right) \mathbf{s}^{\text{(a)}}(t)$, where $M=1$ and $\mathbf{k}_s$ represents the reference wave propagation vector.
	
	\item MIMO-array Radar: \\ $[\mathbf{t}^{\text{(b)}} ]_{n} = \sum_{m=1}^M \exp\left( -j \mathbf{k}_s \|\mathbf{p}_m^f - \mathbf{p}_n^r\|  \right)  \left[ \mathbf{s}^{\text{(b)}}(t) \right]_m$.
	
	\item Hybrid MIMO-phase Radar: \\ $[\mathbf{t}_r^{\text{(c)}} ]_{n} = \sum_{m_r=1}^{M_r}\exp\left( -j \mathbf{k}_s \|\mathbf{p}_{m_r}^f - \mathbf{p}_n^r\|  \right) \left[\mathbf{s}_r^{\text{(c)}}(t)\right]_{m_r}$,	where $\mathbf{p}_{m_r}^f$ denotes the position coordinate of the $m_r$-th feed signal from the $r$-th sub-array. The composite transmission signal is expressed as $\mathbf{t}^{\text{(c)}}= \left[ (\mathbf{t}_1^{\text{(c)}})^T, (\mathbf{t}_2^{\text{(c)}})^T, \cdots, (\mathbf{t}_R^{\text{(c)}})^T \right]^T$.	
\end{itemize}

The established signal models mathematically characterize the capability of PMs to function as standalone sensing transceivers, directly executing active transmission and reception tasks. This analytical framework can be used to evaluate PM-enabled sensing across diverse frequency spectra. Specifically, it lays the foundation for investigating the performance superiority of PMs in mmWave and THz bands.

\subsection{PM-Enabled Multi-Band Sensing}
\begin{figure}[t]
	\centering
	\begin{subfigure}[b]{0.4\textwidth}
		\centering
		\includegraphics[width=0.9\textwidth]{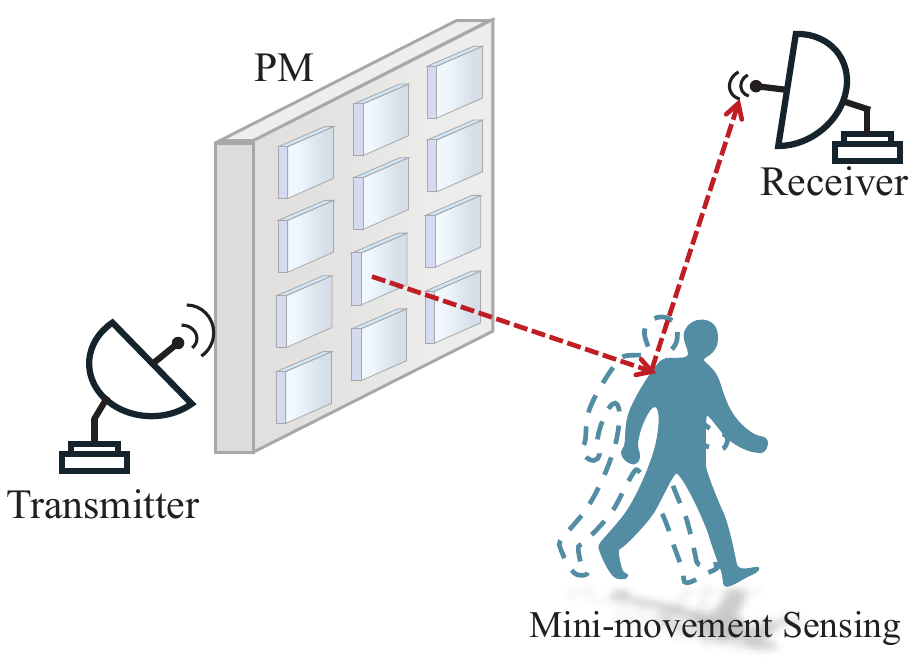}
		\caption{PM-enabled mmWave sensing.}
	\end{subfigure}
	\hfill
	\begin{subfigure}[b]{0.4\textwidth}
		\centering
		\includegraphics[width=0.9\textwidth]{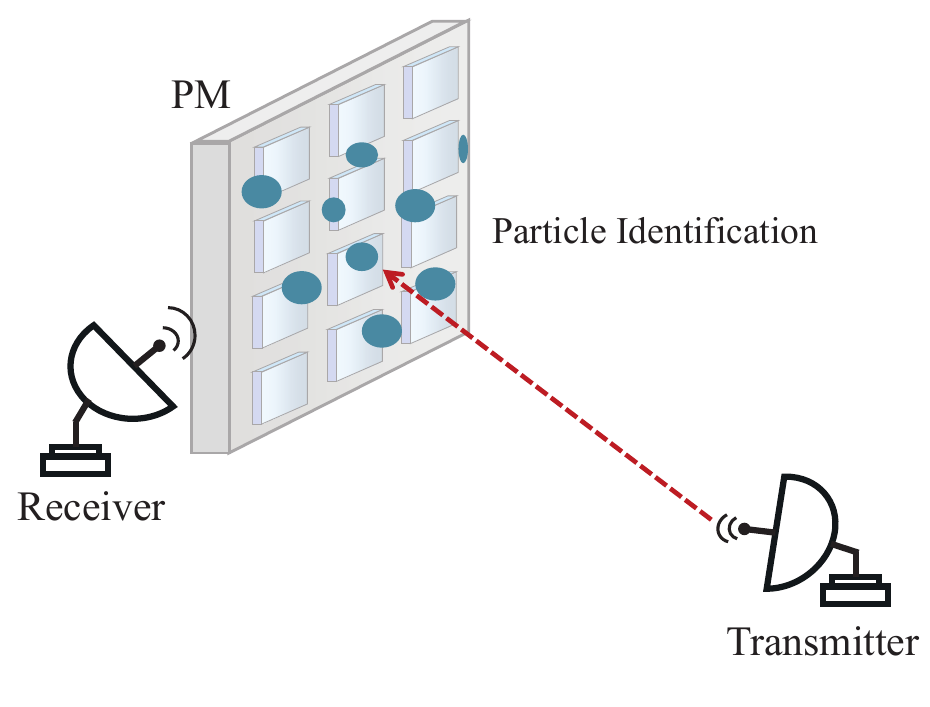}
		\caption{PM-enabled THz sensing.}
	\end{subfigure}
	
	\caption{Illustration of PM-enabled sensing in mmWave and THz frequency bands.}
	\label{fig:sensing_THz}
\end{figure}

As illustrated in Fig.~\ref{fig:sensing_THz}, the functional roles of PM as transmitters and receivers undergo systematic transformation across different frequency regimes. More particularly, for mmWave frequencies, PM-based transmitter enables minute movement detection and precise three-dimensional environment reconstruction. For THz signals, reduced wavelengths and material-specific spectral signatures make PMs ideal front-end receiver components. In the following, we will provide a detailed literature review summarizing PM-enabled sensing under mmWave and THz frequency bands in Table~\ref{tab:sensing_THz}.

\begin{itemize}
	\item \textbf{MmWave Sensing:} Conventional sub-6 GHz WiFi sensing systems utilizing omni-directional antennas frequently experience multi-path interference and limited accuracy in indoor environments. These limitations become particularly problematic for fine-grained applications such as vital sign monitoring, where minute chest movements associated with cardiac activity are substantially smaller than sub-6 GHz wavelengths. To solve these challenges, the authors of \cite{jabbar2025millimeter} demonstrated a practical non-contact vital sign monitoring system employing a dynamic metasurface antenna (DMA) operating at $60$ GHz. Their complete end-to-end framework achieved mean absolute errors below $0.9$ for both respiratory and cardiac measurements. In complementary research, \cite{lan2021metasense} developed MetaSense, an end-to-end dynamic metasurface antenna-based radio frequency sensing system that leveraged antenna pattern diversity to enhance sensing performance. Their methodology utilized high-dimensional uncorrelated channel measurements to create more accurate and robust radio frequency sensing solutions, outperforming non-tunable antennas by $20\%$ across all evaluated scenarios. Advancing imaging applications, the authors of \cite{xue2025chip} proposed a holographic tensor metasurface (HTM) that independently manipulated co-polarized and cross-polarized holograms without inter-channel crosstalk. Integration with a complementary metal-oxide semiconductor chip operating in the millimeter-wave band achieved imaging efficiency exceeding $66.6\%$. Confronting persistent challenges in mmWave imaging, the authors of \cite{wang2025high} addressed two fundamental limitations: constrained antenna counts in commercial hardware and dependence on compressive sensing algorithms requiring sparsity assumptions. Their optimized mmWave metasurface implementation featured joint optimization of metasurface configuration and codebook design to enhance signal quality, complemented by a diffusion-based neural network model that transformed mmWave signals into high-quality images with $0.061$ median root mean square error at $1$ cm resolution.
	
	\item \textbf{THz Sensing:} As operational frequencies progress to the THz regime, PM-enabled sensing and biosensing benefit from attractive properties including non-ionizing radiation and sensitivity to weak molecular interactions. However, THz wavelengths remain large relative to thin-film and biological feature dimensions. Overcoming this scale mismatch, the authors of \cite{beruete2020terahertz} exploited metasurface design freedom to tailor electromagnetic responses and create sub-wavelength details comparable to tested films or microorganisms, producing strong near-resonant field concentration that improved detection accuracy beyond classical terahertz spectroscopy. Their work comprehensively surveyed major advances in THz metasurface sensors from both historical and application-oriented perspectives. Enhancing biochemical specificity, the authors of \cite{jadeja2023detection} introduced a metasurface refractive index sensor demonstrating how variations in inner square width, tail width, and graphene chemical potential yield different optimal configurations for peptide detection. Performance validation incorporated multiple metrics including sensitivity, figure of merit, detection limit, and quality factor. Concurrently, advancing absorptive functionality, the authors of \cite{wang2020properties} proposed all-dielectric metasurface THz absorbers employing single-patterned structures that simplified fabrication while achieving near-perfect absorption across THz bands.
\end{itemize}

\begin{table*}[]
	\centering
	\caption{Summary of existing works on PM-enabled cm-wave, mmWave and THz sensing systems.}
	\label{tab:sensing_THz}
	\resizebox{\textwidth}{!}{
		\begin{tabular}{|>{\arraybackslash}m{2cm}|>{\arraybackslash}m{1cm}|>{\arraybackslash}m{4cm}|>{\arraybackslash}m{14cm}|}
			\hline
			\textbf{Frequency band} &
			\textbf{Ref.} &
			\textbf{PM Type} &
			\textbf{Characteristic/Technique} \\ \hline
			\multirow{4}{2cm}{MmWave Sensing} &
			\cite{jabbar2025millimeter} &
			DMA transmitter &
			Develop DMA-based beam control to provide breathing and heart rate estimation\\ \cline{2-4} 
			&
			\cite{lan2021metasense} &
			DMA transmitter &
			Develop DMA-enabled RF sensing to perform writing motion recognition\\ \cline{2-4} 
			&
			\cite{wang2025high} &
			Metasurface transmitter &
			Propose metasurface modulation to achieve high-quality imaging \\ \cline{2-4} 
			&
			\cite{xue2025chip} &
			HTM transmitter&
			Design HTM to implement independently both the co-polarized and cross-polarized holograms \\ \hline
			\multirow{3}{1.5cm}{THz Sensing} &
			\cite{beruete2020terahertz} &
			Metasurface receiver &
			Review the main advances and applications of THz metesurface sensors \\ \cline{2-4} 
			&
			\cite{jadeja2023detection} &
			Metasurface receiver  &
			Propose metasurface refractive index sensor for highest-sensitivity various peptides detection\\ \cline{2-4} 
			&
			\cite{wang2020properties} &
			Metasurface receiver &
			Propose all-dieletric metasurface THz absorbers for biochemical sensing\\ \hline
		\end{tabular}%
	}
\end{table*}

%% file: 5_computing.tex
\section{SIMs: Intelligent Signal Processing and Computing}\label{sec_computing}
While the preceding sections capitalized on single-layer PMs for communication and sensing, this section now advances to multi-layer architectures designed for high-performance computing. Specifically, this section introduces the signal models of SIMs. Then, it systematically reviews recent advances in SIM-enabled wave-domain analog processing and over-the-air mathematical computing, illustrating a paradigm shift from traditional digital logic to instantaneous electromagnetic calculation.

\begin{figure}[t]
	\centering
	\includegraphics[width=0.45\textwidth]{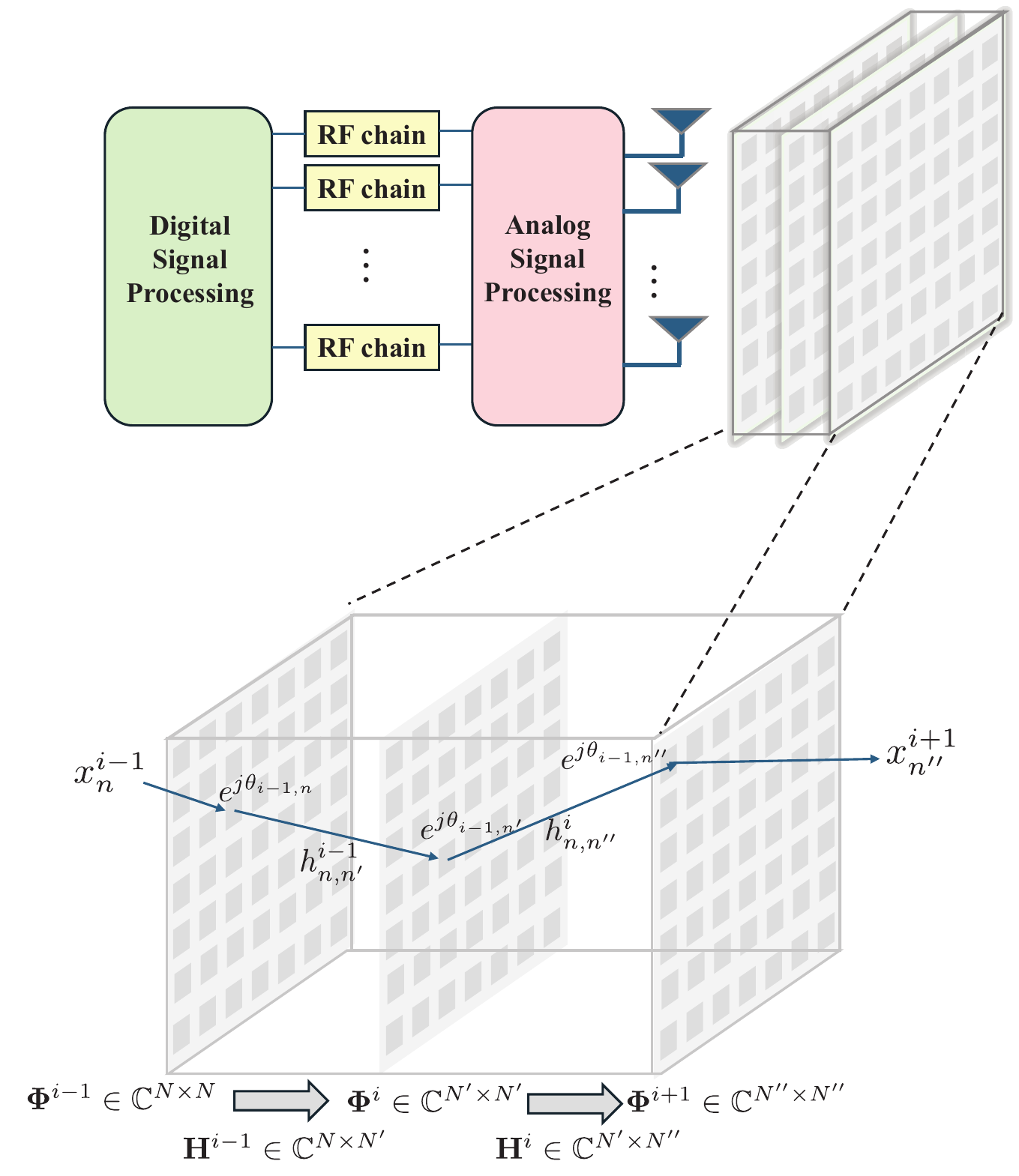}
	\caption{Illustration of the SIM-assisted system and its layered architecture.}
	\label{fig:SIM_computing}
\end{figure}

Specifically, SIM can enhance reconfigurability through cascaded programmable metasurface layers with precisely engineered inter-layer spacing. This multi-layer architecture forms a controllable scattering network that provides substantially higher spatial DoFs in phase, amplitude, and polarization manipulation compared to single-layer configurations\cite{di2025state,an2023stacked}. Through optimized inter-layer design and precise element-level control, SIM architectures can approximate high-rank linear transformations, enabling advanced functionalities including spatial multiplexing, interference shaping, channel equalization, and analog computation for integrated sensing and communication systems. To elucidate the computational capabilities of SIM architectures, fig.~\ref{fig:SIM_computing} illustrates the signal propagation through consecutive metasurface layers. The mathematical representation of signal transmission through the $n$-th element of layer $l-1$, the $n'$-th element of layer $l$, and the $n''$-th element of layer $l+1$ is expressed as
\begin{equation}\label{s5_x1}
	x_{n''}^{l+1} = e^{j \theta_{n''}^{l+1}} h_{n',n''}^l e^{j \theta_{n'}^{l}} h_{n,n'}^{l-1} e^{j \theta_{n}^{l-1}} x_n^{l-1},
\end{equation}
where $ \theta_{n}^{l-1}$, $\theta_{n'}^{l}$, and $ \theta_{n''}^{l+1}$ are the controllable phase shifts of the $n$-th element of the $l-1$-th layer, the $n'$-th element of the $l$-th layer, and the $n''$-th element of the $l+1$-th layer, respectively. This expands the signal processing DoFs compared with conventional RIS-assisted systems. The inter-element channels $h_{n',n''}^l$ and $h_{n,n'}^{l-1}$ can be modeled as free-space LoS channel $h_{n',n''}^l = \sqrt{\frac{c^2}{16 \pi^2 f_c^2}} \frac{e^{-j \frac{2\pi}{\lambda} r^l_{nn'}}}{r^l_{nn'}}$, where $c$ denotes the speed of light, $f_c$ and $\lambda$ represent the carrier frequency and wavelength, respectively, and $r^l_{nn'}$ indicates the distance between the $n$-th element of layer $l-1$ and the $n'$-th element of layer $l$. Remarkably, precise signal manipulation can also be achieved through control of intra-SIM channel characteristics via adjustment of inter-element distances.

\subsection{Signal Models of SIMs}

Considering signal transmission through a complete SIM architecture at the transmitter, let $\mathbf{x}^0 \in \mathbb{C}^{M \times 1}$ denote the transmitted signal from an $M$-antenna base station. After propagation through an $L$-layer SIM, the output signal is mathematically represented as
\begin{equation}\label{s5_x2}
	\mathbf{x}^L =  \bm{\Phi}^L \prod_{l=1}^{L-1} \mathbf{H}^l \bm{\Phi}^l \mathbf{x}^0
\end{equation}
where $\bm{\Phi}^l = \text{diag}\{ e^{j \theta^l_1}, e^{j \theta^l_2},\cdots,e^{j \theta^l_{N_l}}\}$ denotes the phase-shift matrix of the $l$-th SIM layer, and $\mathbf{H}^l \in \mathbb{C}^{N_{l-1} \times N_l}$ represents the channel matrix between layers $l$ and $l+1$ of the SIM. This mathematical formulation in Eq.~\eqref{s5_x2} applies equally to receiver-side SIM architectures. As evidenced by Eq.~\eqref{s5_x1}, the transmitted signal $\mathbf{x}^L$ can be precisely engineered through optimization of multiple SIM parameters, including all element phase shifts, the number of metasurface layers, elements per layer, inter-layer spacing, and layer orientation. This indicates that both the phase response and geometric configuration of each SIM unit contribute to signal control. For a complete system with $L_T$ transmitter SIM layers and $L_R$ receiver SIM layers, the received signal at the base station antennas is given by
\begin{equation}
	\mathbf{y} = \bm{\Phi}_R^{L_R} \prod_{l_2=1}^{L_R-1} \mathbf{H}_R^{l_2} \bm{\Phi}_R^{l_2} \mathbf{G} \bm{\Phi}_T^{L_T} \prod_{{l_1}=1}^{L_T-1} \mathbf{H}_T^{l_1} \bm{\Phi}^{l_1}_T \mathbf{x}^0,
\end{equation}
where $\mathbf{G}$ represents the channel between the transmitter and receiver SIM structures. Through coordinated optimization of the controllable parameter matrices ${ \bm{\Phi}_T^l ,\bm{\Phi}_R^l ,\mathbf{H}_T^l, \mathbf{H}_R^l}$, SIM architectures surpass the signal processing capabilities of conventional RIS, thereby enabling advanced computational functionalities for next-generation wireless systems.

\subsection{SIM-Enabled Wave-Domain Analog Processing}
\begin{table*}[]
	\centering
	\caption{Summary of existing works on SIM-enabled wave-domain analog beamforming.}
	\label{tab:SIM_beamforming}
	\resizebox{\textwidth}{!}{
		\begin{tabular}{|>{\arraybackslash}m{1cm}|>{\arraybackslash}m{6cm}|>{\arraybackslash}m{14cm}|}
			\hline
			\textbf{Ref.} &
			\textbf{PM Type} &
			\textbf{Characteristic/Technique} \\ \hline
			\cite{an2025stacked} &
			SIM transmitter, multi-user receivers &
			Optimize wave-based beamforming at the SIM to maximize the sum rate\\ \hline
			\cite{an2023stacked} &
			SIM transmitter, SIM receiver &
			Joint optimization of TX-SIM and RX-SIM to minimize the beamforming errors \\ \hline
			\cite{li2024stacked} & 
			SIM transmitter, SIM receiver &
			Propose fully-analog wideband beamforming designs for TX-SIM and RX-SIM \\ \hline
			\cite{sun2022energy} &
			SIM transmitter, multi-user receivers &
			Joint design multi-layer analog precoders to achieve energy-efficient design and secure transmission  \\ \hline
			\cite{li2024transmit} & 
			SIM transmitter, multiple users and targets &
			Design SIM to maximize sum rate of users and shape optimal beampattern for detection \\ \hline
			
		\end{tabular}
	}
\end{table*}

Recent research has demonstrated SIM architectures for enhancing wireless system performance through sophisticated wave-based analog beamforming as in Table~\ref{tab:SIM_beamforming}. In \cite{an2025stacked}, the authors pursued sum-rate maximization across all user equipments by developing a joint optimization framework for base station power allocation and SIM-configurated wave-based beamforming, accounting for practical constraints including transmit power budgets and discrete phase shifts. Their proposed computationally efficient algorithm demonstrated remarkable performance gains, achieving approximately $200\%$ sum-rate improvement compared to conventional MISO systems with equivalent antenna configurations. Complementing this direction, the authors of \cite{an2023stacked} investigated channel optimization by minimizing the discrepancy between the actual SIM-parameterized channel matrix and an ideal diagonal structure representing interference-free parallel subchannels. Through coordinated phase shift optimization across both transmitter and receiver SIM layers, their gradient-based algorithm realized a $150\%$ capacity enhancement over traditional MIMO and RIS-assisted systems, while theoretical analysis provided fundamental insights into holographic MIMO capacity bounds. Expanding beyond narrowband applications, the authors of \cite{li2024stacked} developed a fully-analog wideband beamforming architecture utilizing SIM to maintain performance across extended frequency ranges. Their systematic evaluation of system parameters revealed substantial capacity improvements over previous SIM implementations, highlighting both the potential and limitations of frequency-specific optimization strategies. Further advancing system-level optimization, the authors of \cite{sun2022energy} established a comprehensive framework for energy efficiency maximization in hybrid terrestrial-aerial networks through joint optimization of user receive precoders, digital precoders at both terrestrial and aerial platforms, and multi-layer RIS analog precoders. Moreover, to address ISAC requirements, the authors of \cite{li2024transmit} designed multi-layer cascading beamformers that simultaneously maximized user sum rates while optimally shaping beam patterns for detection, enabling precise gradient-level trade-offs between communication and sensing objectives through refined optimization control mechanisms.

\subsection{PM-Enabled Over-the-Air Mathematical Computing}

\begin{figure*}[t]
	\centering
	\begin{subfigure}[b]{0.48\textwidth}
		\centering
		\includegraphics[height=3.2cm]{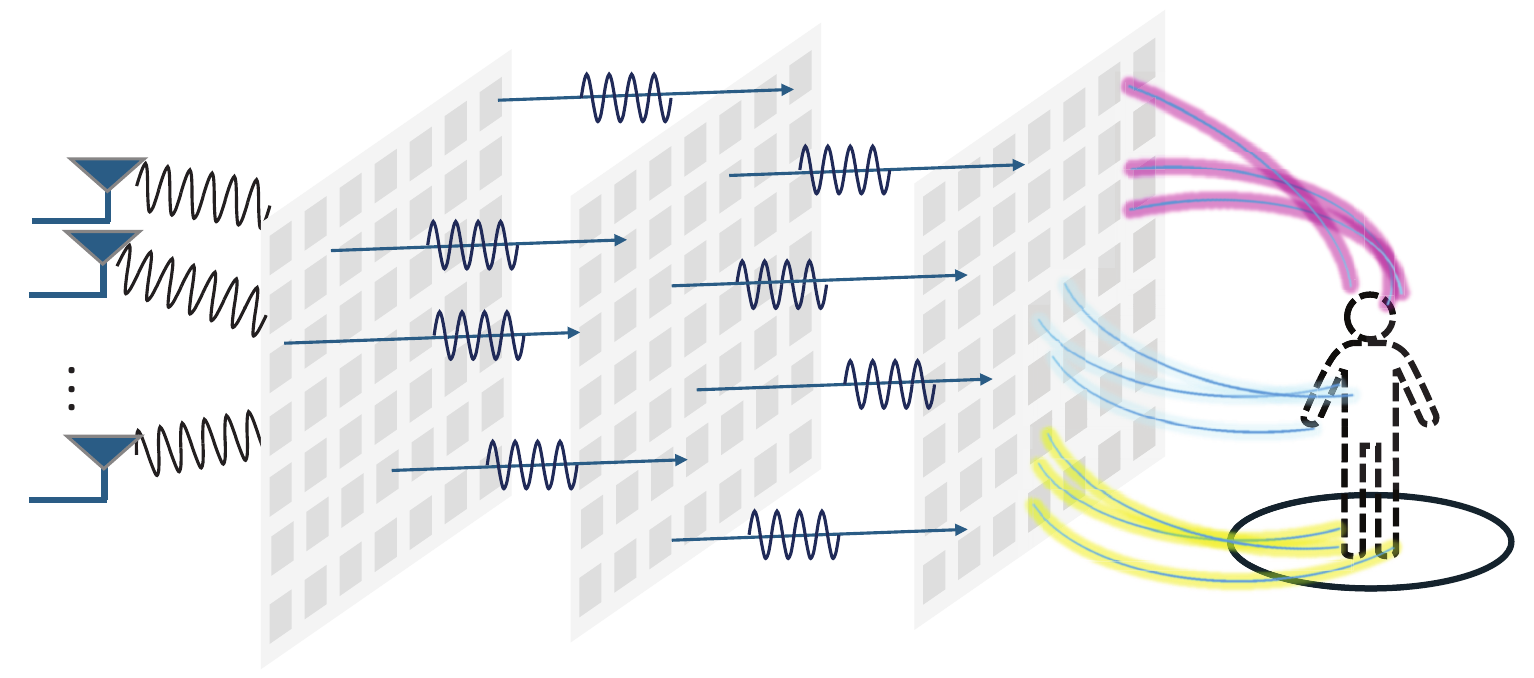}
		\caption{SIM-enabled transmitter for imaging computing.}
	\end{subfigure}
	\hfill
	\begin{subfigure}[b]{0.48\textwidth}
		\centering
		\includegraphics[height=3cm]{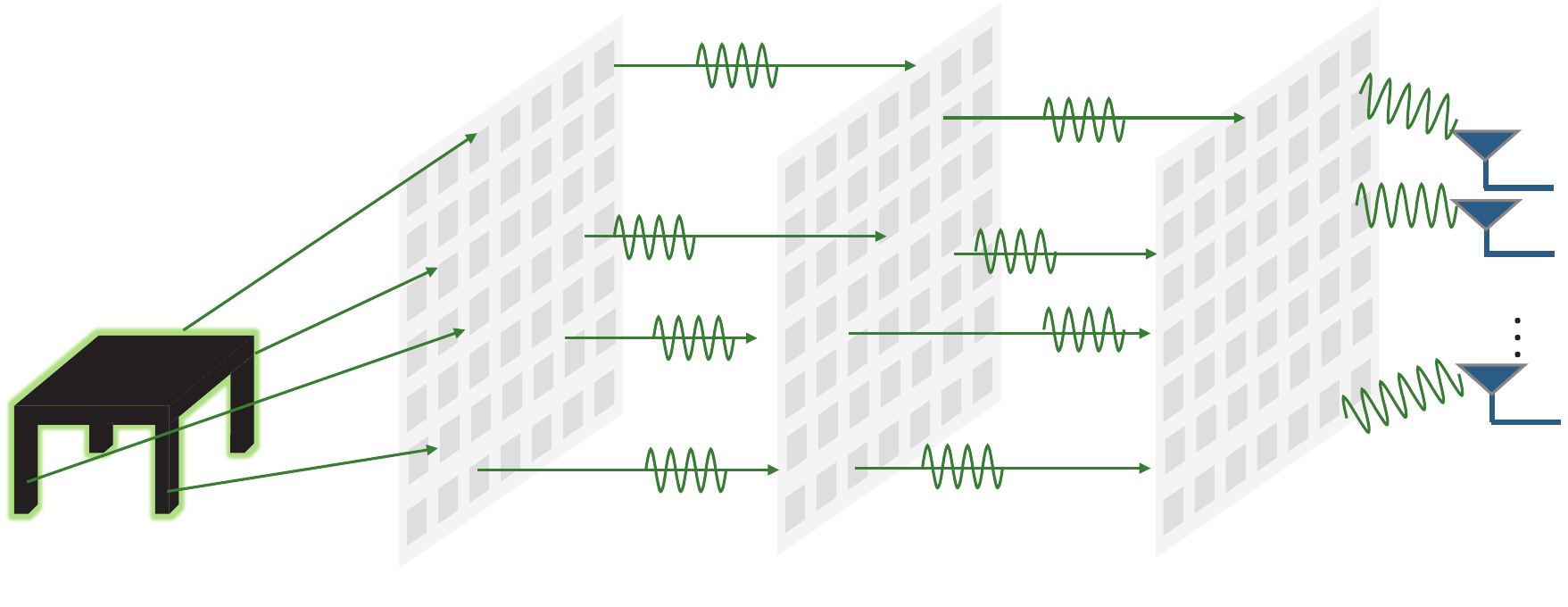}
		\caption{SIM-enabled receiver for target detection computing.}
	\end{subfigure}
	
	\caption{Illustration of SIM-enabled transceiver for imaging computing and target detection.}
	\label{fig:signal_processing}
\end{figure*}

The unique physical properties of SIM have unlocked unprecedented opportunities for implementing intelligent signal processing directly within the electromagnetic domain as in Fig.~\ref{fig:signal_processing}. By harnessing precisely engineered multi-layer structures of the SIMs, these surfaces interact with incident wavefronts to perform sophisticated mathematical operations, including spatial differentiation, convolution, and Fourier transforms, through passive, instantaneous wave-domain interactions. As research progresses, SIM-enabled intelligent signal processing is expected to play a pivotal role in advancing mathematical operations and neural network inference, ultimately bridging the gap between the physical and digital worlds, as in Table~\ref{tab:signal_processing}.

For instance, the authors of \cite{an2024two} implemented a SIM architecture that performed two-dimensional discrete Fourier transforms directly on incident waves, effectively transforming the receiver array into an angular spectrum analyzer that eliminated the need for power-intensive radio frequency chains. Extending this paradigm to multi-modal communications, the authors of \cite{huang2025stacked} developed a semantic communication system where a transmit-side SIM encoded visual information into spatial energy patterns while simultaneously transmitting textual data through conventional modulation, achieving integrated multi-modal transmission through customized gradient optimization. For unmanned aerial vehicle applications, the authors of \cite{lin2025uav} created a hybrid optical-electric neural network incorporating a SIM-based diffractive network that processed signals at light speed alongside digital components, enabling direction-of-arrival estimation using only amplitude measurements. Further advancing computational frameworks, the authors of \cite{stylianopoulos2025integrating} modeled the complete transmitter-channel-receiver chain as an end-to-end deep neural network with trainable SIM parameters, dynamically adjusting transmission power based on user location to balance classification accuracy with energy consumption across diverse scenarios. In parallel hardware development, the authors of \cite{zhang2024radio} designed a radio-frequency convolution layer using sequentially arranged transmissive intelligent surfaces that executed two-dimensional convolution operations directly in the electromagnetic domain, demonstrating the feasibility of offloading computational tasks from digital processors to analog front-ends. 

\begin{table*}[t]
	\centering
	\caption{Summary of PM-enabled intelligent signal processing.}
	\label{tab:signal_processing}
		\resizebox{\textwidth}{!}{
			\begin{tabular}{|>{\arraybackslash}m{1cm}|>{\arraybackslash}m{3cm}|>{\arraybackslash}m{15cm}|}
			\hline
			\textbf{Ref.} &
			\textbf{PM Type} &
			\textbf{Motivation/Technique}  \\ \hline
			\cite{an2024two} &
			SIM receiver &
			Perform 2D DoA estimation by minimizing the fitting error between SIM's response an DFT matrix \\ \hline
			\cite{huang2025stacked} &
			SIM transmitter&
			Design SIM to transmit visual semantic information of complex scene via imaging complex scenes\\ \hline
			\cite{lin2025uav} &
			SIM receiver &
			Design UAV-mounted SIM to enable a hybrid optical-electric neural network for DOA estimation \\ \hline
			\cite{stylianopoulos2025integrating} &
			TX-SIM-RX &
			Design SIM to enable the transmitter-channel-receiver system as an end-to-end DNN \\ \hline
			\cite{zhang2024radio} &
			TX-SIM-RX &
			Propose and validate 2D convolution operation based on three layered TRISs \\ \hline
		\end{tabular}
	}
\end{table*}

%% file: 6_conclusion.tex
\section{Conclusions}\label{sec_conclusion}
In this paper, we have provided a comprehensive overview of multi-functional PMs, covering \emph{full-space communication coverage}, \emph{ubiquitous sensing}, as well as \emph{intelligent signal processing and computing}. The operational principles of STARS were first introduced, followed by their applications in full-space communications, PLS enhancement, UAV networks, and SWIPT. For ubiquitous sensing, we presented the signal modeling of two distinct architectures: PM-assisted and PM-enabled sensing. Within these frameworks, we characterized the performance benefits regarding sensing efficiency, multi-node cooperation, near-field localization, and multi-band operation. Furthermore, we explored the potential of SIMs for layered signal processing, with a specific focus on wave-domain analog processing and over-the-air mathematical computing. We envision that the discussions and insights presented herein will inspire future research efforts to unlock the full potential of multi-functional PMs for future B6G networks.

Complementing the comprehensive review presented in this paper, we identify several critical challenges and envision promising future research directions essential for realizing the full potential of multi-functional PM networks.
\begin{itemize}
	\item \textbf{Accurate Channel Modeling for Multi-functional PMs:} The deliberate manipulation of EM waves in amplitude, phase, and polarization states by PMs necessitates physically consistent channel modeling for precise EM characteristic control. Future research can develop advanced modeling frameworks that capture the intricate relationships between PM configurations and corresponding EM responses, potentially incorporating machine learning techniques to address complex near-field interactions and spatial non-stationarity.

	\item \textbf{Unified Performance Metrics for Multi-functional PM Systems:} The coexistence of communication, sensing, and computing functionalities within integrated systems inevitably creates performance trade-offs across operational domains. The multi-function research requires unified metrics to balance these competing objectives, enabling PM optimization toward Pareto-optimal solutions.This direction demands novel multi-objective optimization frameworks capable of systematically addressing complex inter-dependencies among diverse performance requirements while ensuring practical feasibility.
	
	\item \textbf{Practical Implementation of Multi-functional PMs:} There remains a gap between theoretical analysis and hardware realization of multi-functional PMs. Besides, different applications require specific PM architectures to strike a balance between performance metrics, computational complexity, and hardware implementation costs. To solve these challenges, future research can establish design methodologies that explicitly consider hardware implementation. This effort involves cross-disciplinary collaboration to integrate EM theory with practical manufacturing technologies.
\end{itemize}

%% file: main.bbl
\begin{thebibliography}{100}
\providecommand{\url}[1]{#1}
\csname url@samestyle\endcsname
\providecommand{\newblock}{\relax}
\providecommand{\bibinfo}[2]{#2}
\providecommand{\BIBentrySTDinterwordspacing}{\spaceskip=0pt\relax}
\providecommand{\BIBentryALTinterwordstretchfactor}{4}
\providecommand{\BIBentryALTinterwordspacing}{\spaceskip=\fontdimen2\font plus
\BIBentryALTinterwordstretchfactor\fontdimen3\font minus
  \fontdimen4\font\relax}
\providecommand{\BIBforeignlanguage}[2]{{%
\expandafter\ifx\csname l@#1\endcsname\relax
\typeout{** WARNING: IEEEtran.bst: No hyphenation pattern has been}%
\typeout{** loaded for the language `#1'. Using the pattern for}%
\typeout{** the default language instead.}%
\else
\language=\csname l@#1\endcsname
\fi
#2}}
\providecommand{\BIBdecl}{\relax}
\BIBdecl

\bibitem{clemm2020toward}
A.~Clemm, M.~T. Vega, H.~K. Ravuri, T.~Wauters, and F.~De~Turck, ``Toward truly
  immersive holographic-type communication: Challenges and solutions,''
  \emph{{IEEE} Commun. Mag.}, vol.~58, no.~1, pp. 93--99, 2020.

\bibitem{kanhere2021position}
O.~Kanhere and T.~S. Rappaport, ``Position location for futuristic cellular
  communications: {5G} and beyond,'' \emph{{IEEE} Commun. Mag.}, vol.~59,
  no.~1, pp. 70--75, 2021.

\bibitem{almasan2022network}
P.~Almasan, M.~Ferriol-Galm{\'e}s, J.~Paillisse, J.~Su{\'a}rez-Varela,
  D.~Perino, D.~L{\'o}pez, A.~A.~P. Perales, P.~Harvey, L.~Ciavaglia, L.~Wong
  \emph{et~al.}, ``Network digital twin: Context, enabling technologies, and
  opportunities,'' \emph{{IEEE} Commun. Mag.}, vol.~60, no.~11, pp. 22--27,
  2022.

\bibitem{gonzalez2024integrated}
N.~Gonz{\'a}lez-Prelcic, M.~F. Keskin, O.~Kaltiokallio, M.~Valkama, D.~Dardari,
  X.~Shen, Y.~Shen, M.~Bayraktar, and H.~Wymeersch, ``The integrated sensing
  and communication revolution for {6G}: Vision, techniques, and
  applications,'' \emph{Proc. {IEEE}}, vol. 112, no.~7, pp. 676--723, 2024.

\bibitem{letaief2021edge}
K.~B. Letaief, Y.~Shi, J.~Lu, and J.~Lu, ``Edge artificial intelligence for
  {6G}: Vision, enabling technologies, and applications,'' \emph{{IEEE} J. Sel.
  Areas Commun.}, vol.~40, no.~1, pp. 5--36, 2021.

\bibitem{d20246g}
C.~D’Andrea, J.~M. Jornet, A.~Singh, P.~Sen, Z.~Ghassemlooy, S.~Zvanovec,
  S.~R. Teli, A.~Gholami, C.~D’Andrea, J.~M. Jornet \emph{et~al.}, ``{6G}
  wireless technologies,'' in \emph{The Road towards 6G: Opportunities,
  Challenges, and Applications: A Comprehensive View of the Enabling
  Technologies}.\hskip 1em plus 0.5em minus 0.4em\relax Springer, 2024, pp.
  51--114.

\bibitem{ITU-R-M2160-0}
\BIBentryALTinterwordspacing
\emph{Framework and Overall Objectives of the Future Development of {IMT} for
  2030 and Beyond ({IMT-2030})}, International Telecommunication Union,
  Radiocommunication Sector (ITU-R) Std. M.2160-0, Nov. 2023. [Online].
  Available: \url{https://www.itu.int/rec/R-REC-M.2160-0-202311-I}
\BIBentrySTDinterwordspacing

\bibitem{bjornson2024towards}
E.~Bj{\"o}rnson, C.-B. Chae, R.~W. Heath~Jr, T.~L. Marzetta, A.~Mezghani,
  L.~Sanguinetti, F.~Rusek, M.~R. Castellanos, D.~Jun, and {\"O}.~T. Demir,
  ``Towards {6G} {MIMO}: Massive spatial multiplexing, dense arrays, and
  interplay between electromagnetics and processing,'' \emph{arXiv preprint
  arXiv:2401.02844}, 2024.

\bibitem{bjornson2015optimal}
E.~Bj{\"o}rnson, L.~Sanguinetti, J.~Hoydis, and M.~Debbah, ``Optimal design of
  energy-efficient multi-user {MIMO} systems: Is massive {MIMO} the answer?''
  \emph{{IEEE} Trans. Wireless Commun.}, vol.~14, no.~6, pp. 3059--3075, 2015.

\bibitem{shojaeifard2022mimo}
A.~Shojaeifard, K.-K. Wong, K.-F. Tong, Z.~Chu, A.~Mourad, A.~Haghighat,
  I.~Hemadeh, N.~T. Nguyen, V.~Tapio, and M.~Juntti, ``{MIMO} evolution beyond
  {5G} through reconfigurable intelligent surfaces and fluid antenna systems,''
  \emph{Proc. {IEEE}}, vol. 110, no.~9, pp. 1244--1265, 2022.

\bibitem{basar2019wireless}
E.~Basar, M.~Di~Renzo, J.~De~Rosny, M.~Debbah, M.-S. Alouini, and R.~Zhang,
  ``Wireless communications through reconfigurable intelligent surfaces,''
  \emph{{IEEE} Access}, vol.~7, pp. 116\,753--116\,773, 2019.

\bibitem{di2020smart}
M.~Di~Renzo, A.~Zappone, M.~Debbah, M.-S. Alouini, C.~Yuen, J.~De~Rosny, and
  S.~Tretyakov, ``Smart radio environments empowered by reconfigurable
  intelligent surfaces: How it works, state of research, and the road ahead,''
  \emph{{IEEE} J. Sel. Areas Commun.}, vol.~38, no.~11, pp. 2450--2525, 2020.

\bibitem{yang2016programmable}
H.~Yang, X.~Cao, F.~Yang, J.~Gao, S.~Xu, M.~Li, X.~Chen, Y.~Zhao, Y.~Zheng, and
  S.~Li, ``A programmable metasurface with dynamic polarization, scattering and
  focusing control,'' \emph{Sci. Rep.}, vol.~6, no.~1, p. 35692, 2016.

\bibitem{liang2024filtering}
J.~C. Liang, L.~Zhang, Z.~Luo, R.~Z. Jiang, Z.~W. Cheng, S.~R. Wang, M.~K. Sun,
  S.~Jin, Q.~Cheng, and T.~J. Cui, ``A filtering reconfigurable intelligent
  surface for interference-free wireless communications,'' \emph{Nat. Commun.},
  vol.~15, no.~1, p. 3838, 2024.

\bibitem{fu2020terahertz}
X.~Fu, F.~Yang, C.~Liu, X.~Wu, and T.~J. Cui, ``Terahertz beam steering
  technologies: from phased arrays to field-programmable metasurfaces,''
  \emph{Adv. Opt. Mater.}, vol.~8, no.~3, p. 1900628, 2020.

\bibitem{liu2021reconfigurable}
Y.~Liu, X.~Liu, X.~Mu, T.~Hou, J.~Xu, M.~Di~Renzo, and N.~Al-Dhahir,
  ``Reconfigurable intelligent surfaces: Principles and opportunities,''
  \emph{{IEEE} Commun. Surv. Tut.}, vol.~23, no.~3, pp. 1546--1577, 2021.

\bibitem{pan2021reconfigurable}
C.~Pan, H.~Ren, K.~Wang, J.~F. Kolb, M.~Elkashlan, M.~Chen, M.~Di~Renzo,
  Y.~Hao, J.~Wang, A.~L. Swindlehurst \emph{et~al.}, ``Reconfigurable
  intelligent surfaces for {6G} systems: Principles, applications, and research
  directions,'' \emph{{IEEE} Commun. Mag.}, vol.~59, no.~6, pp. 14--20, 2021.

\bibitem{huang2019reconfigurable}
C.~Huang, A.~Zappone, G.~C. Alexandropoulos, M.~Debbah, and C.~Yuen,
  ``Reconfigurable intelligent surfaces for energy efficiency in wireless
  communication,'' \emph{{IEEE} Trans. Wireless Commun.}, vol.~18, no.~8, pp.
  4157--4170, 2019.

\bibitem{9551980}
X.~Pei, H.~Yin, L.~Tan, L.~Cao, Z.~Li, K.~Wang, K.~Zhang, and E.~Björnson,
  \emph{{IEEE} Trans. Commun.}

\bibitem{10543050}
X.~Gan, C.~Huang, Z.~Yang, C.~Zhong, X.~Chen, Z.~Zhang, Q.~Guo, C.~Yuen, and
  M.~Debbah, ``Bayesian learning for double-{RIS} aided {ISAC} systems with
  superimposed pilots and data,'' \emph{{IEEE} J. Sel. Topics Signal Process.},
  vol.~18, no.~5, pp. 766--781, 2024.

\bibitem{9785633}
Y.~Xu, T.~Zhang, Y.~Liu, D.~Yang, L.~Xiao, and M.~Tao, ``Computation capacity
  enhancement by joint {UAV} and {RIS} design in {IoT},'' \emph{{IEEE} Internet
  Things J.}, vol.~9, no.~20, pp. 20\,590--20\,603, 2022.

\bibitem{liu2022star}
Y.~Liu, X.~Mu, J.~Xu, R.~Schober, Y.~Hao, H.~V. Poor, and L.~Hanzo, ``{STAR}:
  Simultaneous transmission and reflection for 360° coverage by intelligent
  surfaces,'' vol.~28, no.~6, pp. 102--109, 2022.

\bibitem{mu2021simultaneously}
X.~Mu, Y.~Liu, L.~Guo, J.~Lin, and R.~Schober, ``Simultaneously transmitting
  and reflecting ({STAR}) {RIS} aided wireless communications,'' \emph{{IEEE}
  Trans. Wireless Commun.}, vol.~21, no.~5, pp. 3083--3098, 2021.

\bibitem{ahmed2023survey}
M.~Ahmed, A.~Wahid, S.~S. Laique, W.~U. Khan, A.~Ihsan, F.~Xu, S.~Chatzinotas,
  and Z.~Han, ``A survey on {STAR-RIS}: Use cases, recent advances, and future
  research challenges,'' \emph{{IEEE} Internet Things J.}, vol.~10, no.~16, pp.
  14\,689--14\,711, 2023.

\bibitem{liu2025stacked}
H.~Liu, J.~An, X.~Jia, L.~Gan, G.~K. Karagiannidis, B.~Clerckx, M.~Bennis,
  M.~Debbah, and T.~J. Cui, ``Stacked intelligent metasurfaces for wireless
  communications: Applications and challenges,'' vol.~32, no.~4, pp. 46--53,
  2025.

\bibitem{an2023stacked}
J.~An, C.~Xu, D.~W.~K. Ng, G.~C. Alexandropoulos, C.~Huang, C.~Yuen, and
  L.~Hanzo, ``Stacked intelligent metasurfaces for efficient holographic {MIMO}
  communications in {6G},'' \emph{{IEEE} J. Sel. Areas Commun.}, vol.~41,
  no.~8, pp. 2380--2396, 2023.

\bibitem{hassan2024efficient}
N.~U. Hassan, J.~An, M.~Di~Renzo, M.~Debbah, and C.~Yuen, ``Efficient
  beamforming and radiation pattern control using stacked intelligent
  metasurfaces,'' \emph{{IEEE} Open J. Commun. Soc.}, vol.~5, pp. 599--611,
  2024.

\bibitem{an2024two}
J.~An, C.~Yuen, Y.~L. Guan, M.~Di~Renzo, M.~Debbah, H.~V. Poor, and L.~Hanzo,
  ``Two-dimensional direction-of-arrival estimation using stacked intelligent
  metasurfaces,'' \emph{{IEEE} J. Sel. Areas Commun.}, vol.~42, no.~10, pp.
  2786--2802, 2024.

\bibitem{zhang2022active}
Z.~Zhang, L.~Dai, X.~Chen, C.~Liu, F.~Yang, R.~Schober, and H.~V. Poor,
  ``Active {RIS} vs. passive {RIS}: Which will prevail in {6G}?'' \emph{{IEEE}
  Trans. Commun.}, vol.~71, no.~3, pp. 1707--1725, 2022.

\bibitem{zhi2022active}
K.~Zhi, C.~Pan, H.~Ren, K.~K. Chai, and M.~Elkashlan, ``Active {RIS} versus
  passive {RIS}: Which is superior with the same power budget?'' \emph{{IEEE}
  Commun. Lett.}, vol.~26, no.~5, pp. 1150--1154, 2022.

\bibitem{long2021active}
R.~Long, Y.-C. Liang, Y.~Pei, and E.~G. Larsson, ``Active reconfigurable
  intelligent surface-aided wireless communications,'' \emph{{IEEE} Trans.
  Wireless Commun.}, vol.~20, no.~8, pp. 4962--4975, 2021.

\bibitem{yang2023reconfigurable}
B.~Yang, X.~Cao, J.~Xu, C.~Huang, G.~C. Alexandropoulos, L.~Dai, M.~Debbah,
  H.~V. Poor, and C.~Yuen, ``Reconfigurable intelligent computational surfaces:
  When wave propagation control meets computing,'' vol.~30, no.~3, pp.
  120--128, 2023.

\bibitem{chepuri2023integrated}
S.~P. Chepuri, N.~Shlezinger, F.~Liu, G.~C. Alexandropoulos, S.~Buzzi, and
  Y.~C. Eldar, ``Integrated sensing and communications with reconfigurable
  intelligent surfaces: From signal modeling to processing,'' \emph{IEEE Signal
  Process. Mag.}, vol.~40, no.~6, pp. 41--62, 2023.

\bibitem{magbool2024multi}
A.~Magbool, V.~Kumar, A.~Bazzi, M.~F. Flanagan, and M.~Chafii,
  ``Multi-functional {RIS} for a multi-functional system: Integrating sensing,
  communication, and wireless power transfer,'' \emph{IEEE Netw.}, 2024.

\bibitem{li2024reconfigurable}
B.~Li, W.~Xie, and Z.~Fei, ``Reconfigurable intelligent surface for sensing,
  communication, and computation: Perspectives, challenges, and
  opportunities,'' \emph{arXiv preprint arXiv:2407.11402}, 2024.

\bibitem{chen2024metasurfaces}
W.~Chen, L.~Chen, Y.~Zhao, J.~Ren, and X.~Shen, ``Metasurfaces empowering {6G}
  communication and sensing: Opportunities and challenges,'' vol.~32, no.~1,
  pp. 158--164, 2024.

\bibitem{li2022intelligent}
L.~Li, H.~Zhao, C.~Liu, L.~Li, and T.~J. Cui, ``Intelligent metasurfaces:
  control, communication and computing,'' \emph{elight}, vol.~2, no.~1, p.~7,
  2022.

\bibitem{magbool2025survey}
A.~Magbool, V.~Kumar, Q.~Wu, M.~Di~Renzo, and M.~F. Flanagan, ``A survey on
  integrated sensing and communication with intelligent metasurfaces: Trends,
  challenges, and opportunities,'' \emph{{IEEE} Open J. Commun. Soc.}, 2025.

\bibitem{an2025emerging}
J.~An, M.~Debbah, T.~J. Cui, Z.~N. Chen, and C.~Yuen, ``Emerging technologies
  in intelligent metasurfaces: Shaping the future of wireless communications,''
  \emph{IEEE Trans. Antennas Propag.}, 2025.

\bibitem{kishk2020exploiting}
M.~A. Kishk and M.-S. Alouini, ``Exploiting randomly located blockages for
  large-scale deployment of intelligent surfaces,'' \emph{{IEEE} J. Sel. Areas
  Commun.}, vol.~39, no.~4, pp. 1043--1056, 2020.

\bibitem{singh2024performance}
A.~Singh, H.~B. Salameh, M.~Ayyash, and H.~Elgala, ``Performance analysis of
  {OIRS}-aided indoor {VLC} systems under dynamic human blockages and random
  {UE} orientation,'' \emph{{IEEE} Internet Things J.}, vol.~11, no.~20, pp.
  33\,110--33\,119, 2024.

\bibitem{dardari2021nlos}
D.~Dardari, N.~Decarli, A.~Guerra, and F.~Guidi, ``Los/nlos near-field
  localization with a large reconfigurable intelligent surface,'' \emph{{IEEE}
  Trans. Wireless Commun.}, vol.~21, no.~6, pp. 4282--4294, 2021.

\bibitem{fu2024multi}
M.~Fu, W.~Mei, and R.~Zhang, ``Multi-passive/active-{IRS} enhanced wireless
  coverage: Deployment optimization and cost-performance trade-off,''
  \emph{{IEEE} Trans. Wireless Commun.}, vol.~23, no.~8, pp. 9657--9671, 2024.

\bibitem{ben2024design}
G.~Ben-Itzhak, M.~Saavedra-Melo, B.~Bradshaw, E.~Ayanoglu, F.~Capolino, and
  A.~L. Swindlehurst, ``Design and operation principles of a wave-controlled
  reconfigurable intelligent surface,'' \emph{{IEEE} Open J. Commun. Soc.},
  2024.

\bibitem{singh2022noma}
S.~K. Singh, K.~Agrawal, K.~Singh, C.-P. Li, and Z.~Ding, ``{NOMA} enhanced
  hybrid {RIS-UAV}-assisted full-duplex communication system with imperfect
  {SIC} and {CSI},'' \emph{{IEEE} Trans. Commun.}, vol.~70, no.~11, pp.
  7609--7627, 2022.

\bibitem{chauhan2022ris}
A.~Chauhan, S.~Ghosh, and A.~Jaiswal, ``{RIS} partition-assisted non-orthogonal
  multiple access ({NOMA}) and quadrature-{NOMA} with imperfect {SIC},''
  \emph{{IEEE} Trans. Wireless Commun.}, vol.~22, no.~7, pp. 4371--4386, 2022.

\bibitem{wu2025intelligent}
X.~Wu, C.~Liu, L.~Zhao, and J.~Ma, ``Intelligent reflecting surface assisted
  {NOMA} integrated sensing, communication and computation systems,'' in
  \emph{2025 IEEE Wireless Commun. and Netw. Conf. (WCNC)}, 2025, pp. 1--6.

\bibitem{omam2025holographic}
Z.~R. Omam, H.~Taghvaee, A.~Araghi, M.~Garcia-Fernandez, G.~Alvarez-Narciandi,
  G.~C. Alexandropoulos, O.~Yurduseven, and M.~Khalily, ``Holographic
  metasurfaces enabling wave computing for {6G}: Status overview, challenges,
  and future research trends,'' \emph{arXiv preprint arXiv:2501.05173}, 2025.

\bibitem{shabanpour2024multifunctional}
J.~Shabanpour, ``Multifunctional intelligent metamaterial computing system:
  Independent parallel analog signal processing,'' \emph{Advanced Photonics
  Research}, vol.~5, no.~10, p. 2400002, 2024.

\bibitem{10685065}
F.~Yu, C.~Zhang, and T.~Q.~S. Quek, ``{STAR-RIS}-enabled simultaneous
  indoor-and-outdoor communication networks: A stochastic geometry approach,''
  \emph{{IEEE} Trans. Wireless Commun.}, vol.~23, no.~12, pp. 18\,053--18\,069,
  2024.

\bibitem{wu2021coverage}
C.~Wu, Y.~Liu, X.~Mu, X.~Gu, and O.~A. Dobre, ``Coverage characterization of
  {STAR-RIS} networks: {NOMA} and {OMA},'' \emph{{IEEE} Commun. Lett.},
  vol.~25, no.~9, pp. 3036--3040, 2021.

\bibitem{khel2023analytical}
A.~M.~T. Khel and K.~A. Hamdi, ``Analytical performance evaluation of
  {STAR-RIS} assisted terahertz wireless communications,'' \emph{{IEEE} Trans.
  Veh. Technol.}, vol.~73, no.~4, pp. 5500--5515, 2023.

\bibitem{li2024star}
J.~Li, Z.~Song, T.~Hou, C.~Huang, A.~Li, G.~Zhou, and Y.~Liu, ``{STAR-RIS}
  assisted {MISO-NOMA} networks: A simultaneous signal enhancement and
  interference mitigation design,'' \emph{{IEEE} Trans. Commun.}, 2024.

\bibitem{papazafeiropoulos2023cooperative}
A.~Papazafeiropoulos, A.~M. Elbir, P.~Kourtessis, I.~Krikidis, and
  S.~Chatzinotas, ``Cooperative {RIS} and {STAR-RIS} assisted {mMIMO}
  communication: Analysis and optimization,'' \emph{{IEEE} Trans. Veh.
  Technol.}, vol.~72, no.~9, pp. 11\,975--11\,989, 2023.

\bibitem{katwe2024spectrally}
M.~V. Katwe, R.~Deshpande, K.~Singh, C.~Pan, P.~H. Ghare, and T.~Q. Duong,
  ``Spectrally-efficient beamforming design for {STAR-RIS}-aided {URLLC} {NOMA}
  systems,'' \emph{{IEEE} Trans. Commun.}, vol.~72, no.~7, pp. 4414--4431,
  2024.

\bibitem{fang2022energy}
F.~Fang, B.~Wu, S.~Fu, Z.~Ding, and X.~Wang, ``Energy-efficient design of
  {STAR-RIS} aided {MIMO-NOMA} networks,'' \emph{{IEEE} Trans. Commun.},
  vol.~71, no.~1, pp. 498--511, 2022.

\bibitem{alishahi2025efficient}
M.~Alishahi, M.~Zeng, P.~Fortier, O.~Waqar, M.~Hanif, D.~T. Hoang, D.~N.
  Nguyen, and Q.-V. Pham, ``Efficient {STAR-RIS} mode for energy minimization
  in {WPT-FL} networks with {NOMA},'' \emph{{IEEE} Trans. Commun.}, 2025.

\bibitem{wen2024star}
Y.~Wen, G.~Chen, S.~Fang, Z.~Chu, P.~Xiao, and R.~Tafazolli,
  ``{STAR-RIS}-assisted-full-duplex jamming design for secure wireless
  communications system,'' \emph{{IEEE} Trans. Inf. Forensics Secur.}, vol.~19,
  pp. 4331--4343, 2024.

\bibitem{liang2025covert}
Y.~Liang, L.~Yang, I.~Ahmad, and M.~Valkama, ``Covert transmission and
  physical-layer security of {STAR-RIS}-assisted uplink{ SGF-NOMA} systems,''
  \emph{{IEEE} Trans. Commun.}, 2025.

\bibitem{lin2025star}
S.~Lin, Y.~Zou, Z.~Deng, K.~An, Z.~Zheng, L.~Yang, and P.~Yan,
  ``{STAR-RIS}-empowered monitoring over two-way suspicious communications,''
  \emph{{IEEE} Trans. Veh. Technol.}, 2025.

\bibitem{zhou2023robust}
T.~Zhou, K.~Xu, G.~Hu, X.~Xia, W.~Xie, and C.~Li, ``Robust beamforming design
  for {STAR-RIS}-assisted anti-jamming and secure transmission,'' \emph{{IEEE}
  Trans. Green Commun. Netw.}, vol.~8, no.~1, pp. 345--361, 2023.

\bibitem{pala2025robust}
S.~Pala, K.~Singh, O.~Taghizadeh, C.~Pan, O.~A. Dobre, and T.~Q. Duong,
  ``Robust and secure multi-user {STAR-RIS}-aided communications: Optimization
  vs machine learning,'' \emph{{IEEE} Trans. Commun.}, 2025.

\bibitem{xiao2024star}
H.~Xiao, X.~Hu, A.~Li, W.~Wang, Z.~Su, K.-K. Wong, and K.~Yang, ``{STAR-RIS}
  enhanced joint physical layer security and covert communications for
  multi-antenna mmwave systems,'' \emph{{IEEE} Trans. Wireless Commun.},
  vol.~23, no.~8, pp. 8805--8819, 2024.

\bibitem{xiao2025robust}
H.~Xiao, X.~Hu, A.~Li, W.~Wang, and K.~Yang, ``Robust full-space physical layer
  security for {STAR-RIS}-aided wireless networks: Eavesdropper with uncertain
  location and channel,'' \emph{{IEEE} Trans. Wireless Commun.}, 2025.

\bibitem{xie2023physical}
Z.~Xie, Y.~Liu, W.~Yi, X.~Wu, and A.~Nallanathan, ``Physical layer security for
  {STAR-RIS-NOMA}: A stochastic geometry approach,'' \emph{{IEEE} Trans.
  Wireless Commun.}, vol.~23, no.~6, pp. 6030--6044, 2023.

\bibitem{shang2024enhanced}
P.~Shang, M.~Huang, S.~Dang, J.~Li, X.~Wan, and X.~Chen, ``Enhanced security
  index modulation for {STAR-RIS} aided intelligent autonomous transport
  networks,'' \emph{{IEEE} Trans. Intell. Transp. Syst.}, 2024.

\bibitem{dong2024star}
X.~Dong, Z.~Fei, X.~Wang, M.~Hua, and Q.~Wu, ``{STAR-RIS} aided secure {MIMO}
  communication systems,'' \emph{{IEEE} Trans. Veh. Technol.}, vol.~73, no.~10,
  pp. 15\,715--15\,720, 2024.

\bibitem{ye2025physical}
R.~Ye, Y.~Peng, M.~Yue, and J.~Lee, ``Physical layer security for {IoT}
  application with assistance of active {STAR-RIS},'' \emph{{IEEE} Internet
  Things J.}, 2025.

\bibitem{song2025trustworthy}
M.~Song, Y.~Lin, J.~Wang, G.~Sun, C.~Dong, N.~Ma, D.~Niyato, and P.~Zhang,
  ``Trustworthy intelligent networks for low-altitude economy,'' \emph{{IEEE}
  Commun. Mag.}, 2025.

\bibitem{zhang2022joint}
Q.~Zhang, Y.~Zhao, H.~Li, S.~Hou, and Z.~Song, ``Joint optimization of
  {STAR-RIS} assisted {UAV} communication systems,'' \emph{{IEEE} Wireless
  Commun. Lett.}, vol.~11, no.~11, pp. 2390--2394, 2022.

\bibitem{zhao2022simultaneously}
J.~Zhao, Y.~Zhu, X.~Mu, K.~Cai, Y.~Liu, and L.~Hanzo, ``Simultaneously
  transmitting and reflecting reconfigurable intelligent surface ({STAR-RIS})
  assisted {UAV} communications,'' \emph{{IEEE} J. Sel. Areas Commun.},
  vol.~40, no.~10, pp. 3041--3056, 2022.

\bibitem{lei2023noma}
J.~Lei, T.~Zhang, X.~Mu, and Y.~Liu, ``{NOMA} for {STAR-RIS} assisted {UAV}
  networks,'' \emph{{IEEE} Trans. Commun.}, vol.~72, no.~3, pp. 1732--1745,
  2023.

\bibitem{deng2025energy}
X.~Deng, P.~Yang, H.~Lin, L.~Wang, S.~Lin, J.~Gui, X.~Chen, Y.~Qian, and
  J.~Zhang, ``Energy-efficient strategic {UAV}-enabled {MEC} networks via
  {STAR-RIS}: Joint optimization of trajectory and user association,''
  \emph{{IEEE} Internet Things J.}, 2025.

\bibitem{li2024energy}
J.~Li, Y.~Huang, J.~Wu, X.~Wang, and Z.~Ning, ``Energy efficiency maximization
  for {STAR-RIS} and {UAV}-assisted {IUA}: A multi-agent {DRL} approach,''
  \emph{{IEEE} Internet Things J.}, 2024.

\bibitem{zhao2025aerial}
J.~Zhao, Q.~Xu, X.~Mu, Y.~Liu, and Y.~Zhu, ``Aerial active {STAR-RIS}-aided
  {IoT} {NOMA} networks,'' \emph{{IEEE} Internet Things J.}, 2025.

\bibitem{guo2023secure}
L.~Guo, J.~Jia, J.~Chen, and X.~Wang, ``Secure communication optimization in
  {NOMA} systems with {UAV}-mounted {STAR-RIS},'' \emph{{IEEE} Trans. Inf.
  Forensics Secur.}, vol.~19, pp. 2300--2314, 2023.

\bibitem{lukito2024integrated}
W.~D. Lukito, W.~Xiang, P.~Lai, P.~Cheng, C.~Liu, K.~Yu, and X.~Zhu,
  ``Integrated {STAR-RIS} and {UAV} for satellite {IoT} communications: An
  energy-efficient approach,'' \emph{{IEEE} Internet Things J.}, 2024.

\bibitem{khan2025efficient}
N.~Khan, A.~Ahmad, A.~Alwarafy, M.~A. Shah, A.~Lakas, and M.~M. Azeem,
  ``Efficient resource allocation and {UAV} deployment in {STAR-RIS} and
  {UAV}-relay assisted public safety networks for video transmission,''
  \emph{{IEEE} Open J. Commun. Soc.}, 2025.

\bibitem{zhou2024near}
J.~Zhou, Y.~Yang, Z.~Yang, and M.~R. Shikh-Bahaei, ``Near-field extremely
  large-scale star-{RIS} enabled integrated sensing and communications,''
  \emph{{IEEE} Trans. Green Commun. Netw.}, vol.~9, no.~1, pp. 404--416,
  2024.

\bibitem{cai2025energy}
Z.~Cai, M.~Sheng, J.~Liu, J.~Liu, and J.~Li, ``Energy-efficient beamforming for
  near-field {STAR-RIS}-enhanced low-altitude {ISAC} networks,'' \emph{{IEEE}
  Trans. Cogn. Commun. Netw.}, 2025.

\bibitem{xie2024star}
K.~Xie, G.~Cai, J.~He, and G.~Kaddoum, ``{STAR-RIS} aided {MISO} {SWIPT-NOMA}
  system with energy buffer: Performance analysis and optimization,''
  \emph{{IEEE} Internet Things J.}, 2024.

\bibitem{yaswanth2024toward}
J.~Yaswanth, M.~Katwe, K.~Singh, O.~Taghizadeh, C.~Pan, and A.~Schmeink,
  ``Toward green communication: Power-efficient beamforming for
  {STAR-RIS}-aided {SWIPT},'' \emph{{IEEE} Trans. Green Commun. Netw.},
  vol.~8, no.~4, pp. 1545--1563, 2024.

\bibitem{zhu2023robust}
G.~Zhu, X.~Mu, L.~Guo, A.~Huang, and S.~Xu, ``Robust resource allocation for
  {STAR-RIS} assisted {SWIPT} systems,'' \emph{{IEEE} Trans. Wireless Commun.},
  vol.~23, no.~6, pp. 5616--5631, 2023.

\bibitem{luo2024robust}
C.~Luo, W.~Jiang, J.~Nie, D.~Niyato, C.~Pan, Z.~Xiong, and P.~Xiong, ``Robust
  beamforming design for the {STAR-RIS}-assisted secure {SWIPT} system,''
  \emph{{IEEE} Trans. Veh. Technol.}, 2024.

\bibitem{chen2025star}
H.~Chen, L.~Xiao, Y.~Xu, D.~Yang, X.~Mu, and T.~Zhang, ``{STAR-RIS} assisted
  hybrid far-and near-field {SWIPT},'' \emph{{IEEE} Trans. Veh. Technol.},
  2025.

\bibitem{xie2025simultaneous}
W.~Xie, L.~Qin, J.~Wang, W.~Wu, X.~Li, H.~Xu, and L.~Yang, ``Simultaneous
  wireless information and power transfer for {STAR-RIS} assisted {UAV}
  networks,'' \emph{{IEEE} Internet Things J.}, 2025.

\bibitem{zhu2025star}
G.~Zhu, X.~Mu, L.~Guo, A.~Huang, and S.~Xu, ``{STAR-RIS} assisted {SWIPT}
  systems: Active or passive?'' \emph{{IEEE} Trans. Wireless Commun.}, 2025.

\bibitem{yang2024joint}
J.~Yang, X.~Qin, Z.~Jia, and L.~Mao, ``Joint optimization of maximum achievable
  rate in {SWIPT} systems assisted by active {STAR-RIS},'' in \emph{"Proc.
  {IEEE} Int. Conf. Wireless Artif. Intell. Comput. Syst. Appl.
  ({WASA})"}.\hskip 1em plus 0.5em minus 0.4em\relax Springer, 2024, pp.
  387--399.

\bibitem{gao2024power}
C.~Gao, S.~Li, Y.~Wu, S.~Duan, M.~Wei, and B.~Yu, ``Power-efficient resource
  allocation for active {STAR-RIS}-aided {SWIPT} communication systems.''
  \emph{"Future Internet"}, vol.~16, no.~8, 2024.

\bibitem{faramarzi2025energy}
S.~Faramarzi, H.~Zarini, S.~Javadi, M.~R. Mili, R.~Zhang, G.~K. Karagiannidis,
  and N.~Al-Dhahir, ``Energy efficient design of active {STAR-RIS}-aided
  {SWIPT} systems,'' \emph{{IEEE} Trans. Wireless Commun.}, 2025.

\bibitem{buzzi2021radar}
S.~Buzzi, E.~Grossi, M.~Lops, and L.~Venturino, ``Radar target detection aided
  by reconfigurable intelligent surfaces,'' \emph{{IEEE} Signal Process.
  Lett.}, vol.~28, pp. 1315--1319, 2021.

\bibitem{buzzi2022foundations}
------, ``Foundations of {MIMO} radar detection aided by reconfigurable
  intelligent surfaces,'' \emph{{IEEE} Trans. Signal Process.}, vol.~70, pp.
  1749--1763, 2022.

\bibitem{chen2023doa}
H.~Chen, Y.~Bai, Q.~Wang, H.~Chen, L.~Tang, and P.~Han, ``{DOA} estimation
  assisted by reconfigurable intelligent surfaces,'' \emph{IEEE Sens. J.},
  vol.~23, no.~12, pp. 13\,433--13\,442, 2023.

\bibitem{liu2025ris}
Y.~Liu, K.~M. Attiah, and W.~Yu, ``{RIS}-assisted joint sensing and
  communications via fractionally constrained fractional programming,''
  \emph{{IEEE} Trans. Wireless Commun.}, 2025.

\bibitem{wang2023target}
P.~Wang, W.~Mei, J.~Fang, and R.~Zhang, ``Target-mounted intelligent reflecting
  surface for joint location and orientation estimation,'' \emph{{IEEE} J. Sel.
  Areas Commun.}, vol.~41, no.~12, pp. 3768--3782, 2023.

\bibitem{chen2025integrated}
X.~Q. Chen, L.~Zhang, Y.~N. Zheng, S.~Liu, Z.~R. Huang, J.~C. Liang,
  M.~Di~Renzo, V.~Galdi, and T.~J. Cui, ``Integrated sensing and communication
  based on space-time-coding metasurfaces,'' \emph{Nat. Commun.}, vol.~16,
  no.~1, p. 1836, 2025.

\bibitem{wang2024ris}
Z.~Wang, X.~Hu, C.~Liu, and M.~Peng, ``{RIS}-enabled multi-target sensing:
  Performance analysis and space-time beamforming design,'' \emph{{IEEE} Trans.
  Wireless Commun.}, vol.~23, no.~10, pp. 13\,889--13\,903, 2024.

\bibitem{gan2025modeling}
X.~Gan, C.~Huang, Z.~Yang, X.~Chen, F.~Bader, Z.~Zhang, C.~Yuen, Y.~L. Guan,
  and M.~Debbah, ``Modeling and coverage analysis of {RIS}-assisted integrated
  sensing and communication networks,'' \emph{{IEEE} Trans. Wireless Commun.},
  2025.

\bibitem{li2024joint}
J.~Li, G.~Zhou, T.~Gong, N.~Liu, and R.~Zhang, ``Joint active and passive
  beamforming design for {IRS}-aided {MIMO} {ISAC} based on sensing mutual
  information,'' \emph{arXiv preprint arXiv:2407.16543}, 2024.

\bibitem{jin2023ris}
X.~Jin, P.~Zhang, C.~Wan, D.~Ma, and Y.~Yao, ``{RIS} assisted dual-function
  radar and secure communications based on frequency-shifted chirp spread
  spectrum index modulation,'' \emph{China Commun.}, vol.~20, no.~10, pp.
  85--99, 2023.

\bibitem{zou2025target}
H.~Zou, L.~Wu, and Z.~Zhang, ``Target detection for reconfigurable intelligent
  surface assisted {MIMO} radar,'' \emph{{IEEE} Trans. Veh. Technol.}, 2025.

\bibitem{zhang2024target}
X.~Zhang, H.~Zhang, L.~Liu, Z.~Han, H.~V. Poor, and B.~Di, ``Target detection
  and positioning aided by reconfigurable surfaces: Reflective or
  holographic?'' \emph{{IEEE} Trans. Wireless Commun.}, 2024.

\bibitem{wang2025near}
Z.~Wang, P.~Ramezani, Y.~Liu, and E.~Bj{\"o}rnson, ``Near-field localization
  and sensing with large-aperture arrays: From signal modeling to processing,''
  \emph{{IEEE} Signal Process. Mag.}, vol.~42, no.~1, pp. 74--87, 2025.

\bibitem{bjornson2020power}
E.~Bj{\"o}rnson and L.~Sanguinetti, ``Power scaling laws and near-field
  behaviors of massive {MIMO} and intelligent reflecting surfaces,''
  \emph{{IEEE} Open J. Commun. Soc.}, vol.~1, pp. 1306--1324, 2020.

\bibitem{liu2025reconfigurable}
Z.~Liu, B.~Zheng, Q.~Wu, C.~You, W.~Mei, K.~Chen, and J.~Tang, ``Reconfigurable
  intelligent surface-aided near-field localization with array spatial
  correlation,'' \emph{{IEEE} Wireless Commun. Lett.}, 2025.

\bibitem{yuan2024near}
Y.~Yuan, Y.~Chen, X.~Guo, and Y.~Wang, ``Near-field tracking with extremely
  large-scale {RIS}: A sparse learning approach,'' in \emph{Proc. IEEE Wirel.
  Commun. Netw. Conf. (WCNC)}, 2024, pp. 1--6.

\bibitem{zhu2025hybrid}
W.~Zhu, M.~Cao, Y.~Yang, H.~Zhang, W.~Hao, X.~Jia, and H.~Zhang, ``Hybrid
  near-field and far-field localization with multiple reconfigurable
  intelligent surfaces,'' \emph{{IEEE} Commun. Lett.}, 2025.

\bibitem{li2023sensing}
Z.~Li, Z.~Gao, and T.~Li, ``Sensing user's channel and location with terahertz
  extra-large reconfigurable intelligent surface under hybrid-field beam squint
  effect,'' \emph{{IEEE} J. Sel. Topics Signal Process.}, vol.~17, no.~4, pp.
  893--911, 2023.

\bibitem{wu2025hybrid}
Y.~Wu, H.~Chu, H.~Zhou, and X.~Ma, ``A hybrid near-far field channel model for
  {RIS}-assisted healthcare monitoring and precision positioning in internet of
  everything,'' \emph{{IEEE} Internet Things J.}, 2025.

\bibitem{yoo2025ris}
S.~Yoo, J.~Jung, S.~Jeong, J.~Kang, M.~Juntti, and J.~Kang, ``{RIS}-assisted
  {ISAC} systems for industrial revolution 6.0: Exploring the near-field and
  far-field coexistence,'' \emph{arXiv preprint arXiv:2507.07643}, 2025.

\bibitem{kong2024signal}
L.~Kong, X.~Pang, J.~Tang, N.~Zhao, X.~Wang, and N.~Al-Dhahir, ``Signal
  enhancement and suppression schemes for bi-static {ISAC} with {IRS}-mounted
  target,'' \emph{{IEEE} Trans. Commun.}, 2024.

\bibitem{zhang2024intelligent}
Z.~Zhang, W.~Chen, Q.~Wu, Z.~Li, X.~Zhu, and J.~Yuan, ``Intelligent omni
  surfaces assisted integrated multi-target sensing and multi-user {MIMO}
  communications,'' \emph{{IEEE} Trans. Commun.}, vol.~72, no.~8, pp.
  4591--4606, 2024.

\bibitem{yang2024ris}
X.~Yang, Z.~Wei, Y.~Liu, H.~Wu, and Z.~Feng, ``{RIS}-assisted cooperative
  multicell {ISAC} systems: A multi-user and multi-target case,'' \emph{{IEEE}
  Trans. Wireless Commun.}, vol.~23, no.~8, pp. 8683--8699, 2024.

\bibitem{salem2024integrated}
A.~A. Salem, M.~A. Albreem, K.~Alnajjar, S.~Abdallah, and M.~Saad, ``Integrated
  cooperative sensing and communication for {RIS}-enabled full-duplex cell-free
  {MIMO} systems,'' \emph{{IEEE} Trans. Commun.}, 2024.

\bibitem{sankar2023beamforming}
R.~P. Sankar, S.~P. Chepuri, and Y.~C. Eldar, ``Beamforming in integrated
  sensing and communication systems with reconfigurable intelligent surfaces,''
  \emph{{IEEE} Trans. Wireless Commun.}, vol.~23, no.~5, pp. 4017--4031, 2023.

\bibitem{liaquat2025improving}
S.~Liaquat, I.~H. Naqvi, F.~A. Butt, S.~Alawsh, N.~M. Mahyuddin, and A.~H.
  Muqaibel, ``Improving {SNR} for nlos target detection using
  multi-ris-assisted monostatic radar,'' \emph{{IEEE} Open J. Commun. Soc.},
  2025.

\bibitem{li2025multi2}
Y.~Li, X.~Long, M.~Pauli, S.~Tian, X.~Wan, B.~Nuss, T.~Cui, H.~Zhang, and
  T.~Zwick, ``Multi-{RIS} deployment optimization for mmwave {ISAC} systems in
  real-world environments,'' \emph{arXiv preprint arXiv:2508.07226}, 2025.

\bibitem{saigre2023self}
C.~Saigre-Tardif and P.~del Hougne, ``Self-adaptive {RISs} beyond free space:
  convergence of localization, sensing, and communication under rich-scattering
  conditions,'' vol.~30, no.~1, pp. 24--30, 2023.

\bibitem{wang2022location}
Z.~Wang, Z.~Liu, Y.~Shen, A.~Conti, and M.~Z. Win, ``Location awareness in
  beyond {5G} networks via reconfigurable intelligent surfaces,'' \emph{{IEEE}
  J. Sel. Areas Commun.}, vol.~40, no.~7, pp. 2011--2025, 2022.

\bibitem{pang2023cellular}
X.~Pang, W.~Mei, N.~Zhao, and R.~Zhang, ``Cellular sensing via cooperative
  intelligent reflecting surfaces,'' \emph{{IEEE} Trans. Veh. Technol.},
  vol.~72, no.~11, pp. 15\,086--15\,091, 2023.

\bibitem{keykhosravi2021semi}
K.~Keykhosravi, M.~F. Keskin, S.~Dwivedi, G.~Seco-Granados, and H.~Wymeersch,
  ``Semi-passive {3D} positioning of multiple {RIS}-enabled users,''
  \emph{{IEEE} Trans. Veh. Technol.}, vol.~70, no.~10, pp. 11\,073--11\,077,
  2021.

\bibitem{li2025multi}
X.~Li, Q.~Zhu, Y.~Chen, and Y.~Yuan, ``Multi-{BS} cooperative and coordinated
  {ISAC} mechanism with assistance of {STAR-RIS},'' \emph{{IEEE} Commun.
  Lett.}, 2025.

\bibitem{jabbar2025millimeter}
A.~Jabbar, N.~Azam, P.~Ishabakaki, M.~Ur-Rehman, M.~A. Imran, M.~Sevegnani,
  H.~Larijani, S.~A. Shah, Q.~Abbasi, and M.~Usman, ``Millimeter-wave
  contactless vital sign monitoring using dynamic metasurface antenna,''
  \emph{IEEE Antennas Wirel. Propag. Lett.}, 2025.

\bibitem{lan2021metasense}
G.~Lan, M.~F. Imani, Z.~Liu, J.~Manjarr{\'e}s, W.~Hu, A.~S. Lan, D.~R. Smith,
  and M.~Gorlatova, ``Metasense: Boosting {RF} sensing accuracy using dynamic
  metasurface antenna,'' \emph{{IEEE} Internet Things J.}, vol.~8, no.~18, pp.
  14\,110--14\,126, 2021.

\bibitem{xue2025chip}
S.~Xue, Y.~Shen, J.~Guo, Z.~Xu, Y.~Ding, and S.~Hu, ``Chip-integrated
  millimeter-wave imaging system via a low-profile and polarization-multiplexed
  holographic tensor metasurface,'' \emph{AOM}, p. 2500062, 2025.

\bibitem{wang2025high}
Y.~Wang, Y.~Lu, Y.~Zhou, Y.~Shen, L.~Qiu, Z.~Lai, Y.-C. Chen, H.~Pan, J.~Zhou,
  D.~Ding \emph{et~al.}, ``High-resolution mmwave imaging using metasurface and
  diffusion,'' in \emph{Proc. ACM MobiSys}, 2025, pp. 319--332.

\bibitem{beruete2020terahertz}
M.~Beruete and I.~J{\'a}uregui-L{\'o}pez, ``Terahertz sensing based on
  metasurfaces,'' \emph{AOM}, vol.~8, no.~3, p. 1900721, 2020.

\bibitem{jadeja2023detection}
R.~Jadeja, J.~Surve, T.~Parmar, S.~K. Patel, and F.~A. Al-Zahrani, ``Detection
  of peptides employing a {THz} metasurface based sensor,'' \emph{Diam. Relat.
  Mater.}, vol. 132, p. 109675, 2023.

\bibitem{wang2020properties}
Y.~Wang, D.~Zhu, Z.~Cui, L.~Yue, X.~Zhang, L.~Hou, K.~Zhang, and H.~Hu,
  ``Properties and sensing performance of all-dielectric metasurface {THz}
  absorbers,'' \emph{IEEE Trans. Terahertz Sci. Technol.}, vol.~10, no.~6, pp.
  599--605, 2020.

\bibitem{di2025state}
M.~Di~Renzo, ``State of the art on stacked intelligent metasurfaces
  communication, sensing and computing in the wave domain,'' in
  \emph{Proceedings of the 19th European Conference on Antennas and Propagation
  (EuCAP)}, 2025, pp. 1--3.

\bibitem{an2025stacked}
J.~An, M.~Di~Renzo, M.~Debbah, H.~V. Poor, and C.~Yuen, ``Stacked intelligent
  metasurfaces for multiuser downlink beamforming in the wave domain,''
  \emph{{IEEE} Trans. Wireless Commun.}, 2025.

\bibitem{li2024stacked}
Z.~Li, J.~An, and C.~Yuen, ``Stacked intelligent metasurfaces for fully-analog
  wideband beamforming design,'' in \emph{2024 IEEE VTS Asia Pacific Wireless
  Communications Symposium (APWCS)}, 2024, pp. 1--5.

\bibitem{sun2022energy}
Y.~Sun, K.~An, Y.~Zhu, G.~Zheng, K.-K. Wong, S.~Chatzinotas, D.~W.~K. Ng, and
  D.~Guan, ``Energy-efficient hybrid beamforming for multilayer {RIS}-assisted
  secure integrated terrestrial-aerial networks,'' \emph{{IEEE} Trans.
  Commun.}, vol.~70, no.~6, pp. 4189--4210, 2022.

\bibitem{li2024transmit}
S.~Li, F.~Zhang, T.~Mao, R.~Na, Z.~Wang, and G.~K. Karagiannidis, ``Transmit
  beamforming design for {ISAC} with stacked intelligent metasurfaces,''
  \emph{{IEEE} Trans. Veh. Technol.}, 2024.

\bibitem{huang2025stacked}
G.~Huang, J.~An, L.~Gan, D.~Niyato, M.~Debbah, and T.~J. Cui, ``Stacked
  intelligent metasurfaces for multi-modal semantic communications,''
  \emph{arXiv preprint arXiv:2506.12368}, 2025.

\bibitem{lin2025uav}
S.~Lin, J.~An, L.~Gan, and M.~Debbah, ``{UAV}-mounted {SIM}: A hybrid
  optical-electronic neural network for {DoA} estimation,'' in \emph{Proc.
  {IEEE} Int. Conf. Acoust. Speech Signal Process. (ICASSP)}, 2025, pp. 1--5.

\bibitem{stylianopoulos2025integrating}
K.~Stylianopoulos and G.~C. Alexandropoulos, ``Integrating stacked intelligent
  metasurfaces and power control for dynamic edge inference via over-the-air
  neural networks,'' \emph{arXiv preprint arXiv:2509.18906}, 2025.

\bibitem{zhang2024radio}
J.~Zhang, H.~Chen, and D.~M. Blough, ``A radio-frequency-based 2-d
  convolutional layer using transmissive intelligent surfaces,'' in \emph{2024
  IEEE 100th Veh. Technol. Conf. (VTC2024-Fall)}, 2024, pp. 1--7.

\end{thebibliography}
